\documentclass{siamart220329}
\usepackage[utf8]{inputenc}
\usepackage{amsfonts}
\usepackage{amsmath}
\usepackage{amsbsy}
\usepackage{amssymb}
\usepackage{mathtools}
\usepackage{subfig}
\usepackage{arydshln}
\usepackage{algorithm,algpseudocode}
\usepackage[mathscr]{euscript}
\usepackage{verbatim}
\usepackage{xspace}
\usepackage{graphicx}

\newtheorem{remark}{Remark}

\newcommand{\mc}[1]{\mathcal{#1}}

\newcommand{\mb}[1]{\mathbb{#1}}

\newcommand{\Wker}{\bar{\mc{W}}} 
\newcommand{\Wxbar}[1]{\bar{\mc{W}}^{x}_{#1}} 

\newcommand{\eventually}{\Diamond}
\newcommand{\until}{\mathbin{\sf U}}

\newcommand{\word}{\pmb{\pi} }
\newcommand{\letter}{\pi}
\newcommand{\AP}{\mathrm{AP}}

\newcommand{\N}{\mathbb{N}}

\newcommand{\M}{\mathbf{M}}
\newcommand{\X}{\mathbb{X}}
\newcommand{\U}{\mathbb{U}}
\newcommand{\Y}{\mathbb{Y}}
\newcommand{\W}{\mathbb{W}}

\newcommand{\Mh}{\mathbf{\hat{M}}}
\newcommand{\Xh}{\hat{\mathbb{X}}}
\newcommand{\Uh}{\hat{\mathbb{U}}}

\newcommand{\y}{\mathbf{y}}

\newcommand{\x}{\mathbf{x}}

\newcommand{\C}{\mathbf{C}}
\newcommand{\Ch}{\mathbf{\hat{C}}}
\newcommand{\policy}{\boldsymbol{\mu}}

\newcommand{\idf}{\operatorname{idf}}

\newcommand{\DFA}{\mathcal A}

\newcommand{\Bel}{{\mathbf {T}}}

\DeclareMathOperator*{\argmax}{arg\max}

\newcommand{\Mp}{\M_w}
\newcommand{\Rp}{\mc{E}}
\newcommand{\Rm}{\boldsymbol{\mathscr{R}}} 
\newcommand{\Ri}{\mathscr{R}} 
\newcommand{\spg}{s_{fb}}
\newcommand{\sgp}{s_{bf}}
\newcommand{\Xp}{\X_{w}}
\newcommand{\Vp}[1]{V_{{x_w}#1}}
\newcommand{\Vh}[1]{V_{{\hat{x}}#1}}

\newcommand{\syscore}{\texttt{SySCoRe}\xspace}

\newcommand{\sreachtools}{\texttt{SReachTools}\xspace}

\newcommand{\Tr}{\mathbf{t}}

\newcommand{\Birgit}[1]{{\color{blue}#1}}

\newcommand{\new}[1]{{\color{orange}#1}}

\newcommand{\newrr}[1]{{\color{red}#1}}

\definecolor{grayfilling}{gray}{0.95} 
\definecolor{royalblue}{rgb}{0.25, 0.41, 0.88}
\definecolor{lightred}{RGB}{255, 204, 204}
\definecolor{darkred}{RGB}{200, 24, 26}
\definecolor{lightblue}{RGB}{204, 205, 255}
\definecolor{darklavender}{rgb}{0.45, 0.31, 0.59}
\definecolor{lavenderindigo}{rgb}{0.58, 0.34, 0.92}

\usepackage{tikz}
\usetikzlibrary{arrows.meta}
\usetikzlibrary{positioning,calc,patterns,decorations}
\usetikzlibrary{arrows,automata}

\title{Specification-guided temporal logic control for stochastic systems: a multi-layered approach} 

\author{Birgit C.~van Huijgevoort, Ruohan Wang,\\
Sadegh Soudjani and Sofie Haesaert\thanks{This work is supported by
		the Dutch NWO Veni project CODEC (18244) and the Horizon Europe EIC project SymAware (101070802). R. Wang and S. Haesaert are with the Department of Electrical Engineering, TU Eindhoven, The Netherlands. 
		B. C.~van Huijgevoort and S. Soudjani are currently with the Max Planck Institute for software systems. Part of the research was done when B. C.~van Huijgevoort was with TU Eindhoven, The Netherlands. Emails:
		{\tt\small \{r.wang2,s.Haesaert\}@tue.nl, \{bhuijgevoort,sadegh\}@mpi-sws.org.} A subset of the results of this manuscript has been presented as a conference paper entitled ``Multi-layered simulation relations for linear stochastic systems" published by the IEEE 2021 European Control Conference (ECC).
	}%
}
\begin{document}
	\graphicspath{{Figures/}}
\maketitle

\begin{abstract}
Designing controllers to satisfy temporal requirements has proven to be challenging for dynamical systems that are affected by uncertainty.
This is mainly due to the states evolving in a continuous uncountable space, the stochastic evolution of the states, and infinite-horizon temporal requirements on the system evolution, all of which makes closed-form solutions generally inaccessible.
A promising approach for designing provably correct controllers on such systems is to utilize the concept of \emph{abstraction}, which is based on building simplified abstract models that can be used to approximate optimal controllers with provable closeness guarantees. The available abstraction-based methods are further divided into \emph{discretization-based} approaches that build a finite abstract model by discretizing the continuous space of the system, and \emph{discretization-free} approaches that work directly on the continuous state space without the need for building a finite space.

To reduce the conservatism in the sub-optimality of the designed controller originating from the abstraction step, this paper develops an approach that naturally has the flexibility to combine different abstraction techniques from the aforementioned classes and to combine the same abstraction technique with different parameters.
	First, we develop a \emph{multi-layered} discretization-based approach with variable precision by combining abstraction layers with different precision parameters. Then, we exploit the advantages of both classes of abstraction-based methods by extending this multi-layered approach guided by the specification to combinations of layers with respectively discretization-based and discretization-free abstractions. We achieve an efficient implementation that is less conservative and improves the computation time and memory usage. We illustrate the benefits of the proposed multi-layered approach on several case studies.
	%
	
\end{abstract}

\section{Introduction}
\label{sec:intro} 
Stochastic difference equations are often used to model the behavior of complex systems that are operating in an uncertain environment, such as autonomous vehicles, airplanes, and drones. In this work, we are interested in automatically designing controllers with which we can give guarantees on the functionality of the controlled stochastic systems with respect to temporal logic specifications. Such automated control synthesis is often referred to as \emph{correct-by-design} control synthesis \cite{tabuada2009verification,belta2017formal}. 

The techniques in correct-by-design synthesis are divided into two classes: methods that construct a finite-state abstraction of the original continuous-state model by discretizing the state space \cite{belta2017formal,lavaei2022automated}, and methods that do not rely on a space discretization \cite{vinod2019sreachtools,jagtap2020formal}. The former is referred to as \emph{discretization-based} techniques and the latter as \emph{discretization-free} techniques. Selection of a synthesis technique from these classes depends on multiple factors including the complexity of the dynamics, the distribution of the uncertainty, the desired specification, and the labeling function defined on the state space of the system. Any of these techniques also have their own hyper-parameters that result in different computational complexities and different levels of conservatisms in the computed probability of satisfying the satisfaction. These hyper-parameters include among others the diameter of the discretization grid \cite{SA13}, confidence values for sampling-based computations \cite{vinod2019sreachtools}, and the number of basis functions \cite{jagtap2020formal}. A comparison between the current tools handling correct-by-design synthesis of stochastic systems applied to a set of standard benchmarks can be found in the ARCH competition reports \cite{abate2020arch,abate2021arch,abate2022arch}.

\begin{figure}[b!]
	\centering
	\includegraphics[width=0.25\textwidth]{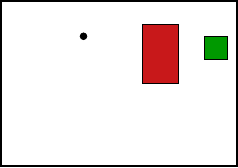}
	\caption{Example of an output space for a reach-avoid specification. The red area is the area to avoid, the green area is the goal region, and the black dot is the initial state $x(0)$.}
	\label{fig:exampleCase}
\end{figure}

As an example, consider a reach-avoid specification as illustrated in Fig.~\ref{fig:exampleCase}, where the objective is to avoid the red area and reach the green area from an arbitrary initial point $x(0)$ in the configuration space expressed in LTL as $\lnot RED \until GREEN$. We are interested in synthesizing a controller, or even simpler, planning a path that satisfies this specification subject to the dynamics of a given system. Since the dynamics of the system are influenced by uncertainty, the objective is to maximize the probability of satisfying this specification, referred to as the \emph{satisfaction probability}. In certain circumstances, the satisfaction probability can be computed accurately by constructing a nominal path surrounded by a tube containing possible paths with a high probability, as is common with discretization-free approaches \cite{vinod2019sreachtools}. Such circumstances are for example 
1) the distribution associated with the uncertainty has a small standard deviation, or
2) a large state space with a small avoid region and a large goal region that is far away from the avoid region.
The accuracy of the approximated satisfaction probability can be improved by extending the usage of a single nominal path to sampling multiple paths in the state space. Still, such sampling-based approaches are only suitable for the circumstances mentioned before, and lack the accuracy to handle more challenging scenarios. 


Discretization-based approaches such as \cite{cauchi2019stochy,cauchi2019efficiency,soudjani2015fau} overcome the aforementioned limitation by (often manually)
refining the grid until the accuracy of the satisfaction probability is sufficient. However, in practice, they are limited by the available memory and computation time. Put differently, when the available memory is limited, large-scale systems pose enormous challenges for discretization-based approaches, and discretization-free approaches are the only option.
In this paper, we are interested in integrating these techniques and developing a \emph{specification-guided} approach that naturally switches between and combines different techniques. More specifically, the approach employs a discretization-free technique whenever it gives acceptable accuracy and uses a discretization-based technique only when necessary, thereby effectively integrating the techniques to compute a single approximate satisfaction probability. The choice of the technique is determined by the specification, hence we refer to our approach as \emph{specification-guided}.

Integrating both abstraction techniques while maintaining the theoretical guarantees is challenging and involves multiple alterations to the current results. A major challenge is the fact that these techniques often have different precisions. Therefore, we initially leave the integration of different abstraction techniques aside and focus on integrating discretization-based techniques that are utilizing approximate simulation relations with different precisions. More specifically, we extend upon the preliminary results in \cite{huijgevoort2022similarity}  and switch between different layers each containing its own simulation relation with specific precision parameters. This is referred to as a \textit{multi-layered approach} with homogeneous layers.
Next, we propose a multi-layered methodology to feature layers with different abstraction techniques that are discretization-based or discretization-free, referred to as a multi-layered approach with heterogeneous layers. 

\subsection{Related Works}
In the area of discretization-based techniques, multiple methods are related to variable precision through non-uniform partitioning of the state space.
For deterministic systems, there exist methods that
construct a non-uniform discretization grid \cite{ren2019dynamic, Tazaki2010}. More specifically, they give an approximate bisimulation relation for variable precision (or dynamic) quantization and develop a method to locally refine a coarse finite-state abstraction based on the system dynamics. Furthermore, for deterministic systems, there also exist methods known as multi-layered discretization-based control synthesis. They focus on maintaining multiple finite-state abstraction layers with different precision, where they use the coarsest finite-state abstraction when possible \cite{camara2011safety, camara2011synthesis, girard2020safety, hsu2018multi}. For stochastic models, non-uniform partitioning of the state space has been introduced for the purpose of verification \cite{Soudjani2013adaptive} 
and for verification and control synthesis in the software tools FAUST$^2$ \cite{soudjani2015fau} and StocHy \cite{cauchi2019stochy,cauchi2019efficiency}. Recent tools developments have considered utilizing interval Markov decision processes \cite{mathiesen2024intervalmdp,wooding2024impact}. In this paper, we consider approximate simulation relations that are used to quantify the similarity between a continuous-state model and its finite-state abstraction \cite{haesaert2017verification,haesaert2020robust}. Hence, we go beyond non-uniform partitioning and achieve a variable precision differently by allowing abstract models builds from both model order reduction methods and discretization of a reduced space.

Among the array of discretization-free techniques available, including control barrier certificate (CBC) \cite{PJP07} and stochastic model predictive control (SMPC) \cite{cominesi2017two,lorenzen2016constraint,mesbah2016stochastic,mesbah2018stochastic}, we emphasize the sampling-based approach \cite{haesaert2014sampling,vasile2013sampling} for its exceptional relevance in our context. CBC seeks to delineate a barrier certificate level set to separate unsafe regions from system trajectories, offering a formal probabilistic safety certificate. Recent works on CBC applied to stochastic systems can be found for example in \cite{clark2021control,jagtap2020formal,santoyo2021barrier}. SMPC relies on iteratively solving constrained optimizations with most of the SMPC approaches handling safety requirements. Recent SMPC methods handles more expressive (finite-horizon) temporal specifications but still require solving stochastic optimizations that are computationally expensive \cite{engelaar2024risk,Farhani_SMPC17}.
In contrast, the sampling-based approach used in this paper presents a distinct advantage by enabling the construction of Probabilistic Roadmaps (PRMs), a versatile tool both for navigating the complexities of stochastic systems, as well as for the integration with dynamic programming technique, which is the core method we use for obtaining the control policy. A sampling-based approach helps to lift the exponential computational complexity from the space discretization, a.k.a. \emph{curse of dimensionality}, by treating a complex problem using random sampling to approximate the solution. The simplicity of implementing a sampling-based approach motivates us to choose it as the discretization-free method in this paper.

\subsection{Contributions}
\begin{figure*}[t]
	\centering
	\includegraphics[width=.9\textwidth]{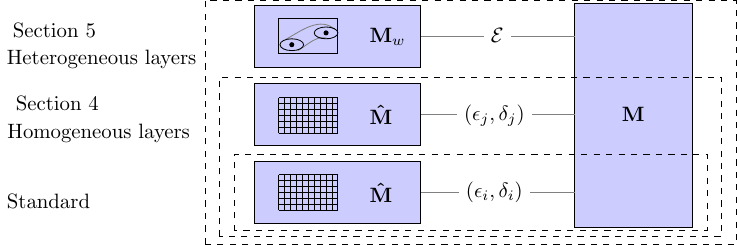}
	\caption{Overview of the structure of this paper with the homogeneous-layered approach (two bottom layers) in Section~\ref{sec:homog} and the heterogeneous-layered approach (all layers) in Section~\ref{sec:MM}. Precision $\epsilon_i$, $\mc{E}$ stand for output deviations, and $\delta_i$ stands for the probabilistic deviations \cite{huijgevoort2022similarity}.}
	\label{fig:overview}
\end{figure*}

We consider abstraction techniques based on approximate simulation relations that allow both model-order reduction and space discretization with quantification of the approximation errors. We start with allowing variable precision by presenting a simulation relation that contains multiple precision layers.
For such relations, we develop an algorithm to determine a switching strategy and a robust dynamic programming approach such that we can compute a lower bound on the probability of satisfying the given specification. We then extend the developed multi-layered methodology to allow for layers with a discretization-free technique and layers with a discretization-based technique, therefore, improving the efficiency and scalability of our method. To this end, we implement a discretization-free technique that constructs a graph based on samples of the state space. The different approaches discussed in this paper are graphically illustrated in Fig.~\ref{fig:overview}. 
More specifically, the approach of this paper has the following contributions.
\begin{itemize}
	\item {\bfseries Homogeneous layers.} First, we introduce a multi-layered approach with variable precision. Here,  we use multiple homogeneous layers, in which \emph{one finite-state model} with multiple layers and \emph{multiple similarity quantifications} is used. 
	We quantify the similarity compared to the original model with different deviation bounds in the output map ($\epsilon$) and in the transition probabilities ($\delta$). By doing so, we achieve a multi-layered simulation relation with variable precision. To perform the value iteration, we modify the standard dynamic programming (DP) operator, such that it suits this multi-layered approach. 
	\item {\bfseries Heterogeneous layers.} Then, we combine the multi-layered approach with a model that is constructed using a discretization-free technique. Hence, we use different abstract models.
	To perform the value iteration with heterogeneous layers, we integrate the DP operator of the discretization-free layer with the DP operator of the multi-layered approach with homogeneous layers.
\end{itemize}

\smallskip

A subset of the results of this paper is presented in \cite{huijgevoort2021multi}. This paper substantially expands the preliminary results of \cite{huijgevoort2021multi} in the following directions:
(a) Different abstraction techniques are integrated into the proposed approach;
(b) The underlying dynamic programming computations are adapted to allow this integration;
(c) A heuristic approach to derive an optimal switch strategy is included; 
(d) Implementation of case studies is expanded and enriched by incorporating more complex and higher-dimensional scenarios, thereby providing a more comprehensive and robust analysis; and
(e) The proofs of statements are provided.

\subsection{Overview}
The main part of this paper is structured as illustrated in Fig.~\ref{fig:overview}. 
The structure of this paper is as follows. In Section~\ref{sec:problem}, we discuss the considered model and specifications, and formulate the problem statement for a general class of nonlinear stochastic difference equations. In Section~\ref{sec:sec3}, we motivate the need for a multi-layered approach by applying a discretization-based and discretization-free approach to a specific example and evaluating the results. Besides that, we give high level details of the abstract models and similarity quantification. In Section \ref{sec:homog}, we discuss the multi-layered approach with homogeneous layers. More precisely, we achieve variable precision by having multiple precision parameters for a single abstract model. In Section \ref{sec:MM}, we extend the multi-layered method to allow discretization-free layers and discretization-based layers resulting in a heterogeneous approach. Finally, in Section \ref{sec:results_Ch4}, we apply our method to multiple case studies.  We end this paper with a conclusion.

\section{Problem formulation}\label{sec:problem}
\textbf{Notation.} We limit ourselves to spaces that are finite, Euclidean, or Polish. The Borel measurable space of a set $\X\subset \mathbb R^n$ is denoted by $(\X,\mathscr{B}(\X))$ with $\mathscr{B}(\X)$ being the Borel sigma-algebra on $\X$. 
A probability measure $\mathbb{P}$ over this space has realization $x \sim \mathbb{P}$, with $x \in \X$. 
Furthermore, a time update of a state variable $x$ is interchangeably denoted by $x(t+1), x_{t+1}$ or $x^+$. 

\textbf{Model.} 
We consider general Markov decision processes (gMDP) as defined in \cite{haesaert2017verification,haesaert2020robust} as follows. 
\begin{definition}[general Markov decision process (gMDP)]
	A gMDP is a tuple $\M=(\X,x_0,\U,\Tr,\Y,h)$ with state space $\X$, initial state $x_0\in\X$, input space $\U$, and with output space $\Y$ and measurable output map $h: \X \rightarrow \Y$. The transitions kernel $\Tr:\X \times \U \times \mathcal{B}(\X) \rightarrow [0,1]$ assigns to each state $x\in\X$ and input $u\in\U$ a probability measure $\Tr( \cdot \mid x,u)$ over $(\X,\mathcal{B}(\X))$.
\end{definition}

We consider output space $\Y$ to be a metric space and denote the class of all models with the same metric output space $(\Y,\textbf{d}_{\Y})$ as $\mathcal{M}_{\Y}$.
The behavior of a gMDP with $n_x$-dimensional state space $\X \subset \mathbb{R}^{n_x}$ can equivalently be described by a  stochastic difference equation. We consider discrete-time systems whose dynamics are described by a stochastic difference equation with Gaussian disturbance
\begin{equation}
	\M: \begin{cases}
		x(t+1) = f(x(t),u(t),w(t)) \\
		y(t) = h(x(t)), \quad \forall t \in \left\{0,1,2, \dots\right\},
	\end{cases}
	\label{eq:model_ML}
\end{equation} with state $x(t)\in\X$, input $u(t)\in\U$, disturbance $w(t)\in\W\subseteq\mb{R}^{n_w}$, output $y(t)\in\Y$, and with measurable functions $f: \X \times \U \times \W \rightarrow \X$ and $h: \X \rightarrow \Y$.  The system is initialized with $x(0)=x_0\in \X$, and $w(t)$ is an independently and identically distributed sequence with realizations $w \sim \mathbb{P}_{w} = \mc{N}(\mu,\Sigma)$. 

A finite path of the model \eqref{eq:model_ML} is a sequence $\boldsymbol{\omega}_{[0,t]}:= x_0, u_0, x_1, u_1, \ldots, x_t$. An infinite path is a sequence $\boldsymbol{\omega}:= x_0, u_0,\ldots $. The paths start at $x_0=x(0)$ and are built up from realizations $x_{t+1}=x(t+1)$ based on \eqref{eq:model_ML} given a state $x(t)=x_t$, input $u(t)=u_t$ and disturbance $w(t)$ for each time step $t$.
We denote the state trajectories as $\x = x_0,x_1,\dots$, with associated suffix $\x_t = x_t,x_{t+1},\dots$. The output $y_t$  contains the variables of interest for the 
performance of the system and for each state trajectory, there exists a corresponding output trajectory $\y = y_0,y_1,\dots$, with associated suffix $\y_t = y_t,y_{t+1},\dots$.

A control strategy is a sequence $\policy = (\mu_0, \mu_1, \mu_2, \dots )$ of maps $\mu_t(\boldsymbol{\omega}_{[0,t]})\in \U$ that assigns for each finite path $\boldsymbol{\omega}_{[0,t]}$ an input $u(t) = u_t$.  The control strategy is a Markov policy if $\mu_t$ only depends on $x_t$, that is $\mu_t: \X \rightarrow \U$. We refer to a Markov policy as \emph{stationary} if 
$\mu_t$ does not depend on the time index $t$, that is $\policy = (\mu, \mu, \dots)$ for some $\mu$. In this work, we are interested in control strategies denoted as $\C$ that can be represented with finite memory, that is, policies that are either time-stationary Markov policies or have finite internal memory. A policy with finite internal memory first maps the finite state execution of the system to a finite set (memory), followed by computing the input as a function of the system state and the memory state. By doing so, we can satisfy temporal specifications on the system trajectories.

\textbf{Specifications.} To express (unbounded time-horizon) temporal logic specifications,  
we use the syntactically co-safe linear temporal logic specifications (scLTL) \cite{belta2017formal,kupferman2001model}. 
Consider atomic propositions $p_1,p_2, \dots p_N$ that are true or false. The set of atomic propositions and the corresponding alphabet are denoted by {\small$\AP=\left\{p_1,\dots,p_N\right\}$} and $2^{\AP}$, respectively. Each letter $\letter\in  2^{\AP}$ contains the set of atomic propositions that are true. A (possibly infinite) string of letters forms a word $\word = \letter_0,\letter_1,\dots$, with associated suffix $\word_t = \letter_t,\letter_{t+1},\dots$. The output trajectory $\y = y_0,y_1,\dots$ of a system \eqref{eq:model_ML} is mapped to the word $\word= L(y_0),L(y_1),\dots$ using labeling function $L:\Y\rightarrow 2^{\AP}$ that translates each output to a specific letter $\letter_t = L(y_t)$. Similarly, suffixes $\y_t$ 
are translated to suffix words $\word_t$. 
By combining atomic propositions with logical operators, the language of scLTL can be defined as follows.
\begin{definition}[scLTL syntax]
	An scLTL formula $\phi$ over a set of atomic propositions $\AP$, with $p\in \AP$ has syntax
	\begin{equation}\label{eq:scLTLspec}
		\phi ::=  p \,|\, \lnot p \,|\, \phi_1 \wedge \phi_2 \,|\, \phi_1 \lor \phi_2 \,|\, \bigcirc \phi \,|\, \phi_1  \until \phi_2.
	\end{equation} 
\end{definition} \noindent 
The semantics of this syntax can be given for the suffixes $\word_t$. An atomic proposition $\word_t \models p$ holds if $p \in \letter_t$, while a negation $\word_t \models \lnot \phi$ holds if $\word_t \not\models \phi$. Furthermore, a conjunction $\word_t \models \phi_1 \wedge \phi_2$ holds if both $\word_t \models \phi_1$ and $\word_t \models \phi_2$ are true, while a disjunction $\word_t \models \phi_1 \lor \phi_2$ holds if either $\word_t \models \phi_1$ or $\word_t \models \phi_2$ is true. Also, a next statement $\word_t \models \bigcirc \phi$ holds if $\word_{t+1} \models \phi$. Finally, an until statement $\word_t \models \phi_1 \until \phi_2$ holds if there exists an $i\in\mathbb{N}$ such that $\word_{t+i} \models \phi_2$ and for all $j \in \mathbb{N}, 0 \leq j < i$, we have $\word_{t+j} \models \phi_1$.
A system satisfies a specification if 
the generated word $\word_0= L(\y_0)$ satisfies the specification, i.e.,  
$\word_0 \models \phi$. 

For control synthesis purposes, an scLTL specification can be written as a deterministic finite-state automaton (DFA), defined by the tuple $\DFA=(Q,q_0,\Sigma_\DFA,\tau_{\DFA},Q_f)$.
Here, $Q$, $q_0$, and $Q_f$ denote the set of states, initial state, and set of accepting states, respectively. Furthermore, $\Sigma_\DFA=2^{\AP}$ denotes the input alphabet and $\tau_\DFA:Q\times \Sigma_\DFA \rightarrow Q$ is a transition function.
For any scLTL specification $\phi$, there exists a corresponding DFA $\DFA_\phi$  such that the word $\word$  satisfies this specification  $\word \models \phi$, iff  $\word$ is accepted by $\DFA_\phi$. 
Here, acceptance by a DFA means that there exists a trajectory $q_0q_1q_2 \dots q_f$, that starts with $q_0$, evolves according to $q_{t+1} = \tau_{\DFA}(q_t,\pi_t)$, and ends at $q_f \in Q_f$.

Correct-by-design control synthesis focuses on designing a controller $\C$, for model $\M$ and specification $\phi$, such that the controlled system $\M\times \C$ satisfies the specification, denoted as $\M\times \C \models \phi$. For stochastic systems, 
we are interested in the satisfaction probability, denoted as $\mathbb{P}(\M \times \C \models \phi)$. The general problem considered in this paper is the following. 

\textbf{Problem.} 
Given model $\M$ as in \eqref{eq:model_ML}, an scLTL specification $\phi$, and a probability $p_\phi \in[0,1],$ find a controller $\C$, such that 
\begin{equation}
	\mathbb{P}(\M\times \C \models \phi) \geq p_\phi.
	\label{eq:ContrProb}
\end{equation}

\section{Motivation for a multi-layered approach}\label{sec:sec3}
In this section, we justify the need for a multi-layered approach with the help of a running example. We first introduce the running example, followed by discussions on discretization-based (DB) techniques and discretization-free (DF) techniques, respectively. Then, we compare the results of applying standard DB and DF approaches on the running example, and summarize their merits and drawbacks. We conclude the section by emphasizing the need for a  multi-layered approach.

\begin{figure}[tbh!]
	\centering
	\includegraphics[width=0.4\textwidth]{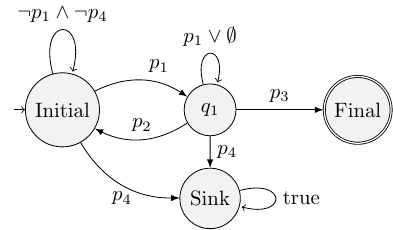}
	\caption{Cyclic DFA corresponding to the specification of the package delivery. Here, $\emptyset$ denotes the \emph{empty set}, which means that all atomic propositions $p \in \AP$ are false.} 
	\label{fig:DFA_PD}
\end{figure}

\noindent \textit{Running example (specification and dynamics).} As a running example, we consider a scLTL-specified scenario that we will refer to as the package delivery scenario. For this scenario, we consider multiple regions defined on the output space $\Y$, namely a pick-up region $P_1= [-4,-3]^2$, a delivery region $P_3= [3,5] \times [-2,-0.5]$, a strict avoid region $P_4 = [0,1] \times [-4,0]$, and a region where we lose a package $P_2= [0, 1] \times [0, 2.5] $. The goal of the controller is to make sure that a package is picked up at $P_1$ and delivered to $P_3$ while avoiding $P_4$. Region $P_2$ is fine to visit without a package, but once crossed while carrying a package, the package is lost and a new package has to be picked up from $P_1$. We label the regions $P_1,P_3,P_4$ and $P_2$ as $p_1,p_3,p_4,p_2$ respectively. The specification of the scenario is written in scLTL as 
\begin{equation}
	\phi_{PD} = \lnot p_4 \until (p_1 \wedge (\lnot (p_4 \lor p_2) \until p_3)). \label{eq:case3_specification}
\end{equation}
The associated DFA is given in Fig. \ref{fig:DFA_PD}.  

The dynamics of the robot carrying the package is a 2-dimensional linear system, as in (\ref{eq:model_ML}):
\begin{equation}
	\M: \begin{cases}
		x(t+1) = 0.9I_2x(t)+0.5I_2u(t)+\sqrt{0.25}I_2w(t) \\
		y(t) = I_2x(t), \quad \forall t \in \left\{0,1,2, \dots\right\},
	\end{cases}
	\label{eq:running_example_original_model}
\end{equation}
with states $x\in \X = [-20,5]^2$, inputs $u \in \U=[-5,5]^2$, outputs $y\in \mathbb{Y}=\X$ and Gaussian disturbance $w(t)\sim \mathcal{N}(\boldsymbol{0},I_2)$.

\subsection{Discretization-based method} \label{subsec:3_DB}
For continuous-state stochastic models, it is computationally hard to design a controller $\C$ and compute the satisfaction probability $\mb{P}(\M \times \C \models \phi)$ \cite{abate2008probabilistic}. We adopt the common solution in correct-by-design control synthesis which is to approximate the continuous-state model by a finite-state version. This step is often called \emph{abstraction}, and we perform it by partitioning or discretization of the state space as in \cite{haesaert2020robust}.
 
\noindent{\bfseries Model.}
Suppose that we have approximated the continuous-state model as given in \eqref{eq:model_ML}, with the following abstract model 
\begin{equation}
	\Mh: \begin{cases}
		\hat{x}(t+1) = \hat{f}(\hat{x}(t),\hat{u}(t),\hat{w}(t)) \\
		\hat{y}(t) = \hat{h}(\hat{x}(t)),
	\end{cases}
	\label{eq:model_MLAbs}
\end{equation}
with state $\hat{x}\in\Xh$ being in a finite space $\Xh$, initialized at $\hat x(0)=\hat{x}_0$ and with input $\hat{u}\in\Uh,$ output $\hat{y}\in\Y$ and disturbance $\hat{w} \in\hat{W}$. Furthermore, we have functions $\hat{f}:\Xh \times\Uh \times \hat{\W} \rightarrow \Xh$ and $\hat{h}: \Xh \rightarrow \Y$, 
and $\hat{w}(t)$ is an independently and identically distributed sequence with realizations $\hat{w} \sim \mathbb{P}_{\hat{w}}$. 
Beyond the given representative points, one generally adds a sink state to both the continuous- and the finite-state model to capture transitions that leave the bounded set of states.

{\bfseries Similarity quantification.} 
We use the abstract model in \eqref{eq:model_MLAbs} to compute a lower bound on the probability of satisfying specification $\phi$. To preserve this lower bound, we need to quantify the similarity between the models \eqref{eq:model_ML} and \eqref{eq:model_MLAbs}. 
%
%
We consider an approximate simulation relation as in Def.~4 in \cite{huijgevoort2022similarity}, since this is suitable for scLTL, which can be unbounded in time. This method for quantifying the similarity is based on an explicit coupling between the models \eqref{eq:model_ML} and \eqref{eq:model_MLAbs}, which allows for analyzing how close the probability transitions are. Def.~4 in  \cite{huijgevoort2022similarity} is based on the following observation. Given a simulation relation $\mathscr{R}\subset \Xh\times \X$, for all pair of states inside this relation $(\hat{x},x) \in\mathscr{R}$, and for all inputs $\hat{u} \in\Uh$ we can quantify a lower bound on the probability that the next pair of states 
is also inside this simulation relation, i.e. $(\hat{x}^+, x^+) \in\mathscr{R}$.
Hence, for all states $(\hat{x},x)\in \mathscr{R}$, $\forall \hat{u} \in\Uh$, we require that
\begin{equation}
(\hat{x}^+,x^+)\in \mathscr{R}
\label{eq:simrelCond3}
\end{equation}
has a lower bound on its probability denoted by $1-\delta$ under the transitions in the coupled model.

The approximate simulation relation in \cite[Def.~4]{huijgevoort2022similarity} does not only quantify the similarity based on the probability deviation $\delta$ but also on output deviation $\epsilon$. Furthermore, if $\Mh$ is $(\epsilon,\delta)$-stochastically simulated by $\M$, then this is denoted as $\Mh \preceq_{\epsilon}^{\delta} \M$.
In \cite{huijgevoort2022similarity}, it has been shown that $\epsilon$ and $\delta$ have a trade-off. Increasing $\epsilon$ decreases the achievable $\delta$ and vice versa.
To compute the deviation bounds $(\epsilon,\delta)$, an optimization problem constrained by a set of parameterized matrix inequalities is given in \cite{huijgevoort2022similarity}.

\noindent \textit{Running example (similarity quantification for DB approaches).} To apply a DB approach, we first construct a finite-state abstraction in the form of $\Mh$ \eqref{eq:model_MLAbs} by gridding the state space $\mb{X}$ of $\M$ in \eqref{eq:running_example_original_model} with $568 \times 563$ grid cells and the input space $\U$ with $14 \times 14$ grid cells. Details on constructing such an abstraction can be found in \cite[Sec.~4]{huijgevoort2022similarity}. To compute the satisfaction probability we have to quantify the similarity. To this end, we choose the interface function $u=\hat{u}$ and relation $\mathscr{R} = \{(\hat{x},x) \in \hat{\mb{X}}\times \mb{X} \mid || x-\hat{x} ||_D \leq \epsilon\}$, with $\epsilon=0.18$ and $D$ a symmetric positive definite weighting matrix. Next, we obtain the approximate simulation relation between the original model $\M$ \eqref{eq:running_example_original_model} and its finite-state abstraction in the form of $\Mh$ \eqref{eq:model_MLAbs}  as $(\epsilon,\delta)=(0.18,0.1217)$ and $D= I_2$ using \syscore tool \cite{huijgevoort2023syscore} based on linear matrix inequalities as in \cite{huijgevoort2022similarity}.

\subsection{Discretization-free method} \label{subsec:3_DF}
{\bfseries Model and similarity quantification.}
For the DF method, we consider an approximate relation similar to the one in DB method, to quantify the similarity between our abstract model of the DF layer and the model (\ref{eq:model_ML}). The reason we emphasize the similarity quantification here is that the output deviation $\epsilon$, i.e. $\epsilon_w$ is encoded in the model construction. Consider a finite set of states taken from the state space of the original model $\M$ as in \eqref{eq:model_ML}, that is $x_w \in \Xp \subset \X$, referred to as waypoints. We associate to each waypoint an ellipsoid 
\begin{equation}\label{eq:epsSet}
	\mc{E}(x_w) := \left\{ x \in \X  \mid ||x-x_w||_{D_w} \leq \epsilon_w \right\}
\end{equation}
that contains the states $x \in \X$ of the original model. Here, $D_w$ is a symmetric positive definite matrix. Note that the center of $\mc{E}$ equals $x_w$, but without loss of generality, we consider its shape the same for all $x_w \in \Xp$.
We now define a state transition function $\Delta_w: \Xp \times \Xp \rightarrow [0,1]$ that gives a lower bound on the probability 
of reaching waypoint $x_w'$ in a finite number of steps $n_s$.
More specifically, for all $x_w \in \mc{E}(x_w)$ there exists a sequence of control strategies, such that with a probability of at least $\Delta_w(x_w,x_w')$, $\exists n_s \in \N$, such that $x_{k+n_s} \in \mc{E}(x_w')$. 
Together, this allows us to define the following waypoint model 
\begin{equation}\label{eq:MPRM}
\Mp = (\Xp, x_{w,0}, \Delta_w, \Y, h_w),
\end{equation}
with 
initial state $x_{w,0} \in\Xp$, 
outputs $y_w\in \Y$, and output map $h_w: \Xp \rightarrow \Y$. 

\begin{figure}[t]
\centering
\includegraphics[width=.3\textwidth]{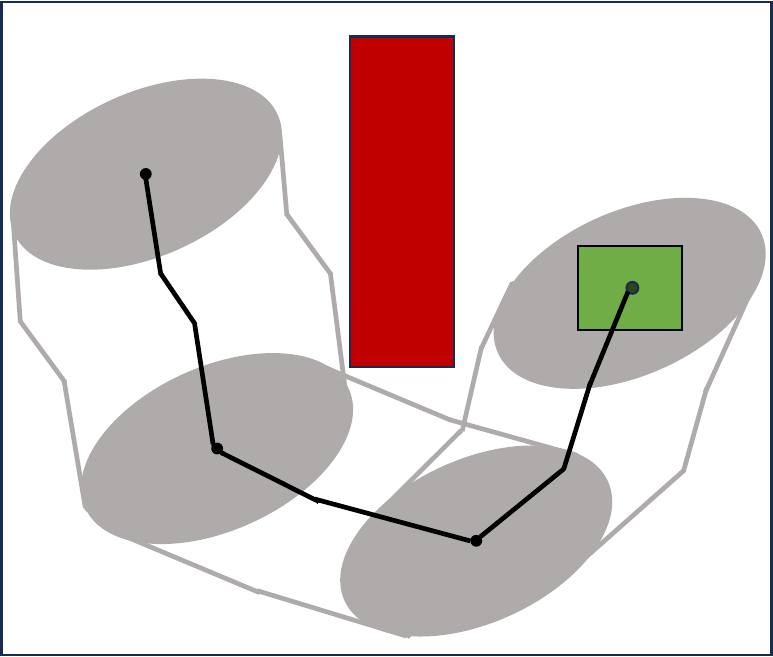}
\caption{Example of a state trajectory (black line) of a waypoint model with the corresponding tube in gray. The red and green regions are respectively an avoid and a goal region projected from the output space to the state space. The black dots are the waypoints $x_w$ and the gray ellipsoids are the sets $\mc{E}$. Note that this waypoint model is not well-posed since not all of the outputs corresponding to the top right ellipsoid have the same label.}
\label{fig:DFtubes}
\end{figure}

{\bfseries Specification.} 
To handle temporal logic specifications based on the model $\Mp$, we require that $\Mp$ is well-posed with respect to labeling $L: \Y \rightarrow 2^{\AP}$, if $\forall x_w \in \Xp:$
\begin{enumerate}
\item all outputs $y$ corresponding to states $x \in \mc{E}(x_w)$ have the same label, and
\item the outputs corresponding to paths from $x_w$ to $x_w'$ either never change label or only once.
\end{enumerate} These two assumptions allow us to keep track of the DFA state in a simple manner, which is necessary when considering temporal logic specifications.

The construction of the waypoint model, 
allows us to handle specifications given using scLTL$\backslash \bigcirc$, where the $\bigcirc$-operator is excluded since the number of time steps between waypoints is usually larger than $1$. An exemplary visualization of a waypoint model is given in Fig.~\ref{fig:DFtubes}.

{\bfseries Available methods.} 
The construction of paths in the waypoint model is in essence the same as solving a (deterministic) reach-avoid problem, for which a variety of methods are available \cite{girard2012controller,althoff2010computing,bogomolov2019juliareach}. For some situations, the transition probability $\Delta_w$ can even be computed directly via a continuous-state (stochastic) model using for example the tool \sreachtools \cite{vinod2019sreachtools}. In this paper, we utilize a sampling-based approach to construct our model of the DF layer. By defining the layer in this manner, we can use dynamic programming (as defined later in this paper) to make a connection between DB and DF methods.

\noindent \textit{Running example (similarity quantification for DF approaches).} To apply the DF approach, we first construct a waypoint model in the form of $\Mp$ \eqref{eq:MPRM}. To this end, we uniformly sample well-posed points in the state space for initializing a directed graph. Vertices and edges of the graph respectively correspond to states $x \in \mc{E}(x_w)$, and possible transitions of $x_w \rightarrow x_w'$. The approach is summarized in Alg.~\ref{alg:alg_ruohan}.
\begin{algorithm}
\caption{Construct waypoint model}
\begin{algorithmic}[1]
	
	\State\textbf{Input:}  $\M$
	
	\State $\Xh_s \leftarrow$  Uniformly sample well-posed points of $\X$
	\State $\mathcal{G}_{\text{PRM}} \leftarrow$ Construct a directed graph with waypoints $x_w \in \Xh_s$ and well-posed edges based on $\M$ 
	\While{$\mathcal{G}_{\text{PRM}}$ is not strongly connected}
	\State $\Xh_s \leftarrow$ Uniformly sample and add new well-posed points of $\X$
	\State  $\mathcal{G}_{\text{PRM}} \leftarrow$ Augment the graph with new well-posed edges
	\EndWhile
	\State Construct $\Mp$ based on $\mathcal{G}_{\text{PRM}}$
	\State \textbf{Output:} $\Mp$
	
\end{algorithmic}  \label{alg:alg_ruohan}
\end{algorithm}  We sample $48$ waypoints in total for our running example and choose $n_s=3$ for the number of steps between waypoints.
To compute the satisfaction probability, we have to quantify the similarity. To this end,
choose the lower bound of the transition probability $\Delta_w=0.0001$. We compute the similarity quantification between the original model $\M$ \eqref{eq:running_example_original_model} and its waypoint model $\Mp$ in the form of \eqref{eq:MPRM} for interface function $u=\hat{u}$, and obtain $\epsilon_w=2.6825$. Details on this computation are given in  Appendix~\ref{app:deriveEps}.

\begin{figure*}[t]
	\centering
	\subfloat[Single discretization-based layer, with $(\epsilon,\delta)=(0.18,0.1217)$.] {\includegraphics[width=0.4\textwidth]{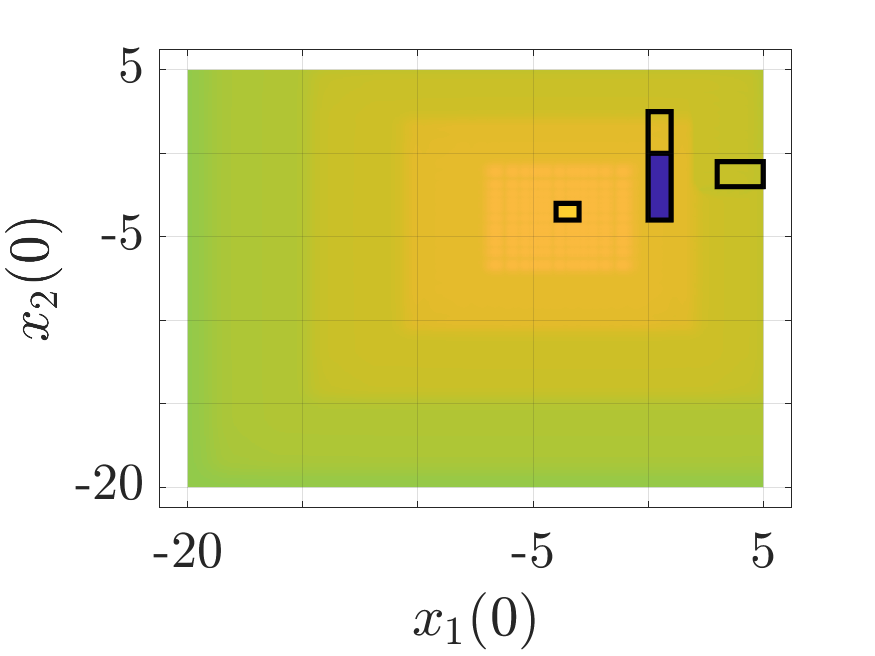}\label{fig:rex_pureDB}} \hfill
	\subfloat[Single discretization-free layer, with $\epsilon_w=2.6825$, $\Delta_w=0.0001$.]  {\includegraphics[width=0.4\textwidth]{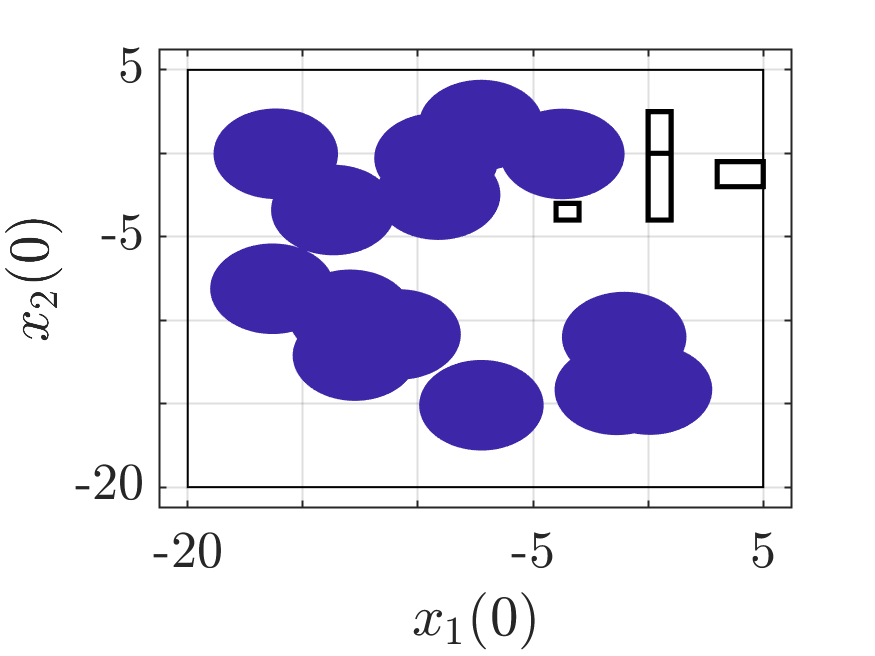}\label{fig:rex_pureSB}}  \hfill 
	\hspace{-1em}
	\includegraphics[width=0.085\textwidth]{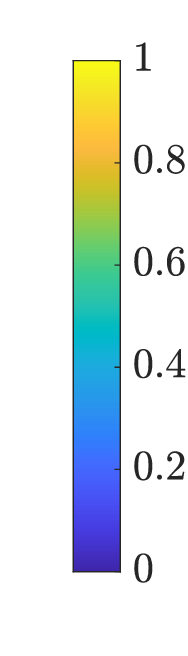}
	\caption{Complete running example with state space $\X=[-20,5]^2$. Robust satisfaction probabilities computed by applying a DB approach in (a), and by applying a DF approach in (b).}  
	\label{fig:rex_comparison}
\end{figure*}

\begin{figure*}[t]
	\centering
	\subfloat[Single discretization-based layer, with $(\epsilon,\delta)=(0.18,0.1217)$ for the smaller state space.] {\includegraphics[width=0.4\textwidth]{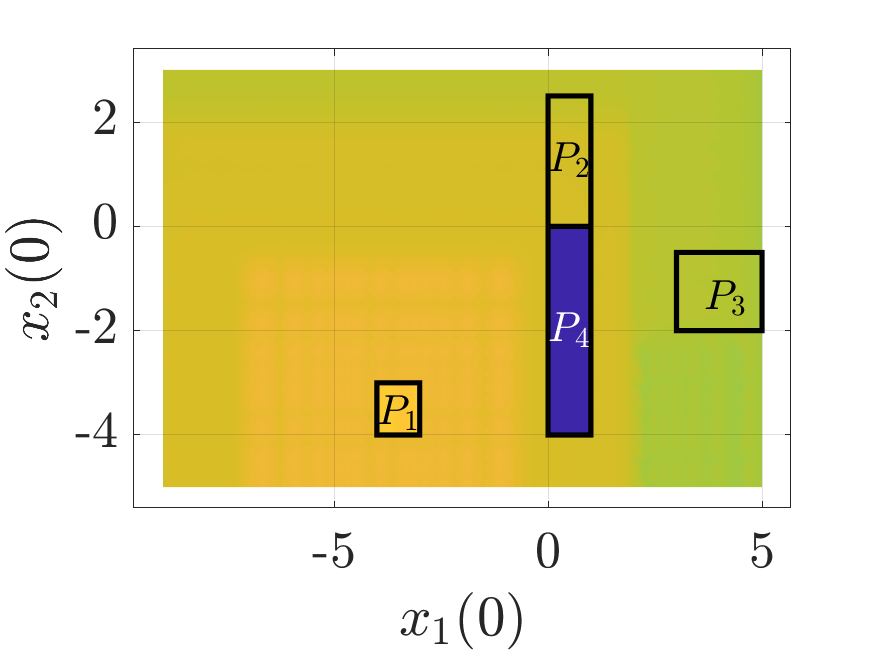}\label{fig:RESmallStateSpace}} \hfill
	\subfloat[Single discretization-free layer, with $\epsilon_w=2.6825$, $\Delta_w=0.0001$.] {\includegraphics[width=0.4\textwidth]{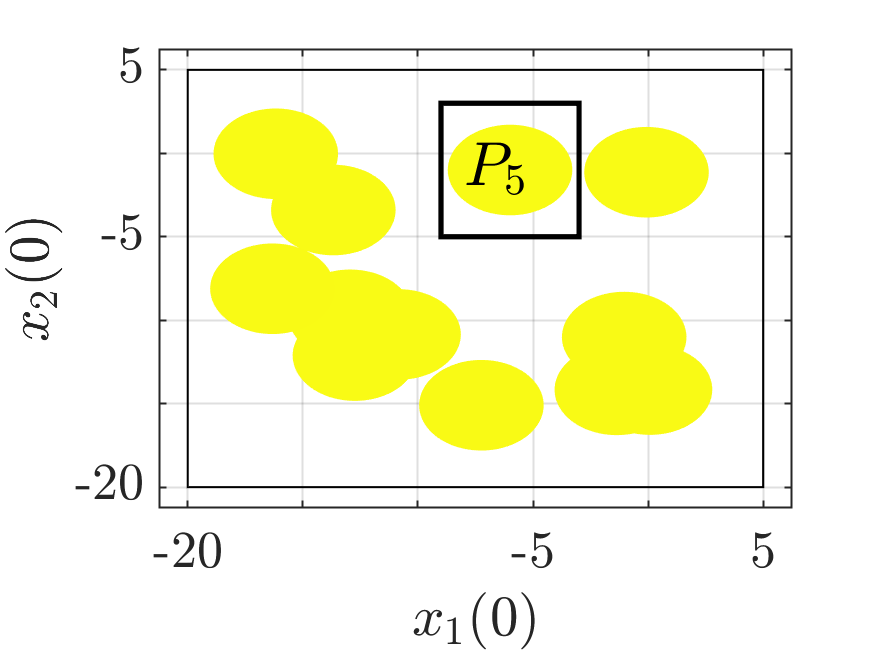}\label{fig:DF_RESmallStateSpace}} \hfill
		\hspace{-1em}
	\includegraphics[width=0.085\textwidth]{Figures/Results/colorbar}
		\caption{Split-up, unconnected running example. Robust satisfaction probabilities computed by applying a DB approach to a specific part of the state space, that is $\X_{\textrm{small}} = [-9,5] \times [-5,3] \subset \X$ in (a), and by applying a DF approach to the complete state space $\X = [-20,5]^2$ to reach the area $P_5 = [-9,-3] \times [-5,3] \subset \X_{\textrm{small}}$ in (b).}
\end{figure*}

\subsection{Analysis of the running example}
In this subsection, we compare computation time, and memory usage, and analyze the robust satisfaction probabilities obtained using different approaches, i.e., the pure DB approach (one single DB layer) and the pure DF approach (one single DF layer). 
The results of applying those approaches to the running example are presented in Fig.~\ref{fig:rex_comparison}. Here, the satisfaction probability is shown for the initial DFA state (labeled \textit{Initial} in Fig.~\ref{fig:DFA_PD}).

The average running time when applying the pure DB approach is $14.44$ seconds, when applying the pure DF approach is $0.3442$ seconds. The corresponding memory usage is equal to $2706$ MB and $0.027$ MB respectively.\footnote{\scriptsize All simulations are performed on a computer with a 3.4 GHz 13th Gen Intel Core i7-13700K processor and 64 GB 3200 MHz memory.}  Computational efficiency of the DF method, (sampling-based approach, in our case) is prominent. However, it can be observed that due to the nature of the pure DF method, in this specific case, for any well-posed waypoint model $\Mp$, the lower bounds on the satisfaction probabilities converge to zero after 3-5 iterations during dynamic programming, while following the DB method the lower bound is a lot higher. This DF approach is not accurate enough for this specific case. In our DF approach, due to the inherent trade-off between $\epsilon_w$ and $\Delta_w$ of our waypoint model $\Mp$, it is impossible to obtain nonzero satisfaction probabilities for our package delivery case. Intuitively, the regions $p_1, \dots, p_4$ are too small for a model with such a (relatively) large stochastic disturbance. Next, we will analyze the results of both approaches in more detail. 

{\bfseries DB approach.} 
In Fig.~\ref{fig:rex_pureDB}, we can see that due to the specification of the running example, the satisfaction probabilities decrease the further you are from the region $P_1$. In our running example, the satisfaction probabilities decrease rapidly due to the relatively large value of $\delta$. 

{\bfseries DF approach.} 
To elaborate on the reasons causing the satisfaction probabilities to (and will always) be $0$ when a DF approach is used, we consider two cases. The first case is to assume we tune $\Delta_w$ so that $\epsilon_w$ is small enough to make sure we have well-posed sample states inside our interested regions, i.e., $P_1$, $P_2$, $P_3$, or $P_4$. By making sure we sample sufficient amounts of well-posed sample states, our transition probability deviation $\Delta_w$ will be way too high consequently such that the robust satisfaction probabilities will drop to $0$ after $1$ or $2$ control actions. The second case is that we maintain our transition probability deviation $\Delta_w$ sufficiently low, which is what we do to obtain the result in Fig.~\ref{fig:rex_pureSB}. The consequence of this choice is that because our interested regions, i.e., $P_1$, $P_2$, $P_3$, and $P_4$ are very small (compared to our ellipsoidal sets $\epsilon_w$ of our waypoint model), we will not be able to sample any well-posed states inside our interested regions, hence in terms of DFA, we are forever stuck in the beginning state $q_0$ and we obtained the trivial lowerbound of $0$ on the satisfaction probability. 

{\bfseries A multi-layered approach.}
Based on comparing the robust satisfaction probabilities and computational efficiency between the DB approach and the DF approach, we conclude that DB approach exceeds DF approach in terms of accuracy, while DF approach surpasses DB approach in terms of efficiency. We, therefore, propose the heterogeneous approach through combining these two methods, to mitigate the drawbacks of applying two approaches separately and alone. To this end, we construct a DF layer for the complete state space and reduce the state space for which we construct the DB layer. By doing so, we exploit the accuracy of the DB approach around the area where the interested regions are clustered, and we exploit the computational efficiency of the DF approach for the “empty” area. 

In this specific case, we would use the DB approach to grid the part of the state space $[-9,5] \times [-5,3]$, since this is where we need its accuracy. To handle the complete state space, we use the DF approach to reach this gridded part of the state space.
To discover the potential benefit of combining a DB and a DF approach, we consider them separately next. First, we applied the DB approach using \syscore to the model \eqref{eq:running_example_original_model} with state space $\X = [-9,5] \times [-5,3]$ by gridding the state space with $318 \times 180$ grid cells and the input space with $14 \times 14$ grid cells. 
We found $(\epsilon,\delta) = (0.18,0.1217)$ and $D = I_2$ for the similarity quantification and obtained the satisfaction probability as in Fig.~\ref{fig:RESmallStateSpace}.  
 Next, we applied the DF approach to the model \eqref{eq:running_example_original_model} with updated specification $\phi = \eventually p_5$, where $p_5$ corresponds to the region $P_5 = [-9,-3] \times [-5,3]$. We obtained a waypoint model by sampling $48$ well-posed points in $\X$ and followed Alg.~\ref{alg:alg_ruohan}. We choose $\Delta_w=0.0001$, which is much smaller than $\delta$, to obtain a high satisfaction probability. We then compute $\epsilon_w=2.6825$. The resulting robust satisfaction probability is given in Fig.~\ref{fig:DF_RESmallStateSpace}. Comparing these results to the original one in Fig.~\ref{fig:rex_comparison}, we can conclude that indeed the DB approach works best for a small state space and can handle small labeled regions, while the DF approach works best for large state spaces with large labeled regions.

The overall goal of this paper is to combine these two results into one while maintaining guarantees on the lower bound of the satisfaction probability. We do this, by introducing different layers each containing its own abstraction approach.
Throughout this paper, we use two running examples. The first running example, introduced in Section~\ref{sec:sec3}, will be used to illustrate our heterogeneous-layer approach. The second running example, referred to as \emph{Running example A}, is a more complex variation of the first and will be used for the demonstration of the homogeneous-layer approach.

\begin{figure*}
	
	\subfloat[]{\includegraphics[width=.46\columnwidth]{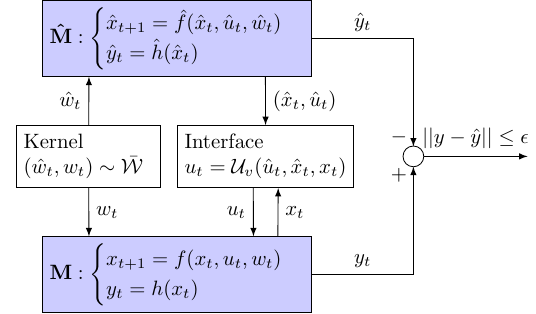} \label{fig:SimQuant}} \hspace{0.02\columnwidth}
	\subfloat[]{\includegraphics[width=.49\columnwidth]{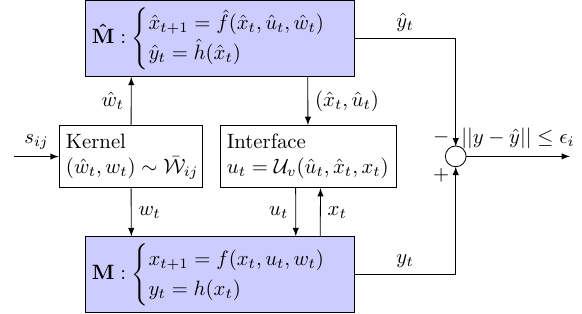} \label{fig:SimQuant_ML}}
	\caption{Similarity quantification between the models $\M$ and $\Mh$. For a standard setting in (a) and for a multi-layered setting with homogeneous layers in (b).}
\end{figure*}

\section{Homogeneous layers with variable precision}
\label{sec:homog} 	
As mentioned before, we quantify the similarity between the original model $\M$ \eqref{eq:model_ML} and abstract model $\Mh$ \eqref{eq:model_MLAbs} by using an approximate simulation relation as in Def.~4 in \cite{huijgevoort2022similarity}. This notion is based on two deviation bounds; output deviation or precision $\epsilon$, and probability deviation or confidence $\delta$. The output deviation is graphically illustrated in Figure~\ref{fig:SimQuant}. Here, we also see coupling between the two models $\M$ and $\Mh$ through a kernel and interface function as detailed in \cite{huijgevoort2022similarity}. The probability deviation $\delta$ associated with the lower bound on the probability of \eqref{eq:simrelCond3} has an inverse variation relation with the output deviation $\epsilon$. Intuitively, the probability that a pair of states remain in the same set is larger if the set itself is larger. In this setting, the (lower bound on the) probability is given by $1-\delta$ and the set by simulation relation $\mathscr{R}$ whose size is directly determined by $\epsilon$. This indicates that a larger output deviation $\epsilon$ gives a smaller probability deviation $\delta$. In the standard (single-layered) setting \cite{huijgevoort2022similarity}, only one $(\epsilon,\delta)$-pair is used to quantify the similarity between the two models $\M$ and $\Mh$, however, this $(\epsilon,\delta)$-pair is not unique.

In this section, we define a 
multi-layered simulation relation that allows variable precision using one abstract model, but different deviation bounds $(\epsilon,\delta)$. 
Furthermore, we detail an appropriate DP approach, discuss multiple computational improvements, and prove the control refinement.

\subsection{Homogeneous-layered simulation relation}
Current methods define one simulation relation for the whole state space, while we 
desire a multi-layered simulation relation $\boldsymbol{\mathscr{R}}=\left\{ \mathscr{R}_i\right\}$ with $i \in \left\{1,2, \dots, N_R\right\}$ that switches between multiple simulation relations to allow for multiple $(\epsilon_i,\delta_i)$ pairs, with $i \in \left\{1,2, \dots, N_R\right\}$ where $N_R$ denotes the number of simulation relations.
Each simulation relation is denoted as $\mathscr{R}_i$
and the corresponding 
precision as $\epsilon_{i}$. 

\begin{figure}[tbh!]
\centering
\includegraphics[width=0.6\columnwidth]{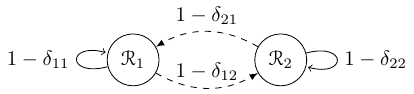}
\caption{Multi-layered simulation relation $\boldsymbol{\mathscr{R}}$ consisting of two simulation relations $\mathscr{R}_1$ and $\mathscr{R}_2$. The edges are labeled with a lower bound on the probability that the transition occurs. }
\label{fig:SimRelModes}
\end{figure}

An example of a multi-layered simulation relation with two simulation relations is given in Fig.~\ref{fig:SimRelModes}.
Here, the self-loops represent remaining in the same simulation relation, and a switch 
is indicated by the dashed arrows. Similar to the invariance requirement in  \eqref{eq:simrelCond3}, we associate a lower bound on the probability of each transition from $\mathscr{R}_i$ to $\mathscr{R}_j$ as $1-\delta_{ij}$, with $ i,j \in \left\{1,2,\dots, N_R\right\}$.
Furthermore, we define ${\boldsymbol{\epsilon}}$, and ${\boldsymbol {\delta}}$ with vector
$\boldsymbol \epsilon = \begin{bmatrix}\epsilon_1, \epsilon_2, \dots, \epsilon_{N_R} \end{bmatrix}$, and matrix $\boldsymbol \delta = \begin{bmatrix} \delta_{ij} \end{bmatrix}_{ij}$ 
In the remainder of this paper, a switch from simulation relation $\mathscr{R}_i$ to  $\mathscr{R}_j$ is denoted by the action $s_{ij}$, and we define the set of all possible switching actions by $\mathbb{S} := \left\{s_{ij} \text{ with } i,j \in \left\{1,2,\dots, N_R\right\} \right\}$.

The similarity quantification between the two models is graphically represented in Fig.~\ref{fig:SimQuant_ML} and is based on coupling the transitions of the models. First, the control inputs $u$ and $\hat{u}$ are coupled through an interface function denoted by
\begin{equation}
\mathcal{U}_v :\Uh\times \Xh \times \X \rightarrow \U.
\label{eq:interface}
\end{equation} 
Next, the 
disturbances $w$ and $\hat{w}$ are coupled through a stochastic kernel  $\Wker_{ij}(\cdot| \hat x, x, \hat u)$ in a similar fashion as in \cite{huijgevoort2022similarity}. However, in this multi-layered setting, we compute multiple stochastic kernels $\Wker_{ij}(\cdot| \hat x, x, \hat u)$ with $i,j \in \left\{1,2,\dots, N_R \right\}$, and use the switching action $s_{ij}$ to select one of them based on the state pair $(\hat{x},q)$ and the current layer $i$.
More specifically, the probability measures $\mathbb{P}_w$ and $\mathbb{P}_{\hat{w}}$ of their disturbances $w$ and $\hat{w}$ are coupled as follows.
	\begin{definition}[Coupling probability measures]\label{def:coupl}
		A coupling \cite{hollander2012probability} of  probability measures $\mathbb{P}_w$ and $\mathbb{P}_{\hat{w}}$ on the same measurable space $(\mathbb{W}, \mathscr{B}(\mathbb{W}))$ is any probability measure $\mathcal{W}_{ij}$ on the product measurable space $(\mathbb{W}\times \mathbb{W}, \mathscr{B}(\mathbb{W}\times \mathbb{W}))$ whose marginals are $\mathbb{P}_w$ and $\mathbb{P}_{\hat{w}}$, that is,
		\begin{align*}
			\mathcal{W}_{ij}(\hat{A}\times \mathbb{W}) &= \mathbb{P}_{\hat{w}}(\hat{A}) \textrm{ for all } \hat{A}\in \mathscr{B}(\mathbb{W}) \\
			\mathcal{W}_{ij}(\mathbb{W} \times A) &= \mathbb{P}_{w}(A) \textrm{ for all } A\in \mathscr{B}(\mathbb{W}). 
			\tag*{\text{\(\Box\)}}
		\end{align*} 
	\end{definition} \noindent
	More information about this state-dependent coupling and its influence on the simulation relation can be found in \cite{haesaert2017verification,huijgevoort2022similarity}.
	
	We can also design $\mc{W}_{ij}$ as a measurable function of the current state pair and actions, similar to the interface function. This yields a Borel measurable stochastic kernel associating to each $(\hat{u},\hat{x},x)$ a probability measure $\Wker_{ij}: \Uh \times \Xh \times \X \rightarrow \mc{P}(\mb{W}^2)$ that couples probability measures $\mathbb{P}_w$ and $\mathbb{P}_{\hat{w}}$ as in Definition~\ref{def:coupl}. 
	
	We can now analyze how close the transitions are (see Fig.~\ref{fig:SimQuant_ML} for visualization) and formally define the multi-layered simulation relation as follows.

\begin{figure*}[t]
\centering
\subfloat[Fixed precision] {\includegraphics[width=0.45\textwidth]{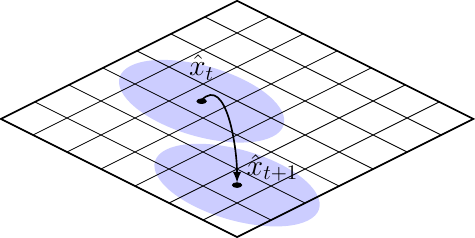}\label{fig:fixPrec}}\quad
\subfloat[Variable precision]  {\includegraphics[width=0.45\textwidth]{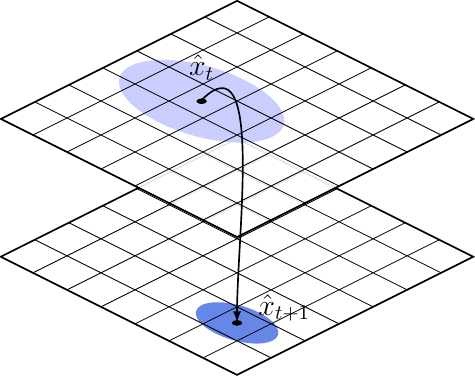}\label{fig:VarPrec}}\quad 
\caption{Graphical representation of a probabilistic transition for a fixed precision in (a) and a variable precision with two discretization-based layers in (b). In the left figure, the light blue ellipsoids contain all states $x_t$ resp. $x_{t+1}$ for which it holds that $(\hat{x}_t,x_t) \in\mathscr{R}_1$ resp. $(\hat{x}_{t+1},x_{t+1}) \in\mathscr{R}_1$. In the right figure, the light blue ellipsoid contains all states $x_t$ for which it holds that $(\hat{x}_t,x_t) \in\mathscr{R}_1$, and the dark blue ellipsoid contains all states $x_{t+1}$ for which it holds that $(\hat{x}_{t+1},x_{t+1}) \in\mathscr{R}_2$. In both figures, the space discretization or grid size is the same, while the simulation relations $\mathscr{R}_1$ and $\mathscr{R}_2$ are ellipsoids with different parameters.}
\label{fig:MultiLayeredSimRel}
\end{figure*}

\begin{definition}[Multi-layered simulation relation] \label{def:HybSimRel}
Let the models $\M, \Mh \in \mathcal{M}_\mathbb{Y}$, 
and the interface function $\mc{U}_v$ \eqref{eq:interface} be given. If there exists measurable relations $\mathscr{R}_i\subseteq \Xh\times \X$ and  Borel measurable stochastic kernels  $\Wker_{ij}$ that couple $\mathbb P_w$ and $\mathbb P_{\hat w}$ for  $i,j \in \left\{1,2,\dots, N_R\right\}$ 
such that 
\begin{enumerate}
\item $\exists i \in \left\{ 1, 2, \dots, N_R\right\}: (\hat{x}_0,x_0) \in \mathscr{R}_i$,
\item $\forall i \in \left\{1,2,\dots,N_R\right\}, \forall(\hat{x},x)\in \mathscr{R}_i:\textbf{d}_{\mathbb{Y}}(\hat{y},y)\leq \epsilon_{i}$, 
\item $\forall i \in \left\{1,2,\dots,N_R\right\}, \forall(\hat{x},x)\in \mathscr{R}_i, \forall \hat{u} \in\Uh: (\hat{x}^+,x^+)\in \mathscr{R}_j$ holds with probability at least $1-\delta_{ij}$ with respect to $\Wker_{ij}$.
\end{enumerate}	
Then $\Mh$ is stochastically simulated by $\M$ in a multi-layered fashion, 
denoted as $\Mh \preceq_{\boldsymbol{\epsilon}}^{\boldsymbol{\delta}} \M$, with precision $\boldsymbol{\epsilon}=[\epsilon_i]_i$ and $\boldsymbol{\delta}=[\delta_{ij}]_{ij}$ for  $i,j \in \left\{1,2,\dots, N_R\right\}$.  \hfill\(\Box\)
\end{definition} \noindent 
\begin{remark}
This simulation relation differs from the original one in \cite[Def.~4]{huijgevoort2022similarity},  
since it contains multiple simulation relations $\mathscr{R}_i$ with different output precision and, therefore, allows for variable precision. Note that for $N_R = 1,$ 
this multi-layered simulation relation becomes equivalent to the simulation relation from \cite[Def.~4]{huijgevoort2022similarity}. The difference between a simulation relation with a fixed precision and a multi-layered simulation relation with a variable precision is illustrated in Fig.~\ref{fig:MultiLayeredSimRel}. Here, we assume ellipsoidal simulation relations, however, the method described in this paper is not restricted to relations with this specific shape.
\end{remark}

The implementation of computing the similarity quantification for LTI systems is detailed in Appendix~\ref{sec:Impl}.

\noindent \textit{Running example A (dynamics and simulation relation)}
Here we present the simulation relations of the homogeneous approach (consisting of 2 DB layers) for the package delivery scenario with specification \eqref{eq:case3_specification} and corresponding regions as before. The dynamics of the car are modeled using an LTI stochastic difference equation as in \eqref{eq:model_MLLTI} (Appendix~\ref{sec:Impl}) with $A=0.9I_2, B=0.5I_2, B_w=\sqrt{0.25}I_2$, and $C=I_2$. We used states $x\in \X = [-5,5]^2$, inputs $u \in \U=[-1.25,1.25]^2$, outputs $y\in \mathbb{Y}=\X$ and Gaussian disturbance $w \sim \mathcal{N}(0,I_2)$. Note that this running example is more challenging than the original running example, as introduced in \eqref{eq:running_example_original_model} since the input space is reduced.

We construct abstract model $\Mh$ in the form of \eqref{eq:model_MLAbsLTI} by partitioning the state space 
	with $283 \times 283$ regions and the input space with $3 \times 3$ regions. We choose interface function $u=\hat{u}$ and simulation relations $\mathscr{R}_i = \{(\hat{x},x) \in \hat{\mb{X}}\times \mb{X} \mid || x-\hat{x} ||_D \leq \epsilon_i\}$, $i \in \{1,2\}$.

Next, we set the first layer with $\mathscr{R}_1$, where $\epsilon_{1}=0.5$ and the second layer with $\mathscr{R}_2$, where $\epsilon_{2} =0.3$. Given that we have $\boldsymbol{\epsilon} = \begin{bmatrix}
	0.5 & 0.3
\end{bmatrix}$, we compute $\boldsymbol{\delta}$ based on Lemma \ref{lem:epsdelBounds} in Appendix~\ref{sec:Impl}. We obtain the weighting matrix $D = I_2$ and deviation bounds as 
\begin{align*}
	\boldsymbol{\epsilon} = \begin{bmatrix}
		0.5 & 0.3
	\end{bmatrix}, \quad &  \boldsymbol{\delta} = \begin{bmatrix}
		0 &0.1586  \\ 
		0 & 0.0160
	\end{bmatrix}.
\end{align*}

\subsection{Homogeneous-layered dynamic programming}\label{sec:DP} 
After quantifying the similarity, we can synthesize a robust controller and compute a lower bound on the satisfaction probability based on the deviation bounds $\boldsymbol{\epsilon}$ and $\boldsymbol{\delta}$. Since we consider specifications written using scLTL, we can formulate this problem as a reachability problem that can be solved as a dynamic programming problem. 

\textbf{scLTL satisfaction as a reachability problem.}
By converting the scLTL specification $\phi$ to a DFA $\DFA_\phi$ such that the word $\word$  satisfies this specification  $\word \models \phi$, iff  $\word$ is accepted by $\DFA_\phi$, we can reason about the satisfaction of specifications over $\M$ by analyzing its product composition with $\DFA_{\phi}$ \cite{tkachev2013quantitative} denoted as $\M\otimes \DFA_\phi$. This composition yields a stochastic system with states $(x_t, q_t)\in \X\times Q$ and input $u_t$.  Given input $u_t$, the stochastic transition  from $x_t$ to $x_{t+1} $ of $\M$  is extended to the transition from $(x_t,q_t)$ to $(x_{t+1},q_{t+1})$ with $q_{t+1} = \tau_{\DFA_\phi}(q_t,L( h(x_t)))$.  
Hence, by computing a Markov policy over $\M\otimes \DFA_\phi$ that solves this reachability problem, we obtain a controller 
with respect to specification $\phi$, 
that has a finite memory \cite{belta2017formal}. In \cite{huijgevoort2022similarity}, it is shown that for any gMDP (and other equivalent representations), this reachability problem can be rewritten as a dynamic programming (DP) problem.

{\bfseries Standard robust dynamic programming.}
The standard robust DP approach detailed in \cite{huijgevoort2022similarity} with a single layer is as follows.
Given Markov policy $\policy = (\mu_0, \mu_1, \dots, \mu_N)$ with time horizon $N$ for  $\M\otimes \DFA_\phi$, define the time-dependent value function $V_N^{\policy}$  as 
\begin{align}\label{eq:V}
	& \textstyle   V^{\policy}_N(x,q)= \mathbb E_{\policy} \bigg[ \sum\limits_{ k=1}^N \mathbf{1}_{Q_f}( q_k) \prod \limits_{\mathclap{j = 1}}^{k-1}\mathbf{1}_{ Q \setminus Q_f}( q_j ) \bigg|( x_0,q_0)=(x,q) \bigg]
\end{align}
with indicator function $\textbf{1}_{E}(q)$ equal to $1$ if $q \in E$ and $0$ otherwise.
Since $ V^{\policy}_N(x, q)$ expresses the probability that a trajectory generated under $\policy: \X \times Q \rightarrow \U$ starting from $(x,q)$ will reach the target set $Q_f$ within the time horizon $[1,\ldots,N]$, it also expresses the probability that specification $\phi$ will be satisfied in this time horizon.
Next, define the associated DP operator
\begin{equation}
	\Bel^{u}(V)({x},q) :=  \mathbb{E}_{u} \big( \max\left\{ \boldsymbol{1}_{Q_f}(q^+),V({x}^+,q^+ )\right\}\big),
	\label{eq:Tmu}
\end{equation} with $u=\mu(x,q)$, $x^+:=x(t+1)$ evolving according to \eqref{eq:model_ML}, and 
with the implicit transitions  $q^+ =\tau_{\DFA_\phi}(q,L(h(x^+)))$. Consider a policy $\policy_k=\cramped{(\mu_{k+1}, \ldots \mu_N)}$ with time horizon $N-k$, then it follows
that $\cramped{ V^{{\policy}_{k-1}}_{{N-k}+1}}(x,q) = \cramped{\Bel^{{ \mu}_k}}\cramped{( V^{{\policy}_k}_{N-k})}(x,q)$, $\forall (x,q) \in \M\otimes \DFA_{\phi}$. Thus if $\cramped{ V^{{\policy}_k}_{N-k}}$ expresses the probability of reaching $Q_f$ within $N-k$ steps, then $ \Bel^{\mu_k} \cramped{ (V^{{\policy}_k}_{N-k}) }$ expresses the probability of reaching $Q_f$ within $N-k+1$ steps with policy ${\policy}_{k-1}$. It follows that for a stationary policy ${\policy} $, the infinite-horizon value function can be computed as $ V_\infty^{\policy} = \lim_{N\rightarrow \infty} (\Bel^{\policy})^N  V_0$ with ${V}_0 \equiv 0$. Furthermore, the optimal DP operator $\Bel^\ast(\cdot) := \sup_{\mu} \Bel^\mu(\cdot)$ can be used to compute the optimal converged value function $V^\ast_\infty$. The corresponding satisfaction probability can now be computed as 
\begin{equation*}
	\mathbb{P}^{\policy^\ast} := \max(\boldsymbol{1}_{Q_f}(\bar{q}_0),V^{\ast}_{\infty}(x_0,\bar{q}_0)),
\end{equation*}
with $\bar{q}_0=\tau_{\DFA_\phi}(q_0,L(h(x_0))$ and is also denoted as $\mb{P}(\M \times \C \models \phi)$.
When the policy $\policy^\ast$, or equivalently the controller $\C$, 
yields a satisfaction probability higher than $p_\phi$, then \eqref{eq:ContrProb} is satisfied and the synthesis problem is solved.

Due to the continuous states of $\M$, the DP formulation above cannot be computed for the original model $\M$, instead, we use abstract model $\Mh$. As a consequence, we have to take deviation bounds $\epsilon$ and $\delta$ into account. For a fixed precision, this robust DP operator is introduced by \cite{haesaert2020robust}. We repeat it here for completeness:
\begin{align}
	&\Bel^{\hat{u}}(V)(\hat{x},q) = 
	\textbf{L}\big( \mathbb{E}_{\hat{u}} \big( \min_{q^+ \in Q_{\epsilon}^+} \max\left\{ \boldsymbol{1}_{Q_f}(q^+),V(\hat{x}^+,q^+)\right\}\big) -\delta  \big),
	\label{eq:Bell-orig}
\end{align}  with $\hat u=\mu(\hat x,q)$ and $\textbf{L}:\mathbb{R}\rightarrow [0,1]$ being a truncation function defined as $\textbf{L}(\cdot):=\min(1,\max(0,\cdot))$, and with 
\begin{align}
	Q_{\epsilon}^+(q,\hat{y}^+) := \left\{\tau_{\DFA} (q,  L(y^+)) \mid 
	||y^+-\hat{y}^+|| \leq \epsilon \right\}.
	\label{eq:Qplus-orig}
\end{align} Note that compared to \eqref{eq:Tmu}, in \eqref{eq:Bell-orig} we subtract $\delta$ from the expected value to take the deviation in probability into account. Similarly, the deviation in the outputs $\epsilon$ can give different labels for the outputs $y$ and $\hat{y}$. Therefore, we minimize the probability with respect to the possible DFA states associated with these labels. This is done through $Q^+_\epsilon$ as in \eqref{eq:Qplus-orig}. A graphical representation of the value iteration using this robust DP operator is given in Fig.~\ref{fig:VIstandard}.

\begin{figure*}[t]
	\centering
	\begin{tikzpicture}[auto,node distance=2cm, every node/.style = {inner sep = 0, outer sep =0}]
		
		\node[state, draw = none] (V0) {$V_{0}^{\policy_N}$};
		\node[state,draw=none] (V1) [left of = V0] {$V_1^{\policy_{N-1}}$}; 
		\node[state, draw=none] (Vnk) [left of = V1, xshift=-.5cm]{$V_{N-k}^{\policy_k}$};
		\node[state, draw=none] (Vnk1) [left of = Vnk]{$V_{N-k+1}^{\policy_{k-1}}$};
		\node[state, draw=none] (Vn1) [left of = Vnk1, xshift=-.5cm]{$V_{N-1}^{\policy_{1}}$};
		\node[state, draw=none] (Vn) [left of = Vn1]{$V_{N}^{\policy_{0}}$};
		
		\node at ( $ (Vnk)!0.5!(V1) $ )  (arrowR) {$\hdots$};
		\node at ( $ (Vnk1)!0.54!(Vn1) $ )  (arrowL) {$\hdots$}; 
		
		\node at ( $ (Vnk)!0.59!(V1) $) (arrowR2) {};
		\node at ( $ (Vnk)!0.4!(V1) $) (arrowR3) {};
		\node at ( $ (Vnk1)!0.64!(Vn1) $) (arrowL2) {};
		\node at ( $ (Vnk1)!0.45!(Vn1) $) (arrowL3) {};
		
		\path[-latex] (V0) edge node[above, align = center, yshift=1pt] {\eqref{eq:Bell-orig}} (V1);
		\path[-latex] (Vnk) edge node[above, align = center, yshift=1pt] {\eqref{eq:Bell-orig}} (Vnk1);
		\path[-latex] (Vn1) edge node[above, align = center, yshift=1pt] {\eqref{eq:Bell-orig}} (Vn);
		\path[-latex] (V1) edge (arrowR2);
		\path[-latex] (arrowR3) edge (Vnk);
		
		\path[-latex] (Vnk1) edge (arrowL3);
		\path[-latex] (arrowL2) edge (Vn1);
	\end{tikzpicture}
	\caption{Graphical representation of the value iteration with a standard robust DP operator. 
		The arguments $(\hat{x},q)$, respectively the abstract state and DFA state, of the value functions are omitted for simplicity.}
	\label{fig:VIstandard}
\end{figure*} 

\textbf{Switching strategy.} For the multi-layered approach, we introduce multiple layers. To compute the value function, we need to keep track of the layers. To this end, we use a \emph{switching strategy}.
Consider a switching strategy that associates a switching action $s_{ij}$ to state pairs $(\hat{x},q) \in \Xh \times Q$ depending on the layer $i$. More specifically, we extend the input action $\hat{u}$ to $(\hat{u},s_{ij})$ and split up the corresponding policy as $\policy = (\policy^u, \policy^s)$, with $\policy^u =  (\mu_0^u, \mu_1^u, \dots, \mu_N^u)$ and $\policy^s =  (\mu_0^s, \mu_1^s, \dots, \mu_N^s)$ determining respectively the input action and switching action, with respectively the mappings $\mu^u_k: \Xh \times Q \times \left\{1,2,\dots,N_R\right\} \rightarrow \Uh$, and  $\mu^s_k: \Xh \times Q \times \left\{1,2,\dots,N_R\right\} \rightarrow \mb{S}$, $\forall k \in [0,\dots, N]$. Note that for $\policy$, we still have $\policy = (\mu_0, \mu_1, \dots, \mu_N)$, with $N$ the time horizon. For the DP, we are now interested in computing both $\hat{u} = \mu^u_k(\hat x, q, i)$ and $s_{ij} = \mu^s_k(\hat{x},q,i)$ for all $k \in [0, \dots, N]$. 

\begin{figure}
	\centering
	\begin{tikzpicture}[auto,node distance=1.5 cm,
		scale = 0.4]
		\draw[dashed, draw=gray, fill=grayfilling] (-6.8,-5.3) rectangle (-1.5,1.5);
		
		\node[state, draw=none] (V1) {$V_{i, N-k}^{\policy_k}$};
		\node[state,draw=none] (V2) [below of = V1] {$V_{j, N-k}^{\policy_k}$};
		\node[state, draw=none] (V1next) [left of = V1, xshift=-2cm]{$V_{i, N-k+1}^{\policy_{k-1}}$};
		\node[state, draw=none] (V2next) [left of = V2, xshift=-2cm]{$V_{j,N-k+1}^{\policy_{k-1}}$};
		
		\path[-latex] 	(V1) edge node [above, align=center]  {$\Bel_{s_{ii}}^{\hat u}$} (V1next);
		\path[-latex] 	(V2) edge node [below, align=center]  {$\Bel_{s_{jj}}^{\hat u}$} (V2next);
		\path[-latex, royalblue] (V1) edge node [above, align=center]  {$\Bel_{s_{ij}}^{\hat u}$} (V2next);
		\path[-latex, darkred] 	(V2) edge node [below, align=center]  {$\Bel_{s_{ji}}^{\hat u}$} (V1next);
		
		\node[draw=none] at (-4.15,-6.2) {\eqref{eq:Tmu-epsdel}};
		
		\draw[fill = black] (2.85,-1.9) circle (3pt);
		\draw[fill = black] (3.35,-1.9) circle (3pt);
		\draw[fill = black] (3.85,-1.9) circle (3pt);
		
		\draw[fill = black] (-11.15,-1.9) circle (3pt);
		\draw[fill = black] (-11.65,-1.9) circle (3pt);
		\draw[fill = black] (-12.15,-1.9) circle (3pt);
	\end{tikzpicture}
	\caption{Graphical representation of the value iteration for homogeneous layers, where all possible switching actions $s_{ij} \in \mb{S}$ are considered. All operations in the gray box can be performed using \eqref{eq:Tmu-epsdel}. The arguments $(\hat{x},q,i)$ and $(\hat{x},q,j)$ of the value functions in respectively the top and bottom rows are omitted for simplicity. Note that the layer is indicated as a subscript instead. The dots indicate that the full value iteration follows the sequence as in Fig.~\ref{fig:VIstandard}.}
	\label{fig:VI_ML}
\end{figure}
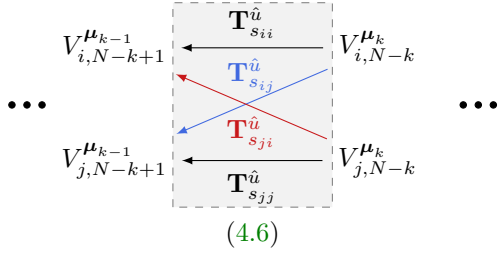

\textbf{Homogeneous-layered DP.}
To implement a layered DP approach, 
we extend the value function to include the different relations $\mathscr{R}_i$ as different layers. Hence, to each layer, we assign a value function $V(\hat{x},q,i)$.
This value function 
defines a lower bound on the probability that specification $\phi$ will be satisfied in the time horizon $\left[1, \dots, N \right]$. We define a new robust operator $\Bel^{\hat{u}}_{s_{ij}}$
as
\begin{align}
\label{eq:Tmu-epsdel}
	&\Bel^{\hat{u}}_{s_{ij}}(V)(\hat{x},q,i) =
	\textbf{L}\big( \mathbb{E}_{\hat{u}} \big( \min_{q^+ \in Q_{\epsilon_{j}}^+} \max\left\{ \boldsymbol{1}_{Q_f}(q^+),V(\hat{x}^+,q^+,j)\right\}\big) -\delta_{ij}  \big), \notag \\
& \text{with }
	Q_{\epsilon_{j}}^+(q,\hat{y}^+) := \left\{\tau_{\DFA} (q,  L(y^+)) \mid 
	||y^+-\hat{y}^+|| \leq \epsilon_{j} \right\}.
\end{align} 
For given policy $\mu = (\mu^u, \mu^s)$, 
we define
$\Bel^{\mu} (V)(\hat{x},q,i)= \Bel^{\hat{u}}_{s_{ij}}(V)(\hat{x},q,i)$ with $\hat{u} = \mu^u(\hat{x},q,i)$, and $s_{ij}=\mu^s(\hat x, q, i).$
Consider a policy $\boldsymbol{\mu}_k=(\mu_{k+1}, \dots, \mu_N)$,  
then for all $(\hat{x},q, i)$
we have that $V_{N-k+1}^{\boldsymbol{\mu}_{k-1}}= \Bel^{\mu_k}(V_{N-k}^{\boldsymbol{\mu}_k})$, initialized with ${V}_0 \equiv 0$. Fig.~\ref{fig:VI_ML} gives a graphical representation of the multi-layered value iteration for two layers and all possible switching actions. As before, for a stationary policy $\boldsymbol{\mu}$, the infinite-horizon value function for all layers 
can be computed as $ V_\infty^{\boldsymbol{\mu}} = \lim_{N\rightarrow \infty} (\Bel^{\boldsymbol{\mu}})^N  V_0$ with ${V}_0 \equiv 0$. 
\begin{remark}
	The DP operator in \eqref{eq:Tmu-epsdel} is an adaptation of the operator in \eqref{eq:Bell-orig} 
	to a multi-layered setting. For a fixed precision, that is one value for $\epsilon$ and one for $\delta$, the DP operators in \eqref{eq:Tmu-epsdel} and \eqref{eq:Bell-orig} become equivalent, hence we retrieve the DP operator for a fixed precision. 
\end{remark}

Following \cite{haesaert2020robust}, to use the DP operator to compute a lower bound on the satisfaction probability, it should satisfy some properties. The first property is that it is monotonically increasing. Note that a Bellman-operator $\Bel(\cdot)$ is \emph{monotonically increasing} if for any two functions $V$ and $W$ for which we have that $\forall (\hat x,q,i) \in \Xh \times Q \times \left\{1,2,\dots,N_R\right\}$: $V(\hat x,q,i) \geq W(\hat x,q,i)$, it holds that $\forall j \in \left\{1,2,\dots,N_R\right\}$:
\begin{equation*}
	\Bel^{\hat u}_{s_{ij}}(V)(\hat x,q,i) \geq 	\Bel^{\hat u}_{s_{ij}}(W)(\hat x,q,i).
\end{equation*} 
We can now conclude the following.
\begin{theorem}\label{prop:properties}
	The robust DP operator in \eqref{eq:Tmu-epsdel} is monotonically increasing, and the series $\left\{(\Bel^{\hat{u}}_{s_{ij}})^k(V_0)\right\}_{k \geq 0}$ with $V_0 \equiv 0$ is monotonically increasing and \emph{point-wise converging}. Furthermore, the fixed point equation $V_\infty^{\hat{u}} = \Bel^{\hat{u}}_{s_{ij}}(V_ \infty^{\hat{u}})$ has a unique solution for $\delta_{ij} >0$, which is $V_\infty^{\hat{u}} = \lim_{N \rightarrow \infty} (\Bel^{\hat{u}}_{s_{ij}})^N V_0$ with $V_0 \equiv 0$.
\end{theorem}
\begin{proof}
The proof of this theorem follows along the same lines as Lemma~1 in \cite{haesaert2020robust}, however, in this case, we have an adaptive robust DP operator. The difference is the introduction of the layers and the different values for $\epsilon_j$ and $\delta_{ij}$. Since this does not impact the properties in the proof, we can follow the proof of Lemma~1  in \cite{haesaert2020robust} to show that the properties in this theorem still hold.
\end{proof}

The value function gives the probability of satisfying the specification in $1$ to $\infty$ time step, by including the first time instance based on $x_0$, we can compute the robust satisfaction probability, that is
\begin{equation} \label{eq:R}
	\mathbb{R}^{\policy} := \max(\boldsymbol{1}_{Q_f}(\bar{q}_0),\max_{i \in \left\{1, 2, \dots, N_R\right\}}V^{\policy}_{\infty}(\hat{x}_0,\bar{q}_0,i)),
\end{equation} with $\bar{q}_0 = \tau_{\DFA_{\phi}}(q_0,L(h(x_0)))$ and $\hat{x}_0 \in \mathscr{R}^{-1}(x_0)$. 
The robust satisfaction probability  as in \eqref{eq:R} is also denoted as $\mb{R}_{\boldsymbol{\epsilon,\delta}}^{\policy}(\Mh \times \Ch \models \phi)$

To compute the optimal value function $V^\ast_\infty$, we use the optimal robust operator 
\begin{equation}
\label{eq:Bell_optS}
	\Bel^\ast(\cdot)_{s_{ij}} := \sup_{\mu} \Bel^{\mu}_{s_{ij}}(\cdot).
\end{equation}
The operator $\sup_{\mu}(\cdot)$ implicitly optimizes both over $\hat{u} \in \Uh$ and $j\in\{1, 2, \dots, N_R\}$, which in practice is a very expensive operation with respect to computation time and memory usage. This is especially the case when the number of abstract inputs or the number of layers is large.  

\subsection{Efficient implementation of homogeneous-layered dynamic programming}
In this section, we detail two implementation approaches that improve the computational efficiency. 

\textbf{Optimize the switching strategy using a surrogate model.}
To reduce the computation time and memory usage, we split the computation of the optimal input policy and switching strategy. To this end, we use the DP operator in \eqref{eq:Tmu-epsdel} in two different ways. First to compute a switching strategy, then to compute the abstract input and the robust satisfaction probability. In this section, we detail a practical approach to determine the switching strategy that is close to optimal, while also being efficient with respect to computation time and memory usage.

In \eqref{eq:Bell_optS}, we determine the optimal action pair $(\hat{u},s_{ij})$, by considering 
all possible input actions $\hat{u} \in \Uh$ and all possible switching actions $s_{ij} \in \mathbb{S}$ at all states and layers. 
This computation requires a lot of memory and computation power, therefore we pre-compute an approximately optimal switching strategy. To save computational power, we only compute the switching strategy for an abstract model with an extra coarsely partitioned state space, referred to as the \emph{surrogate model} $\Mh_s$ with state space $\Xh_s$.  
Note that this surrogate model is a finite-state abstraction of the original model $\M$ that is constructed in the exact same way as finite-state abstraction $\Mh$, however, the grid is a lot coarser for $\Mh_s$ compared to $\Mh$.

The approach to compute the switching strategy using this surrogate model is summarized in Alg.~\ref{alg:SS}. After constructing the surrogate model by partitioning the state space with a finite number of grid cells, we compute a suitable switching action using the $(\boldsymbol{\epsilon,\delta})$ values corresponding to the actual abstract model $\Mh$. More specifically, we compute the switching action at $(\hat{x},q,i)$ as 
\begin{equation} \label{eq:sij_comp}
(\hat{u}^\ast,s_{ij}^\ast) = 
\argmax_{(\hat{u},s_{ij})} \Bel^{\hat{u}}_{s_{ij}}(V)(\hat{x},q,i), 
\end{equation} with operator $\Bel^{\hat{u}}_{s_{ij}}(\cdot)$ as in \eqref{eq:Tmu-epsdel}, and with $j \in \left\{1, 2, \dots, N_R \right\}$. This optimal switching action $s_{ij}^\ast = \mu^s_k(\hat{x},q,i)$ is computed at each time step $k \in [0, \dots, N]$, with $N$ the time horizon. Hence, it is used to build up the optimal switching strategy $\policy^s = (\mu_0^s, \mu_1^s, \dots, \mu_N^s)$.

\begin{algorithm} 
\caption{Optimize switching strategy}
\begin{algorithmic}[1]
\State \textbf{Input:} $\M$, $\boldsymbol{\epsilon}$, $\boldsymbol{\delta}$, $\DFA_\phi$
\State $\Mh_s \leftarrow$ Construct surrogate model $\Mh_s$ using a coarse grid.
\State $ \policy^s(\hat{x}_s,q,i) \leftarrow$ Perform dynamic programming through \eqref{eq:Tmu-epsdel} over all possible switching actions in $\mathbb{S}$ based on model $\Mh_s$ from step 2 and inputs $(\boldsymbol{\epsilon},\boldsymbol{\delta})$. After convergence is reached, compute optimal switching action using \eqref{eq:sij_comp}.  
\State $\policy^s(\hat{x},q,i) \leftarrow$Translate switching strategy for $\Mh_s$ from step 3 to switching strategy for $\Mh$.
\State \textbf{Output:} $\policy^s(\hat{x},q,i)$
\end{algorithmic}  \label{alg:SS}
\end{algorithm}

Since the DP algorithm determines the best switching action $s_{ij}$ for each abstract state of the surrogate model, it is optimal with respect to this surrogate model. 
Finally, in step 4, we translate the switching strategy from the surrogate model to the actual abstract model $\Mh$. More specifically, for each $\hat{x} \in \Xh$ we determine the closest (with respect to the $L_2$-norm) state $\hat{x}_s \in \Xh_s$ and associate the same switching action $s_{ij}$ to it. 

After computing the switching strategy $\policy^s$, the control policy $\policy^u$  is determined for the actual abstract model $\Mh$ together with the robust satisfaction probability. The complete approach is summarized in Alg.~\ref{alg:controlSyn}. 

\begin{figure*}[t]
	\centering
	\subfloat[Switching strategy for DFA state $q_0$ and layer 1. The blue and red dots corresponds to $s_{11}$ and $s_{12}$ respectively.] {\includegraphics[width=0.4\textwidth]{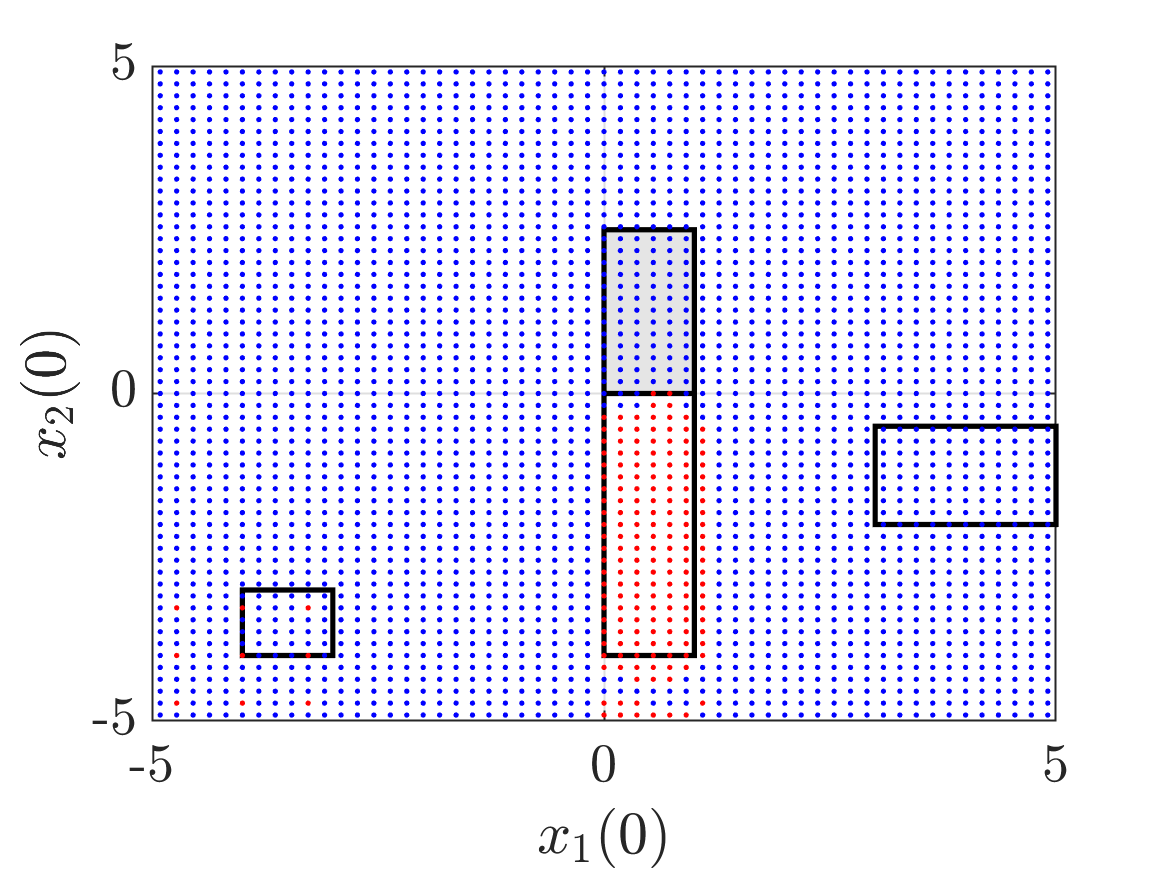}\label{fig:PD_SS_q0_lay1}}\quad
	\subfloat[Switching strategy for DFA state $q_0$ and layer 2. The blue and red dots corresponds to $s_{21}$ and $s_{22}$ respectively.]  {\includegraphics[width=0.4\textwidth]{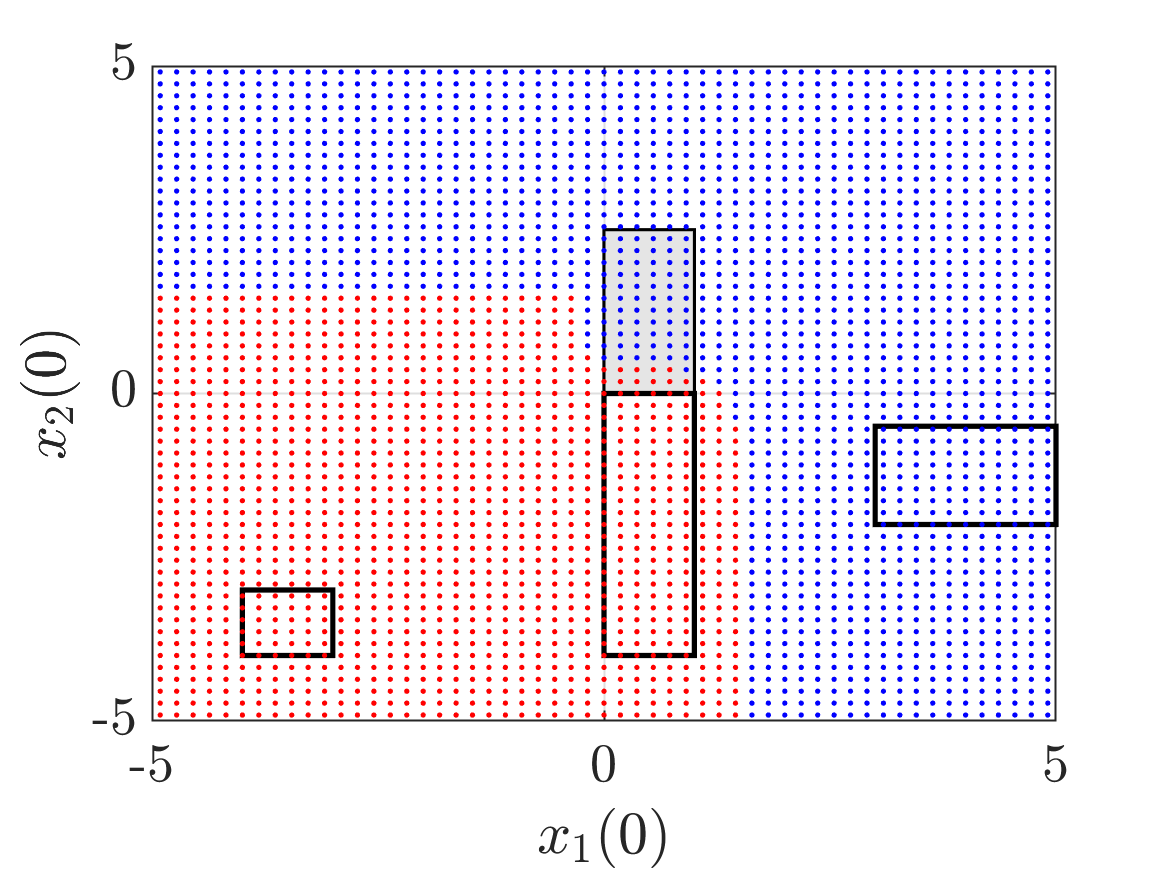}\label{fig:PD_SS_q0_lay2}} 
	 
	\subfloat[Switching strategy for DFA state $q_1$ and layer 1. The blue and red dots corresponds to $s_{11}$ and $s_{12}$ respectively.] {\includegraphics[width=0.4\textwidth]{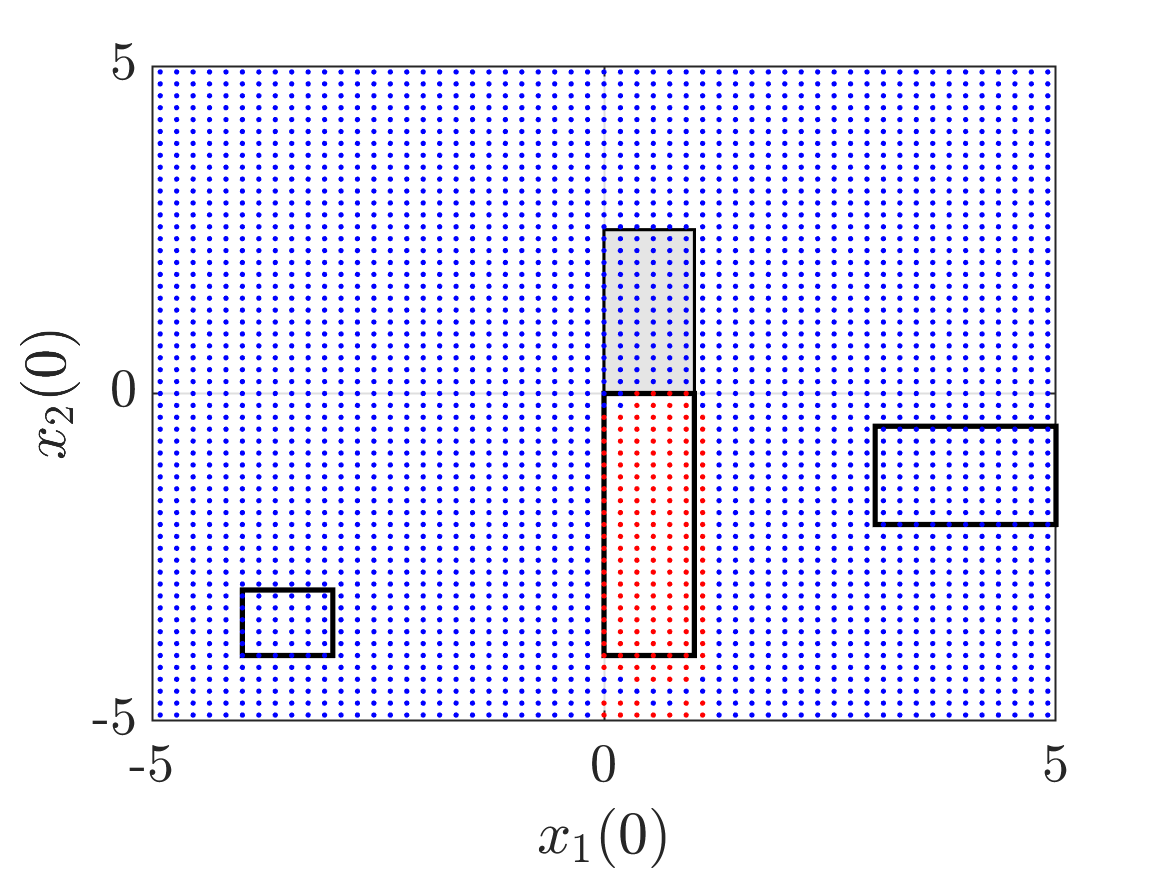}\label{fig:PD_SS_q1_lay1}}\quad
	\subfloat[Switching strategy for DFA state $q_1$ and layer 2. The blue and red dots corresponds to $s_{21}$ and $s_{22}$ respectively.]  {\includegraphics[width=0.4\textwidth]{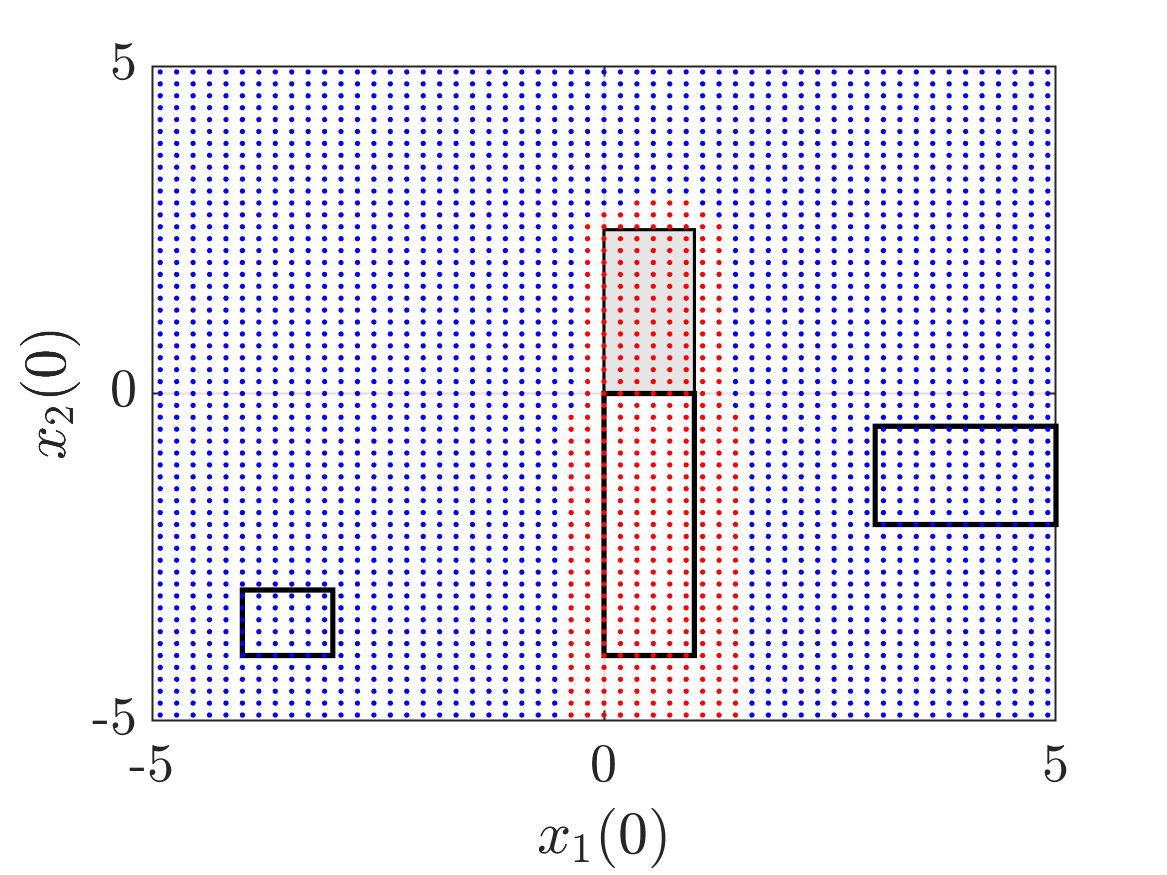}\label{fig:PD_SS_q1_lay2}}
	
	\caption{Switching strategy for the states of the surrogate model of running example A for DFA states $q_0$ in (a), (b) and $q_1$ in (c), (d). Here, 
		a blue resp. red dot represents switching to layer 1 resp. layer 2. The black boxes indicate regions $P_1$, $P_2$, $P_3$, and $P_4$. Region $P_2$ is colored gray.} 
	\label{fig:SS_PD}
\end{figure*}

\noindent \textit{Running example A (Switching strategy based on the surrogate model $\Mh_s$).} 
	To compute the switching strategy, 
we construct a surrogate model $\Mh_s$ by partitioning with $55 \times 55$ grid cells. This is substantially less than the $283 \times 283$ grid cells of the abstract model $\Mh$. By following Alg~\ref{alg:SS}, we obtain the optimal switching strategy corresponding to this grid as illustrated in Fig.~\ref{fig:SS_PD}. Here, the blue and red dots correspond to switching to layer 1 and layer 2 respectively. From this figure, we can see that this approach is indeed guided by the specification, since depending on the DFA state, the \emph{role} of the  $P_2$ (either \emph{not relevant} or \emph{try to avoid}) is different and the switching strategy changes accordingly.

\begin{algorithm} 
\caption{Control synthesis}
\begin{algorithmic}[1]
\State \textbf{Input:} $\M$, $\Mh$, $\DFA_\phi$
\State $(\boldsymbol{\epsilon},\boldsymbol{\delta}) \leftarrow$ Compute $(\boldsymbol{\epsilon},\boldsymbol{\delta})$ such that $\Mh \preceq_{\boldsymbol{\epsilon}}^{\boldsymbol{\delta}} \M$ holds with relation $\Rm$.
\State $\policy^s(\hat{x},q,i) \leftarrow$ Alg.~\ref{alg:SS}.
\State $\policy^u(\hat{x},q,i), \mb{R}_{\boldsymbol{\epsilon,\delta}}^{\policy}(\Mh \times \Ch \models \phi) \leftarrow$ Compute robust controller and satisfaction probability given the switching strategy from step 3.
\State \textbf{Output:}  $\policy = (\policy^u, \policy^s)$, and $\mb{R}_{\boldsymbol{\epsilon,\delta}}^{\policy}(\Mh \times \Ch \models \phi)$
\end{algorithmic}  \label{alg:controlSyn}
\end{algorithm}

\begin{figure}[tbh!]
\centering
\includegraphics[width = 0.4\textwidth ]{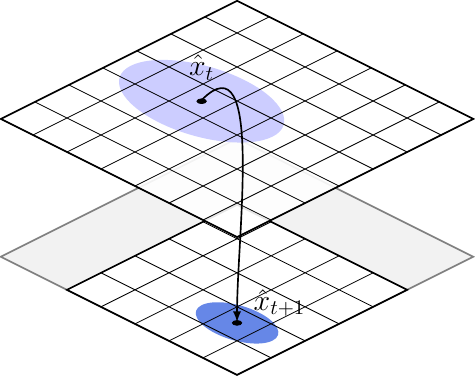}
\caption{Homogeneous layers with a variable precision, where the value function in the high-precision layer is not fully computed. This part of the high-precision layer (bottom) is given in gray.}
\label{fig:partialCov}
\end{figure}

\textbf{Partial value iteration.} 
To improve the computational efficiency of our approach even further, we consider layers where we only partially compute the value function. An example is illustrated in Fig.~\ref{fig:partialCov}, where for the gray part of the state space in the high-precision layer, we fix the value function to zero and do not update it. 
Therefore, the total computation time and memory usage is lower than when performing the value iteration fully. This is especially the case for high-dimensional systems, models with a large number of layers, and models with a large state space.

\subsection{Homogeneous-layered control refinement}\label{sec:contrRef}
The abstract controller $\Ch$ is designed based on the abstract model $\Mh$ in \eqref{eq:model_MLAbs}. Since we take the similarity quantification into account during the control design, we can refine the abstract controller $\Ch$ to a controller $\C$ that can be deployed on the original model $\M$. In this section, we show that with the given DP approach, such a control refinement is indeed possible and the computed satisfaction probability is valid.

\begin{figure*}
\subfloat[]{\includegraphics[width=.48\columnwidth]{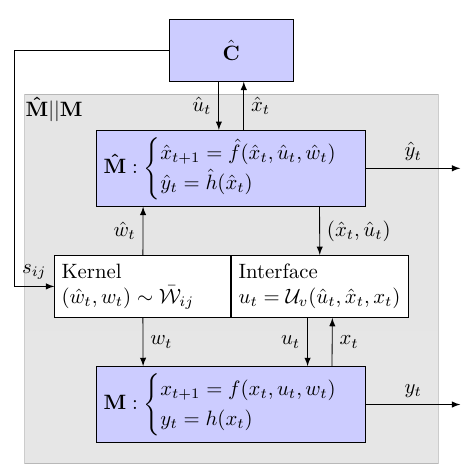} \label{fig:coupledModel_ML}}  \hspace{0.01\columnwidth}
\subfloat[]{\includegraphics[width=.46\columnwidth]{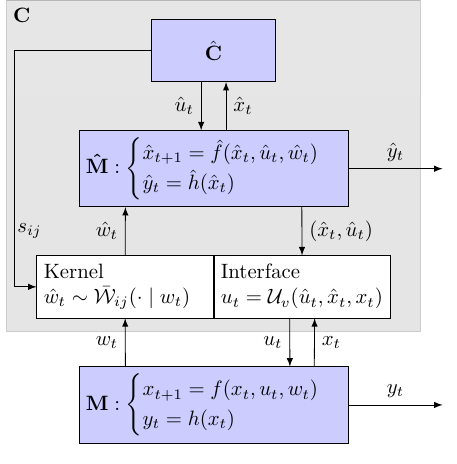} \label{fig:refinement_ML}}
\caption{In (a), designing an abstract controller $\Ch$ for coupled transitions of $\Mh|| \M$, where the switching action $s_{ij}$ determines the stochastic kernel $\Wker_{ij}$. In (b), refined controller $\C$ deployed on the model $\M$ with \emph{conditional} kernel $\Wker_{ij}$. }
\end{figure*}

We design the abstract control $\Ch$ based on the abstract model $\Mh$ that is coupled through an interface function $\mc{U}_v$ as in \eqref{eq:interface} and a stochastic kernel $\Wker_{ij}$ as in \eqref{eq:combeqwiths}. Due to this coupling, we can define the combined stochastic difference equation, with finite mode $\sigma_t$ that keeps track of the layer as
\begin{align}\label{eq:combeqwiths}
\Mh\| \M:	\left\{\begin{array}{ll}\begin{pmatrix}
\hat{x}_{t+1} \\
x_{t+1} 
\end{pmatrix} &= \begin{pmatrix}
\hat{f}(\hat{x}_t,\hat{u}_t,\hat{w}_t) \\
f(x_t, \mathcal{U}_v(\hat{u}_t,\hat{x}_t,x_t), w_t)
\end{pmatrix}\hspace{-.5cm} \\
\hat{y}_t & = \hat{h}(\hat x_t)\\
y_t &= h(x_t) \\
\sigma_{t+1} & = j \text{ if } \sigma_t = i, \text{ and } s_t = s_{ij}\\
(\hat{w}_t,w_t) &\sim \Wker_{ij}(\cdot| \hat x, x, \hat u) \text{ if } s_t = s_{ij},
\end{array}
\right.
\end{align} 
with pair of states $(\hat{x},x)\in \Xh\times \X$, control input $\hat{u}\in\Uh$, outputs $\hat y, y \in \mathbb{Y}$, and  coupled disturbance $(\hat{w},w)$. The input space of this combined system has been extended; that is, next to control input $\hat{u}_t $ we also have a switching input $s_{t}$. More specifically, 
since the disturbances of the combined transitions \eqref{eq:combeqwiths} are generated from the stochastic kernel $\Wker_{ij}$, 
\eqref{eq:combeqwiths} holds, with $(\hat w , w)\sim \Wker_{ij}$ if $s_t= s_{ij}$.  The combined system is illustrated in the gray box in Fig.~\ref{fig:coupledModel_ML}. Given the coupled stochastic difference equation, we can analyze how close the transitions are and therefore, quantify the similarity between models \eqref{eq:model_ML} and \eqref{eq:model_MLAbs}. Furthermore, the additional switching action allows us to do this in a multi-layered manner, as defined in Definition~\ref{def:HybSimRel}.

Consider a control strategy $\policy$ for 
$\Mh$, indicated by $\Ch$ in Fig.~\ref{fig:coupledModel_ML}. This strategy can also be implemented on the combined model $\Mh \| \M$ and we denote the value function of the combined model as $V_c(\hat{x},x,q,i)$. The control strategy for the combined model can be refined to a control strategy of the original model $\M$ \eqref{eq:model_ML}, as depicted in Fig.~\ref{fig:refinement_ML}. Although $V_c(\hat{x},x,q,i)$  expresses the probability of satisfaction, it cannot be computed directly, instead we can compute $V(\hat{x},q,i)$ over the abstract model $\Mh$ using \eqref{eq:Tmu-epsdel}.

To prove that for any control strategy $\policy$ (or controller $\Ch$) for $\Mh\| \M$ there still trivially exists a controller $\C$ 
for $\M$ that preserves the lower bound on the satisfaction probability, it is sufficient to formulate the following lemma.

\begin{lemma}\label{lem:ValueFunc}
Suppose $\Mh \preceq_{\boldsymbol{\epsilon}}^{\boldsymbol{\delta}} \M$ with a multi-layered simulation relation $\boldsymbol{\mathscr{R}}$
is given. Let $V(\hat{x},q,j) \leq V_c(\hat{x},x,q,j)$ for all $(\hat{x},x)\in \mathscr{R}_j$, then
\begin{equation}
\Bel^{\hat{u}}_{s_{ij}}(V)(\hat{x},q,i) \leq \Bel_{s_{ij}}^{\hat{u}}(V_c)(\hat{x},x,q,i) \quad \forall (\hat{x},x) \in \mathscr{R}_i,
\end{equation} where $\Bel^{\hat{u}}_{s_{ij}}(V)(\hat{x},q,i)$ is the $(\boldsymbol{\epsilon},\boldsymbol{\delta})$-robust operator \eqref{eq:Tmu-epsdel} with respect to stochastic transitions of $\Mh$ and $\Bel_{s_{ij}}^{\hat{u}}(V_c)(\hat{x},x,q,i)$ is the exact recursion \eqref{eq:Tmu} extended to the combined stochastic transitions \eqref{eq:combeqwiths}.  
\end{lemma}
The proof of Lemma~\ref{lem:ValueFunc} follows along the same lines as the proof of Lemma~3 
in the extended version of  \cite{haesaert2020robust}. It is given here for completeness.

\begin{proof}
The expected-value part of the operator $\Bel^{\hat{u}}_{s_{ij}}(V)(\hat{x},q,i)$ in
\eqref{eq:Tmu-epsdel} equals
\begin{equation}
\mathbb{E}_{\hat{u}} \big( \min_{q^+ \in Q_{\epsilon_{j}}^+} \max\left\{ \boldsymbol{1}_{Q_f}(q^+),V(\hat{x}^+,q^+,j)\right\}\big) -\delta_{ij}.
\label{eq:Bel_proof1}
\end{equation}

This expected value can be rewritten to an integral and multiplication with the transition kernel. Hence, \eqref{eq:Bel_proof1} is equivalent to
\begin{equation}
\int\limits_{\Xh} \min_{q^+ \in Q_{\epsilon_j}} \max\left\{ \boldsymbol{1}_{Q_f}(q^+),V(\hat{x}^+,q^+,j)\right\} \Tr(\text{d}\hat{x}^+ \mid \hat{x},\hat{u}) -\delta_{ij}.
\label{eq:Bel_proof2}
\end{equation}

Denote the transition kernel of the combined model $\Mh || \M$ as $\Wxbar{ij}(\cdot \mid \hat{x},x,\hat u): \mc{B}(\Xh \times \X) \rightarrow [0,1]$, which is equivalent to the evolution of the state pair of $\Mh || \M$ as in \eqref{eq:combeqwiths}, that is 
\begin{align*}
&	\begin{pmatrix}
\hat{x}_{t+1} \\
x_{t+1} 
\end{pmatrix} = \begin{pmatrix}
\hat{f}(\hat{x}_t,\hat{u}_t,\hat{w}_t) \\
f(x_t, \mathcal{U}_v(\hat{u}_t,\hat{x}_t,x_t), w_t),
\end{pmatrix}
\end{align*} with $(\hat{w}_t, w_t) \sim \Wker_{ij}(\cdot \mid \hat{x},x,\hat u): \mc{B}(\W \times \W) \rightarrow [0,1]$, written as a transition kernel. 
For $(\hat{x},x) \in \mathscr{R}_i$ and input $\hat{u}$ applied to the 
combined model $\Mh || \M$, the integral in \eqref{eq:Bel_proof2} is equivalent to the integral
\begin{equation*}
\int\limits_{\Xh \times \X} \min_{q^+ \in Q_{\epsilon_j}} \max\left\{ \boldsymbol{1}_{Q_f}(q^+),V(\hat{x}^+,q^+,j)\right\} \Wxbar{ij}(d\zeta | z)   -\delta_{ij},
\end{equation*} with  abbreviation $\Wxbar{ij}(d\zeta | z) : = \Wxbar{ij}(\text{d}\hat{x}^+ \times \text{d}x^+ | (\hat{x},x,\hat{u}))$.
For $(\hat{x},x) \in \mathscr{R}_i$, we can split this integral up using the simulation relation $\mathscr{R}_i \subseteq \Xh \times \X$ and find the following upper bound
\begin{align*}
&	\int\limits_{\mathscr{R}_i} \min_{q^+ \in Q_{\epsilon_j}} \max\left\{ \boldsymbol{1}_{Q_f}(q^+),V(\hat{x}^+,q^+,j)\right\} \Wxbar{ij}(d\zeta | z)  \\
& \!\!	+ \!\!\!\!  \int\limits_{\Xh \times \X \backslash \mathscr{R}_i} \!\!\!\!\!\! \min_{q^+ \in Q_{\epsilon_j}} \!\!\! \max\left\{ \boldsymbol{1}_{Q_f}(q^+),V(\hat{x}^+, q^+, j)\right\}\Wxbar{ij}(d\zeta | z) 
-\delta_{ij} \\
&	 \leq 	\int\limits_{\mathscr{R}_i} \min_{q^+ \in Q_{\epsilon_j}} \max\left\{ \boldsymbol{1}_{Q_f}(q^+),V(\hat{x}^+,q^+,j)\right\} \Wxbar{ij}(d\zeta | z).
\end{align*} Here, we used that  $\Wxbar{ij} ((\Xh\times \X) \backslash \mathscr{R}_i \mid \hat x, x, \hat u) \leq \delta_{ij}$.

Following the assumption of the lemma, we have that for all $(\hat{x}^+,x^+) \in \mathscr{R}_j$, it holds that $V(\hat{x}^+,q^+,j) \leq V_c(\hat{x}^+, x^+, q^+,j)$. Furthermore, for all $(\hat{x}^+,x^+) \in \mathscr{R}_j$, we have $q^+ = \tau_{\DFA}(q,L(y^+)) \in Q^+_{\epsilon_j}$. Hence, we can rewrite the last integral over $\mathscr{R}_i$ and find the upper bound
\begin{align*}
&	\int\limits_{\mathscr{R}_i} \max\left\{ \boldsymbol{1}_{Q_f}(q^+),V_c(\hat{x}^+, x^+,q^+,j)\right\} \Wxbar{ij}(d\zeta | z) \\
& \leq \int\limits_{\Xh \times \X} \max\left\{ \boldsymbol{1}_{Q_f}(q^+),V_c(\hat{x}^+, x^+,q^+,j)\right\} \Wxbar{ij}(d\zeta | z),
\end{align*}  where the last integral is equal to $ \Bel_{s_{ij}}^{\hat{u}}(V_c)(\hat{x},x,q,i)$.

By taking the truncation $\mathbf{L}$ to the $[0,1]$ interval over \eqref{eq:Bel_proof1}, we obtain operator $\Bel^{\hat{u}}_{s_{ij}}(V)(\hat{x},q,i)$ as in \eqref{eq:Tmu-epsdel}. This truncation operation does not alter the steps in the proof, and since  $ \Bel_{s_{ij}}^{\hat{u}}(V_c)(\hat{x},x,q,i)$ naturally falls within the interval $[0,1]$, Lemma~\ref{lem:ValueFunc} is proven.
\end{proof}

As the combined model represents the extension of $\Ch$ to $\C$, as depicted in Fig.~\ref{fig:refinement_ML}, we can guarantee that the robust satisfaction probability computed using \eqref{eq:R}, gives a lower bound on the actual satisfaction probability of $\M \times \C$. To recap,  we quantify their similarity between models $\M$ in \eqref{eq:model_ML} and $\Mh$ in \eqref{eq:model_MLAbs} through a multi-layered simulation relation as in Definition~\ref{def:HybSimRel}. We use the abstract model $\Mh$ to compute 1) a policy $\policy$ or equivalently an abstract controller $\Ch: (\hat{u}, s_{ij}) = \policy(\hat{x},q,i)$, and 2) the robust satisfaction $\mb{R}_{\boldsymbol{\epsilon,\delta}}(\Mh \times \Ch \models \phi)$ as in \eqref{eq:R}. In this computation, we take the deviation bounds $(\boldsymbol{\epsilon},\boldsymbol{\delta})$, such that we have  $\Mh \preceq_{\boldsymbol{\epsilon}}^{\boldsymbol{\delta}} \M$, into account. Based on Lemma~\ref{lem:ValueFunc}, we can now conclude that this robust satisfaction probability 
gives a lower bound on the actual satisfaction probability. 

\begin{theorem} \label{prop:controlRef}
Given models $\M$ in \eqref{eq:model_ML}, $\Mh$ in \eqref{eq:model_MLAbs}, DFA $\DFA_{\phi}$, and stationary Markov policy $\policy$. If $\Mh \preceq_{\boldsymbol{\epsilon}}^{\boldsymbol{\delta}} \M$ holds, then there exists $\C$, such that $\mb{P}(\M \times \C \models \phi) \geq \mb{R}_{\boldsymbol{\epsilon,\delta}}^{\policy}(\Mh \times \Ch \models \phi)$.  
\end{theorem}
\begin{proof}
See proof of Theorem~4 in  \cite{haesaert2020robust}, but now with operator $\Bel_{s_{ij}}^{\hat{u}}$. Since this robust DP operator has the same properties as the DP operator in  \cite{haesaert2020robust}, the proof follows along the same lines.
\end{proof}
	


\section{Heterogeneous layers} \label{sec:MM}
In this section, we first introduce the switching condition between heterogeneous layers, which lays the foundation for integrating the DF method with the DB method. 
	Then we review the standard dynamic programming technique for one DF layer, based on which we describe the adjusted heterogeneous layered dynamic programming.
	
\subsection{Switching between heterogeneous layers}
In this subsection, we are interested in switching between DF layers and DB layers. 
Denote the switching actions from a DF layer to a DB layer and vice versa as respectively $\spg$ and $\sgp$.
For ease of notation, we explain everything for only one DF layer and one DB layer, and we remark on how to deal with multiple DF and DB layers. 
%

In the previous section on the multi-layered approach, we use the inherent contraction of the state dynamics to switch to a layer with higher precision. For heterogeneous layers, this is not possible, since the states of the original model that are associated with the DF layer might not cover the complete state space. More specifically, for the DB layers, all states $x\in\X$ can be mapped to a representative state $\hat{x} \in \Xh$ through the operator $\Pi:\X \rightarrow \Xh$. This does not necessarily hold for the DF layer, due to the sparsity of the sample states $x_w$. Hence, we have to define when a switch between heterogeneous layers is allowed. 

A switch from the DB layer to the DF layer is always from one state $\hat{x} \in \Xh$ to one state $x_w \in \Xp$, but a switch from the DF layer to the DB layer is from one state $x_w$ to a set of states in $\Xh$, denoted as $\hat{A}(x_w) \subseteq \Xh$.
We define the conditions of switching between the DF layer and the DB layer as follows.
\begin{definition}[Conditions for switching] 
Given models $\M$,  $\Mp$ \eqref{eq:MPRM} and $\Mh$, 
ellipsoidal sets $\mc{E}$, and 
simulation relation  
$\Ri \subset \Xh \times \X$, such that we have   
$\Mh \preceq_{\epsilon}^{\delta} \M$.
\begin{enumerate}
\item A switch $\sgp$  
from state $\hat{x}$ to state $x_w$ is possible, if $\forall x \in \Ri(\hat{x}): 	(x_w,x) \in \Rp$ holds.
\item A switch $\spg$  
from state $x_w$ to subset $\hat{A}(x_w)$ is possible, if $\forall x \in \Rp(x_w), \exists \hat{x} \in \hat{A}(x_w): (\hat{x},x)\in \Ri$ holds.
\end{enumerate} 
\label{def:switch}
\end{definition}

\begin{remark}
When there are multiple DB layers with different precision, the conditions in Definition~\ref{def:switch} are determined for each DB layer.
\end{remark}

\begin{figure*}[t]
\centering
\subfloat[
For a specific state $\hat{x}$ the states $x \in \Ri(\hat{x})$ are indicated by the light blue ellipsoid, which projected on the DF layer yields the dashed ellipsoid.  The states $(x_w,x) \in \Rp$ are shown in red.]{\includegraphics[width = 0.4\textwidth]{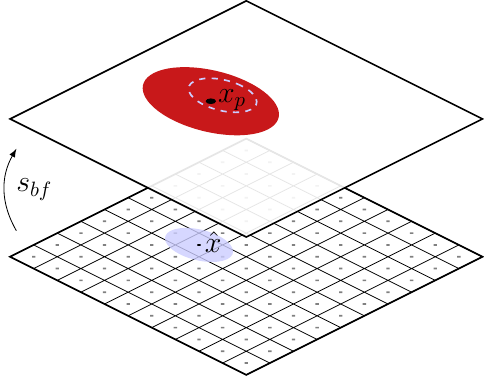}} \quad 
\subfloat[For a specific state $x_w$ the states $x \in \Rp(x_w)$ are indicated by the red ellipsoid, which projected on the DB layer yields the dashed ellipsoid).  The states $(\hat{x},x) \in \Ri$ for $\hat{x} \in \hat{A}$ are shown in light blue. The set $\hat{A}$ consists of the states represented by black dots.]{\includegraphics[width = 0.4\textwidth]{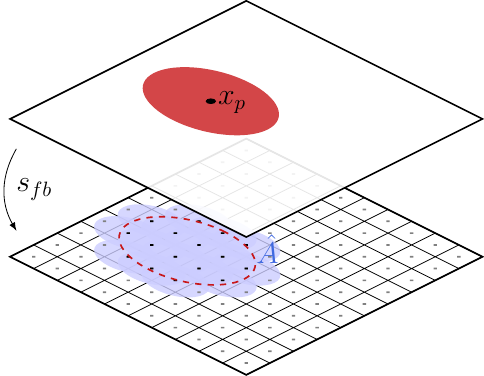}}
\caption{Following respectively conditions 1 and 2 of Definition~\ref{def:switch}, a possibility of switching action $\sgp$ from state $\hat{x}$ towards the state $x_w$ is shown in (a). A possibility of switching action $\spg$ from state $x_w$ towards a subset $\hat{A}(x_w)$ is shown in (b). }
\label{fig:MMswitch}
\end{figure*}

The conditions of Definition~\ref{def:switch} can equivalently be written as 1. $\Ri(\hat{x}) \subseteq \Rp(x_w)$ and 2. $\Rp(x_w) \subseteq \Ri(\hat{A}(x_w))$. In Fig.~\ref{fig:MMswitch} we illustrate these conditions for a specific state and subset $\hat{A}(x_w)$. 

\subsection{Heterogeneous layered dynamic programming}
Let us first introduce the robust DP approach for a single sampling-based layer.
 
\noindent{\bfseries Sampling-based dynamic programming.}
Denote the value function of the DF layer as $\Vp{}: \Xp \times Q \rightarrow [0,1]$. Consider time horizon $[1,\dots,N]$ and a policy $\policy_k=\cramped{(\mu_{k+1}, \ldots \mu_N)}$ with time horizon $N-k$, and with $\mu_k: \Xp \rightarrow \Xp$ that chooses a waypoint $x_w'$ that is reachable from waypoint $x_w$. We compute the value function at the next iteration as $\Vp{}^{\text{next}}= \Vp{, N-k+1}^{{\policy}_{k-1}} = \Bel_{x_w}^{ \mu_k}( \Vp{, N-k}^{\policy_k})$, with the Bellman operator defined as
\begin{align}\label{eq:Bel_PRM}
	&\Bel_{x_w}^{x_w'}(\Vp{})(x_w,q) =
	\textbf{L}\big(  \max\left\{ \boldsymbol{1}_{Q_f}(q'),\Vp{}(x_w',q')\right\} - (1-\Delta_w(x_w,x_w')) \big), \notag 
\end{align}  with $q'$ the DFA state after $n_s$ time steps, that is $q'= q(t+n_s)$, and with $\Delta_w(x_w,x_w')$ the probability of reaching $x_w'$ from $x_w$.  
Comparing this operator for a sampling-based layer to the one for a discretization-based layer as in \eqref{eq:Bell-orig}, we can see that in this case, the Bellman operator does not contain an expected value operator anymore. Hence, the value function is computed in a deterministic fashion.

{\bfseries Heterogeneous layered dynamic programming.}
To distinguish between the value function of the different layers, we denote the value function of the DF layer and DB layer respectively as $\Vp{}: \Xp \times Q \rightarrow [0,1]$ and $\Vh{}: \Xh \times Q \rightarrow [0,1]$. Consider time horizon $[1,\dots,N]$, then we initially update the value functions of both layers \emph{separately}. To this end, define the updated value functions as $\Vp{}^{\text{next}} := \Bel_{x_w}^{ \mu_k}( \Vp{, N-k}^{\policy_k})$ and $\Vh{}^{\text{next}}:= \Bel_{\hat{x}}^{ \mu_k}( \Vh{, N-k}^{\policy_k})$ with the operators defined in respectively \eqref{eq:Bel_PRM} and  \eqref{eq:Bell-orig}. 
After computing the next iteration of the value function for both the DF layer and DB layers, we optimize the switching strategy for each layer. More specifically, we take the maximum of either staying in the same layer or switching to the other layer to implicitly determine the switching strategy. Furthermore, we take the conditions as defined in Definition~\ref{def:switch} into account. 

For the DF layer, we 
compute the value function based on $\Vp{}^{\text{next}}$ and $\Vh{}^{\text{next}}$ 
Since, it is not known towards which specific state $\hat{x} \in \hat{A}(x_w) $ we switch, to preserve a lower bound on the satisfaction probability, we consider the worst-case possibility. The value function that takes a possible switch into account can now be given as
\begin{align}
\label{eq:V_switch1}
& \Vp{, N-k+1}^{\policy_{k-1}} (x_w,q) = \max \left\{\Vp{}^{\text{next}}(x_w,q), \min_{\hat{x}\in\hat{A}(x_w)} \{ \Vh{}^{\text{next}}(\hat{x},q) \} \right\}, 
\end{align} with $\hat{A}(x_w) \subset \Xh$ being the set as defined in point 2 of Definition~\ref{def:switch} if a switch is possible and the empty set, denoted as $\emptyset$ otherwise. If a switch is not possible, hence $\hat{A}(x_w) = \emptyset$, then $\min_{\hat{x}\in\hat{A}(x_w)} \{ \Vh{}^{\text{next}}(\hat{x},q)\}$ equals zero. 

\begin{remark}
When it is allowed to switch from the DF layer towards multiple DB layers  
according to Definition~\ref{def:switch}, an additional $\max$-operator is required to determine the optimal DB layer to switch towards.
\end{remark}

For the DB layer, we compute the value function as 
\begin{equation}
\Vh{, N-k+1}^{\policy_{k-1}}(\hat{x},q) = \max \{\Vh{}^{\text{next}} (\hat{x},q),\Vp{}^{\text{next}}(\mc{S}_{bf}(\hat{x}),q) \}. \label{eq:V_switch2}
\end{equation} Here, $\mc{S}_{bf}: \Xh \rightarrow \Xp \cup x_{\emptyset}$, with $x_{\emptyset}$ an auxiliary state, for which we get  
$\Vp{}^{\text{next}}(x_w,q) = \Vp{}^{\text{next}}(x_\emptyset,q) \equiv 0$,  is a function defined as follows
\begin{equation*}
\mc{S}_{bf}(\hat{x}) = \begin{cases}
x_w \text{ if $\sgp$ is possible from $\hat{x}$ to $x_w$ (see Def.~\ref{def:switch})}, \\
x_\emptyset \text{ otherwise.}
\end{cases}
\end{equation*}

It is trivial to prove that the operations in \eqref{eq:V_switch1} and \eqref{eq:V_switch2} are \emph{monotonically increasing} and \emph{upper bounded by one}. First of all, we can see that \eqref{eq:V_switch1} and \eqref{eq:V_switch2} consist of $\max$- and $\min$-operators and the operators defined in \eqref{eq:Bel_PRM} and  \eqref{eq:Bell-orig}. Since \eqref{eq:Bel_PRM} is a minor adaptation of the standard DP operator in \eqref{eq:Bell-orig}, we can follow Proposition~\ref{prop:properties} to conclude that it is monotonically increasing. Next, we can follow the proof of Lemma~1  in \cite{haesaert2020robust} and show that all of the operators that build up these operations preserve an inequality, such as $V(x,q) \geq W(x,q)$.

The overall value function, that is the value function of the total model with multiple heterogeneous layers is denoted as $\bar{V}: (\Xp \sqcup \Xh) \times Q \rightarrow [0,1]$, with $\sqcup$ the disjoint union. This operator differs from the normal union operator $\bigcup$, by keeping the original set membership as a distinguishing characteristic of the union set. Given policy $\bar{\policy}_k = (\bar{\mu}_{k+1}, \dots, \bar{\mu}_N)$, with $\bar{\mu}: (\Xp \sqcup \Xh) \times Q \rightarrow \Xp \times \Uh$, the overall value function is computed iteratively as $\bar{V}^{\bar{\policy}_{k-1}}_{N-k+1} = \bar{\Bel}^{\bar{\mu}_k}(\bar{V}^{\bar{\policy}_k}_{N-k})$. Here, the overall Bellman-operator $\bar{\Bel}^{\bar{\mu}_k}(\bar{V}^{\bar{\policy}_k}_{N-k})$ is defined as
\begin{equation}\label{eq:overallBell}
\bar{\Bel}^{\bar{\mu}_k}(\bar{V}^{\bar{\policy}_k}_{N-k}) =
\begin{cases}
\eqref{eq:V_switch1}  \text{ for }  x_w \in \Xp  \\
\eqref{eq:V_switch2}  \text{ for }  \hat{x} \in \Xh. 
\end{cases}
\end{equation}

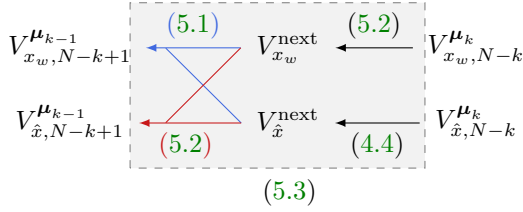
\begin{figure}
\centering
\begin{tikzpicture}[auto,node distance=1 cm,
scale = 0.4]
\draw[dashed, draw=gray, fill=grayfilling] (-11.5,-4) rectangle (-1.7,1.5);
\node[state, draw=none] (V1) {$\Vp{, N-k}^{\policy_k}$};
\node[state,draw=none] (V2) [below of = V1] {$\Vh{, N-k}^{\policy_k}$};
\node[state, draw=none] (V1next) [left of = V1, xshift=-1.5cm]{$\Vp{}^{\text{next}}$};
\node[state, draw=none] (V2next) [left of = V2, xshift=-1.5cm]{$\Vh{}^{\text{next}}$};
\node[state, draw=none] (V1plus) [left of = V1next, xshift=-1.9cm]{$\Vp{, N-k+1}^{\policy_{k-1}}$};
\node[state, draw=none] (V2plus) [left of = V2next, xshift=-1.9cm]{$\Vh{, N-k+1}^{\policy_{k-1}}$};
\path[-latex] 	(V1) edge node [above, align=center]  {\eqref{eq:Bel_PRM}} (V1next);
\path[-latex] 	(V2) edge node [below, align=center]  {\eqref{eq:Bell-orig}} (V2next);
\node[draw=none] (V1temp) [left of = V1next, xshift=-.75cm,yshift=.12cm] {};
\path[-latex, royalblue] 	(V1next) edge node [above, align=center]  {\eqref{eq:V_switch1}} (V1plus);
\draw[royalblue] (V2next.west) -- (V1temp);
\node[draw=none] (V2temp) [left of = V2next, xshift=-.75cm,yshift=-.12cm] {};
\path[-latex, darkred] 	(V2next) edge node [below, align=center]  {\eqref{eq:V_switch2} } (V2plus);
\draw[darkred] (V1next.west) -- (V2temp);
\node at (-6.2,-4.8) {\eqref{eq:overallBell}};
\end{tikzpicture}
\caption{Graphical representation of the value iteration for heterogeneous layers. The operator in \eqref{eq:overallBell} is equivalent to the complete operation in the gray box. The arguments $(x_w,q)$ and $(\hat{x},q)$ of the value functions in respectively the top and bottom rows are omitted for simplicity.
}
\label{fig:VItotal}
\end{figure}

The complete procedure of the value iteration is illustrated in Fig.~\ref{fig:VItotal}.
\begin{remark}
When there are multiple DB layers with different precision, Fig.~\ref{fig:VItotal} becomes more evolved, and among other changes, equation \eqref{eq:Tmu-epsdel} is used instead of \eqref{eq:Bell-orig} to take into account switching between the DB layers.
\end{remark}

We can now conclude the following about the overall Bellman-operator in \eqref{eq:overallBell}.
\begin{theorem}
Given any policy $\bar{\policy}$. 
Suppose that the 
operations in  \eqref{eq:V_switch1} and \eqref{eq:V_switch2} are \emph{monotonically increasing} and \emph{upper bounded by one}, then also the overall Bellman-operator $\bar{\Bel}^{\bar{\mu}_k}(\bar{V}^{\bar{\policy}_k}_{N-k})$ as in \eqref{eq:overallBell} is \emph{monotonically increasing} and \emph{upper bounded by one}.
\end{theorem}
\begin{proof}
The overall Bellman-operator $\bar{\Bel}^{\bar{\mu}_k}(\bar{V}^{\bar{\policy}_k}_{N-k})$ as in \eqref{eq:overallBell} 
is either computed as \eqref{eq:V_switch1} or \eqref{eq:V_switch2} depending on the considered state. Since both operations in  \eqref{eq:V_switch1}, and \eqref{eq:V_switch2} 
are monotonically increasing and upper bounded by one, this also holds for the overall Bellman-operator $\bar{\Bel}^{\bar{\mu}_k}(\bar{V}^{\bar{\policy}_k}_{N-k})$.
\end{proof}

As before, the value function gives the probability of satisfying the specification in $1$ to $\infty$ time step, by including the first time instance based on $x_0$, we can compute the robust satisfaction probability, that is
\begin{equation} \label{eq:R_hetero}
\bar{\mathbb{R}}^{\bar{\policy}} := \max\left(\boldsymbol{1}_{Q_f}(\bar{q}_0),\bar{V}^{\bar{\mu}}_{\infty}\big((x_{w,0},\hat{ x}_0),\bar{q}_0\big)\right),
\end{equation} with $\bar{q}_0 = \tau_{\DFA_{\phi}}(q_0,L(h(x_0)))$, and with $x_{w,0} \in \mc{E}^{-1}(x_0)$ and $\hat{x}_0 \in \mathscr{R}^{-1}(x_0)$.

Following the same reasoning as in Section~\ref{sec:contrRef}, 
we can now conclude that the robust satisfaction probability in \eqref{eq:R_hetero} computed through the overall Bellman-operator $\bar{\Bel}^{\bar{\mu}_k}(\bar{V}^{\bar{\policy}_k}_{N-k})$ in \eqref{eq:overallBell}  provides a lower bound on the actual satisfaction probability $\mb{P}(\M \times \C \models \phi)$.

\section{Results}\label{sec:results_Ch4} 
To show the benefits of the multi-layered approach, we consider several case studies. We start with homogeneous layers, where we consider two case studies with increasing complexity. Next, we show that the multi-layered approach with heterogeneous layers outperforms similar single-layered approaches by applying it to a complex case study. All simulations are performed on a computer with a 3.4 GHz 13th Gen Intel Core i7-13700K processor and 64 GB 3200 MHz memory, and use the toolbox \syscore \cite{huijgevoort2023syscore} as a basis. For each case study, we mention the computation time and memory usage. The computation time is averaged over $5$ computations, where we observed a maximum $9\%$ standard deviation. The memory usage is computed as the size of the matrices stored in the workspace.

\begin{figure}[t]
	\centering
	\includegraphics[width=.3\linewidth]{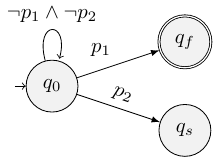}
	\caption{DFA $\DFA_{\phi_{park}}$ associated with specification $\phi_{park}= \lnot p_2 \until p_1$. The initial, final, and sink state are denoted by respectively $q_0, q_f$, and $q_s$.}
	\label{fig:DFA}
\end{figure}

\subsection{Simple reach-avoid specification}
Consider a simple reach-avoid specification to park a car in a 
two-dimensional (2D) space. 
The goal of the controller is to guarantee that the car parks in area $P_1$, without going through area $P_2$. This specification can be written as 
\begin{equation} \label{eq:phi_park}
	\phi_{park}= \lnot p_2 \until p_1,
\end{equation}
where the labels $p_1$, and $p_2$ correspond to respectively regions $P_1$ and $P_2$, and can be represented by the DFA in Fig.~\ref{fig:DFA}.

The dynamics of the car are modeled using an LTI stochastic difference equation as in \eqref{eq:model_MLLTI} with $A=0.9I_2, B=0.5I_2$ and $B_w=C=1$. We used states $x\in \X = [-5,5]^2$, inputs $u \in \U=[-1,1]^2$, outputs $y\in \mathbb{Y}=\X$ and Gaussian disturbance $w \sim \mathcal{N}(0,0.5)$. 
Furthermore, we have regions $P_1 = [3,5] \times [-1,0]$ and $P_2 = [3,5] \times [0,1]$ defined on the output space $\Y$ and labeled as respectively $p_1$ and $p_2$. 

\begin{figure*}[t]
\centering
\subfloat[Switching strategy for layer 1. The blue and red dots correspond to $s_{11}$ and $s_{12}$ respectively.] {\includegraphics[width=0.4\textwidth]{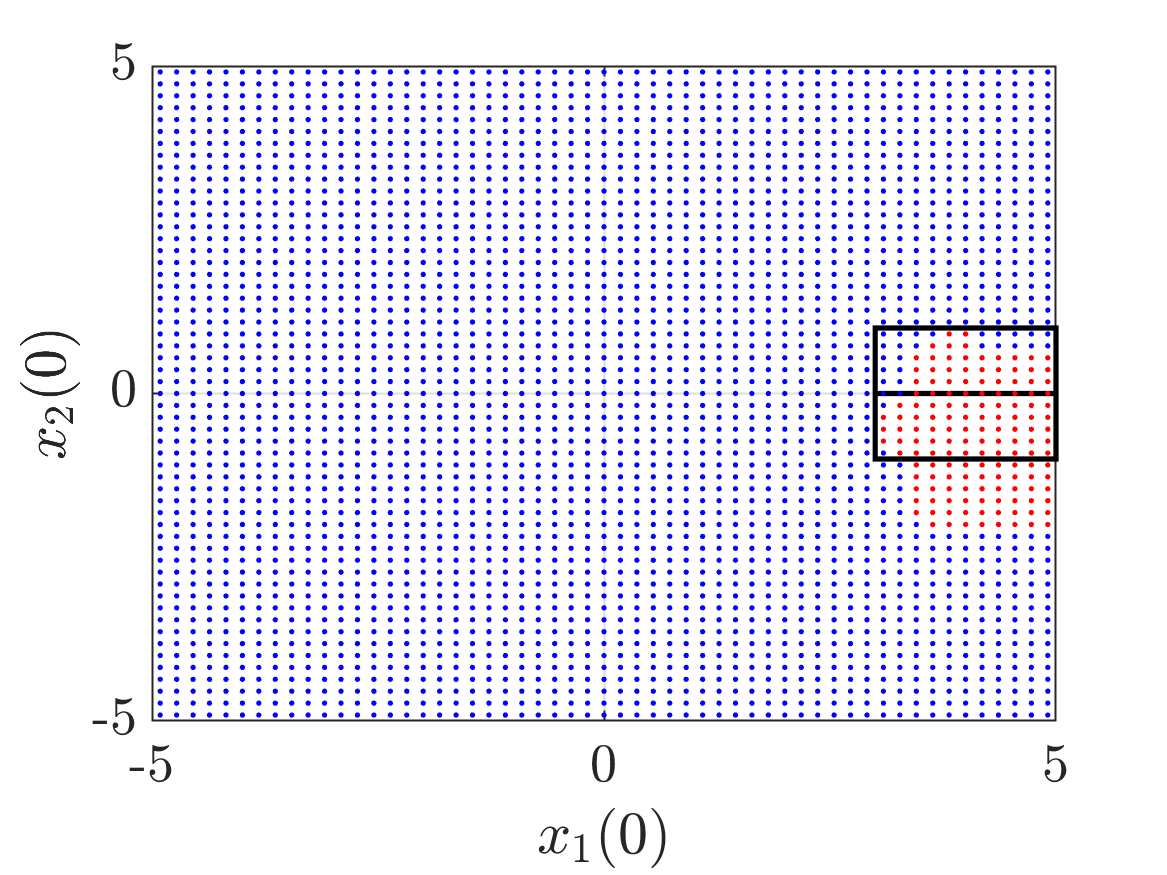}\label{fig:CP2_SS_lay1}}\qquad
\subfloat[Switching strategy for layer 2. The blue and red dots correspond to $s_{21}$ and $s_{22}$ respectively.]  {\includegraphics[width=0.4\textwidth]{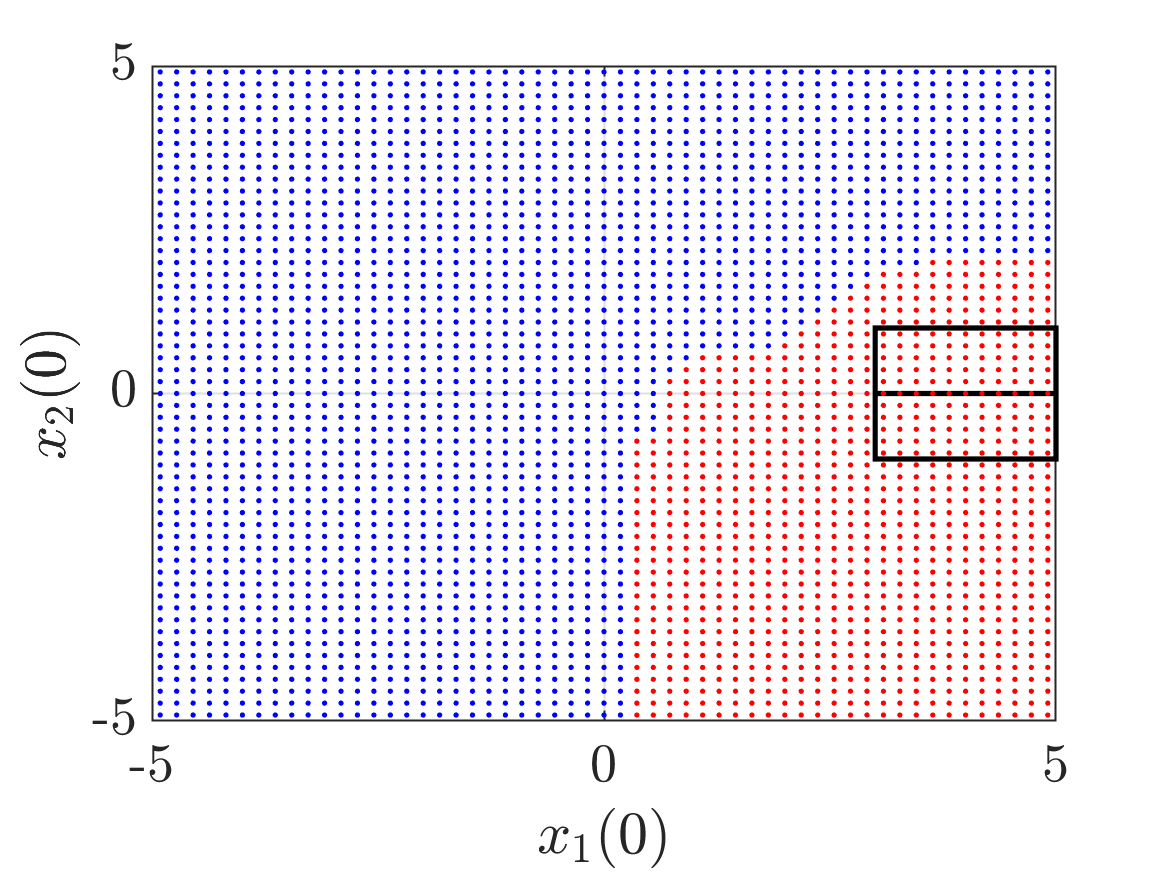}\label{fig:CP2_SS_lay2}}
\caption{Switching strategy for the states of the surrogate model of the 2D car park case study. Here, 
a blue resp. red dot represents switching to layer 1 resp. layer 2. The black boxes indicate regions $P_1$ (bottom) and $P_2$ (top). } 
\label{fig:SS_carPark2D}
\end{figure*}

We computed deviation bounds $(\boldsymbol{\epsilon}, \boldsymbol{\delta})$ that satisfy Lemma~\ref{lem:epsdelBounds} as
\begin{align*}
\boldsymbol{\epsilon} = \begin{bmatrix}
0.5 & 0.2
\end{bmatrix}, \quad &  \boldsymbol{\delta} = \begin{bmatrix}
0 &0.168  \\ 
0 & 0.0169
\end{bmatrix}.
\end{align*}
Next, we obtained a surrogate model by partitioning with $55 \times 55$ grid cells and found the optimal switching strategy\footnotemark[1]  corresponding to this grid as illustrated in Fig.~\ref{fig:SS_carPark2D}. 
\begin{figure*}[t]
\centering
\subfloat[
Single layer, with $(\epsilon_1,\delta_1) = (0.5,0)$.] {\includegraphics[width=0.3\textwidth]{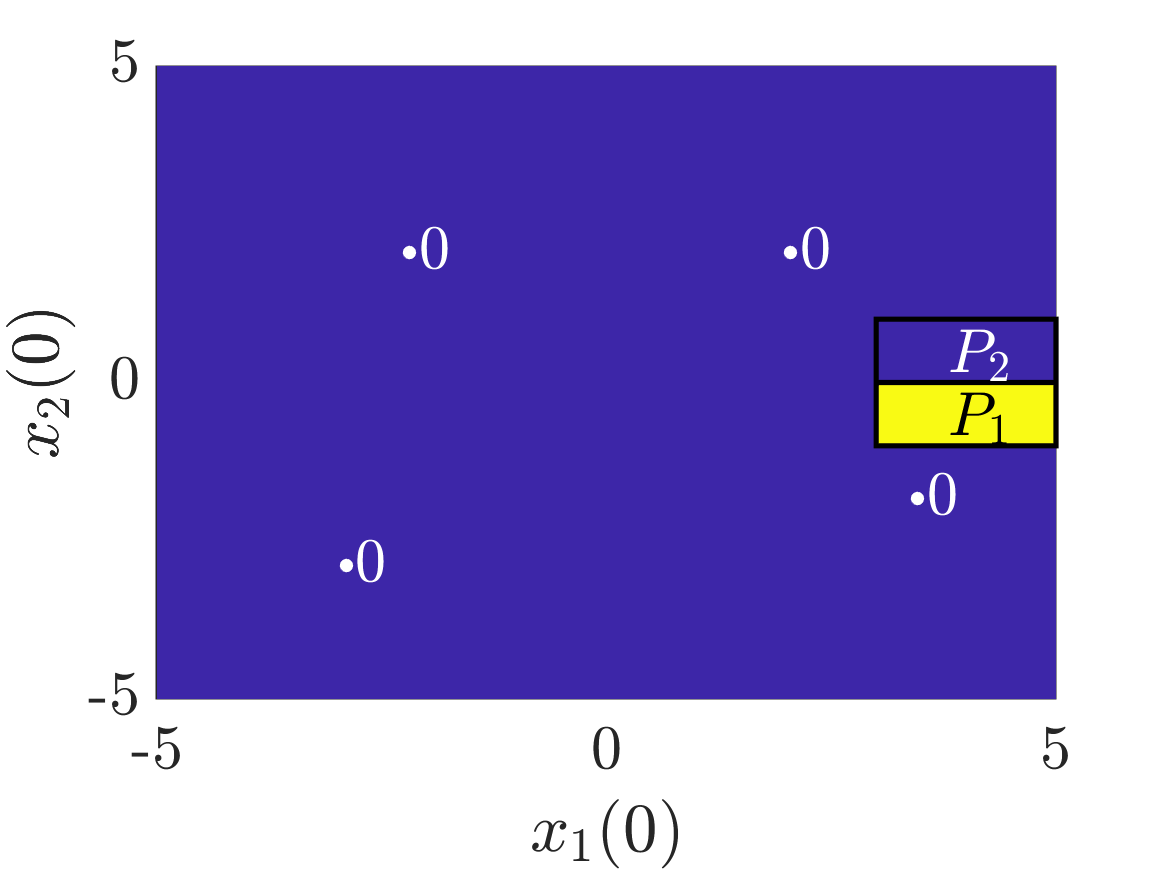}\label{fig:CP2_satProb_R1}} \hfill
\subfloat[
Single layer, with $(\epsilon_2,\delta_2) = (0.2,0.0169)$.]  {\includegraphics[width=0.3\textwidth]{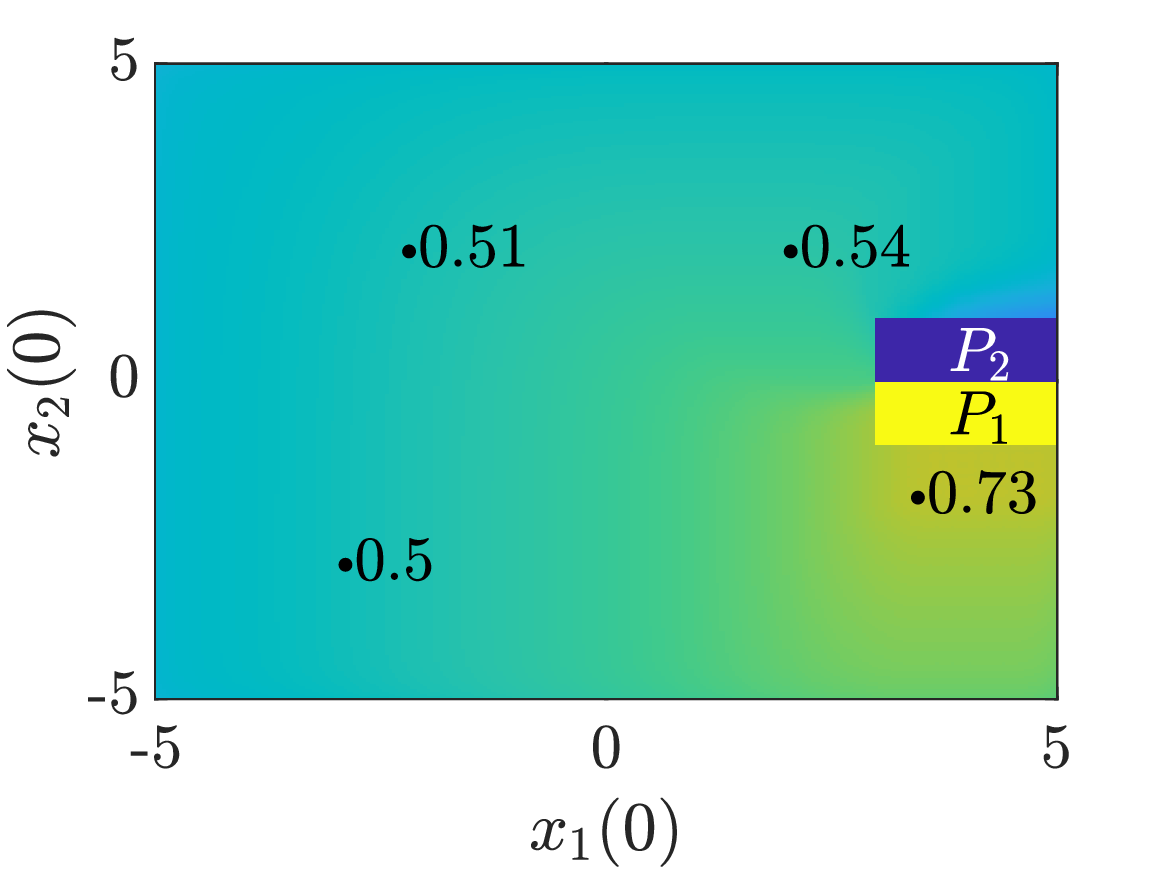}\label{fig:CP2_satProb_R2}}  \hfill 
\subfloat[
Multiple layers with  simulation relation~$\boldsymbol{\mathscr{R}}$.] {\includegraphics[width=0.3\textwidth]{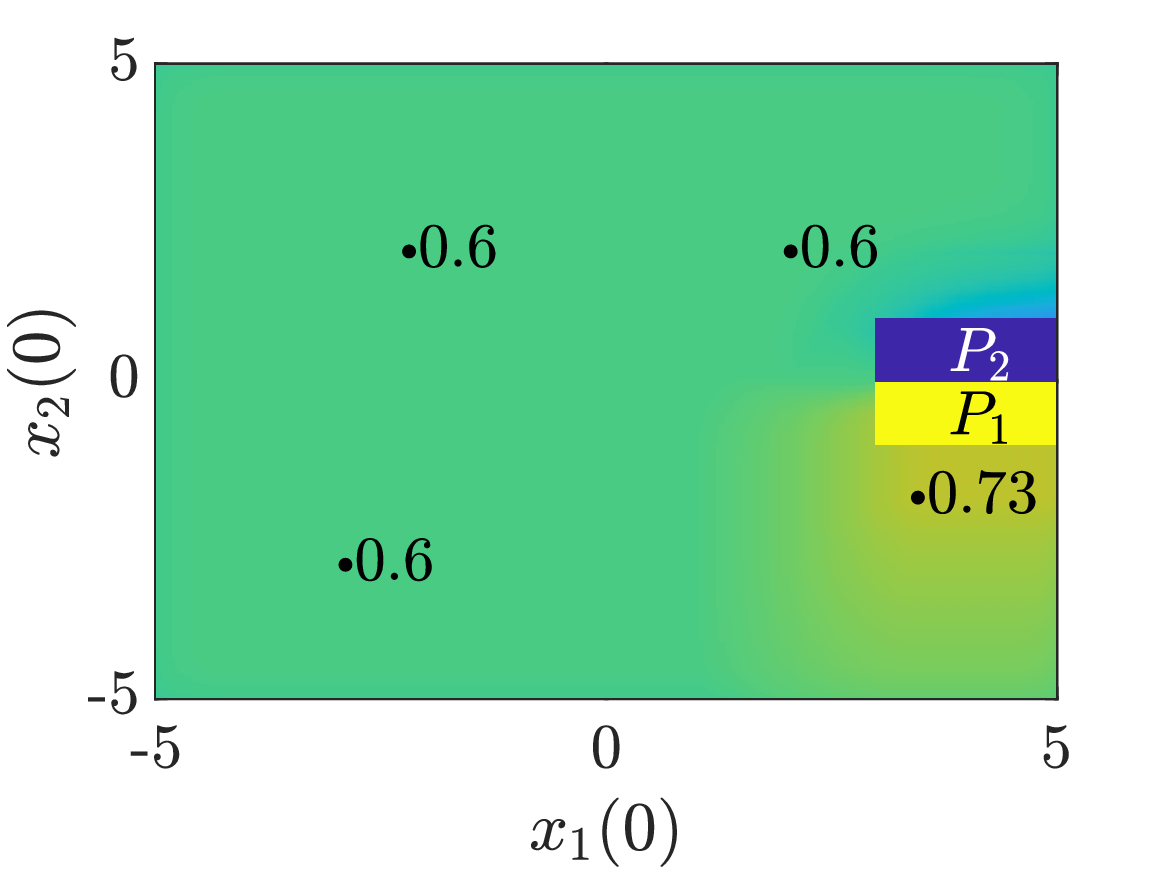}\label{fig:CP2_satProb_ML}} \hspace{-1em}
\includegraphics[width=0.06\textwidth]{Figures/Results/colorbar}
\caption{Robust satisfaction probability of the 2D car park case study where a single layer is used in (a) and (b). A multi-layered simulation relation, with switching strategy as in Fig.~\ref{fig:SS_carPark2D} is used in~(c).} 
\label{fig:CP2_satProb}
\end{figure*}

Then, we constructed abstract model $\Mh$ in the form of \eqref{eq:model_MLAbsLTI} by partitioning 
with $283 \times 283$ regions leading to $\mathscr{B}=[-0.0353,0.0353]^2$. 
We quantified the accuracy of $\Mh$ with a bi-layered simulation relation $\boldsymbol{\mathscr{R}}$ and 
obtained the robust satisfaction probability 
in Fig.~\ref{fig:CP2_satProb_ML}. The average 
computation time is $34.4$ seconds while using a memory of $243$ MB.
The robust satisfaction probability of the multi-layered approach is higher everywhere compared to only using a single layer, with either $\mathscr{R}_1,$ with $(\epsilon_1,\delta_1) = (0.5,0)$ or $\mathscr{R}_2$, with $(\epsilon_2,\delta_2) = (0.2,0.0169)$ as can be seen in Fig.~\ref{fig:CP2_satProb}.

\subsection{Package delivery specification}\label{sec:caseB}
\begin{figure*}[t]
\centering
\subfloat[
Single layer with $(\epsilon_1,\delta_1) = (0.5,0)$. DFA state $q_0$.] {\includegraphics[width=0.3\linewidth]{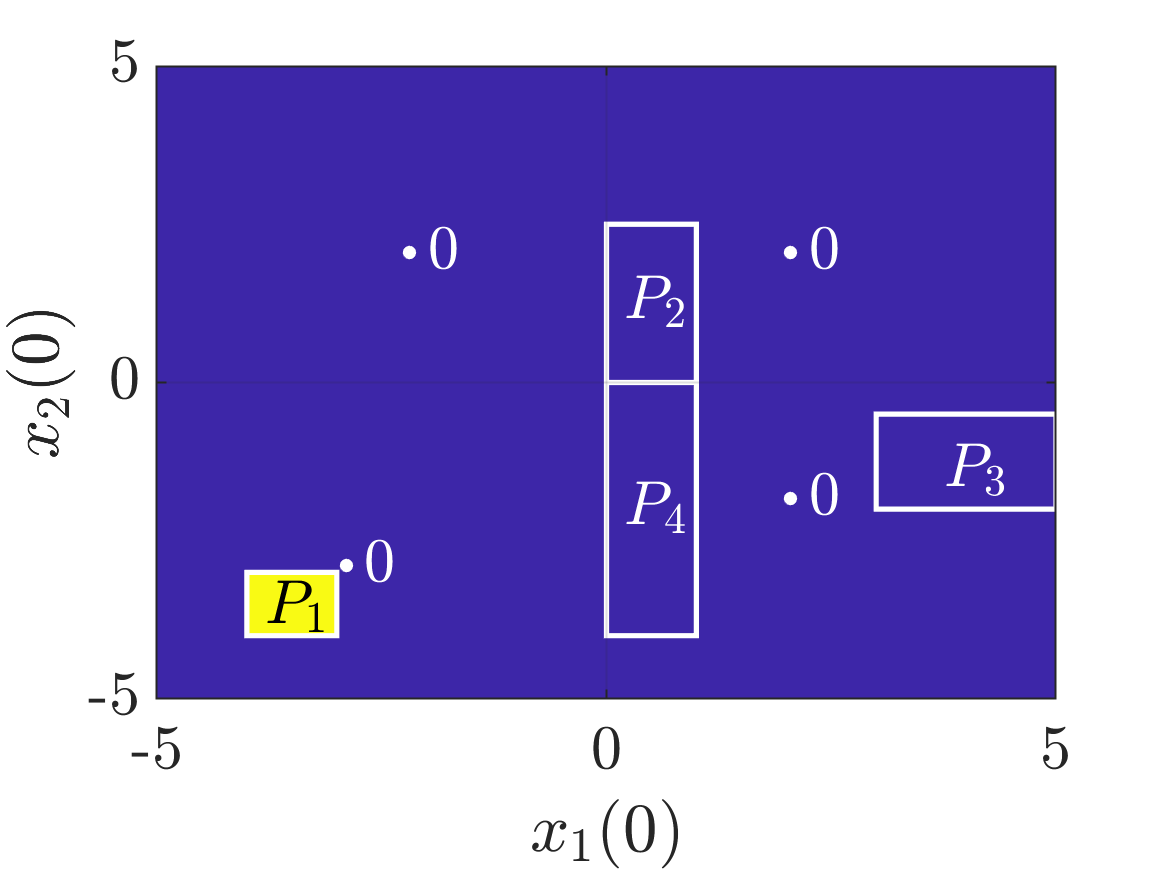}\label{fig:PD_satprob_R1}}\hfill
\subfloat[
Single layer with $(\epsilon_2,\delta_2) = (0.3,0.0160)$. DFA state $q_0$.]  {\includegraphics[width=0.3\linewidth]{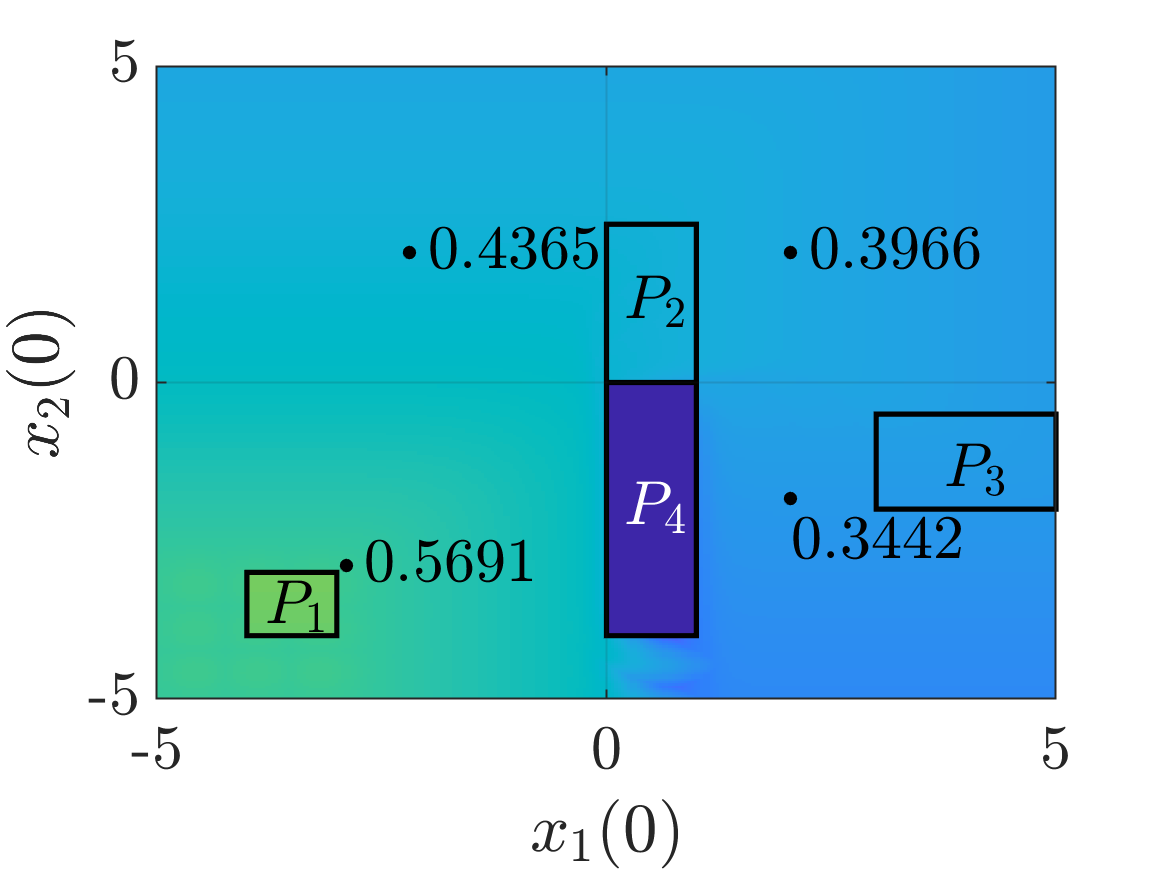}\label{fig:PD_satprob_R2}} \hfill 
\subfloat[
Multiple layers with  simulation relation~$\boldsymbol{\mathscr{R}}$. 
DFA state~$q_0$.]  {\includegraphics[width=0.3\linewidth]{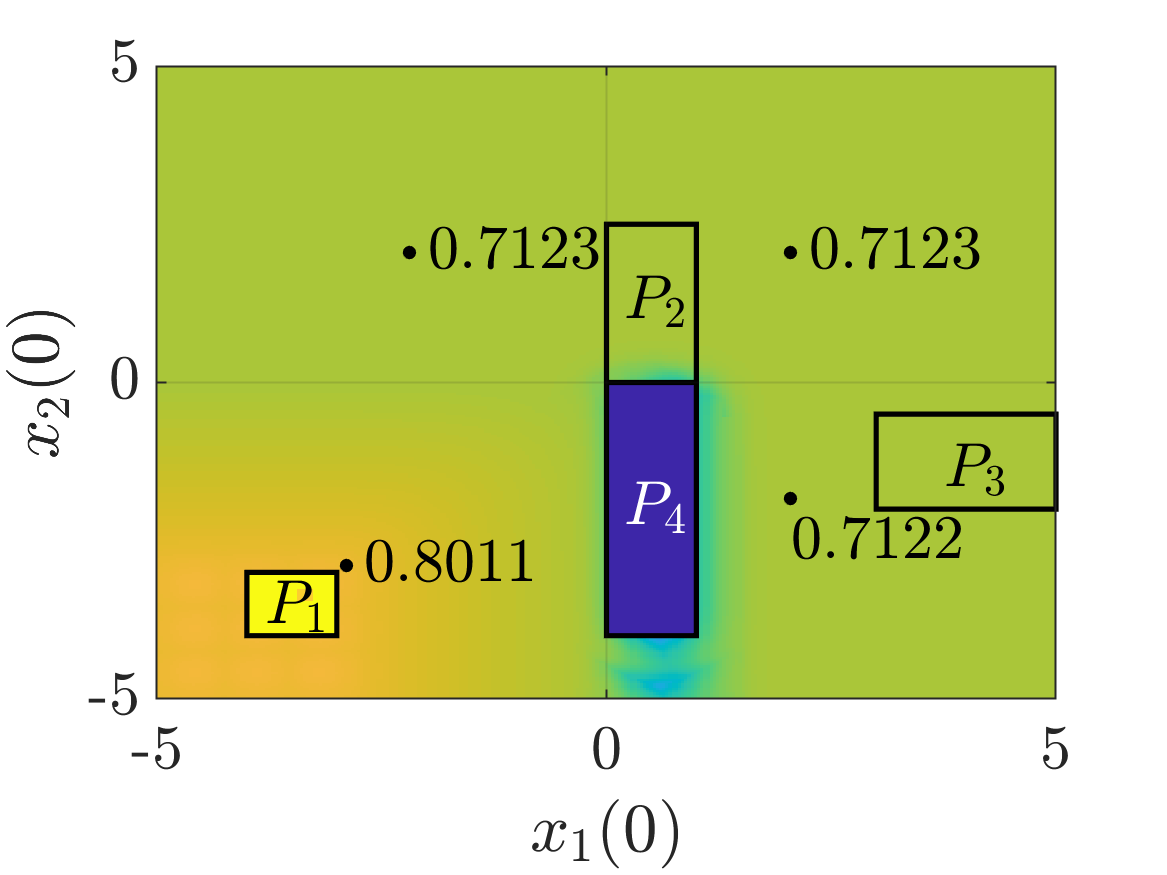}\label{fig:PD_satprob_ML}} 
\hfill
\includegraphics[width=0.06\linewidth]{Figures/Results/colorbar}

\subfloat[
Single layer with $(\epsilon_1,\delta_1) = (0.5,0)$. DFA state $q_1$.] {\includegraphics[width=0.3\linewidth]{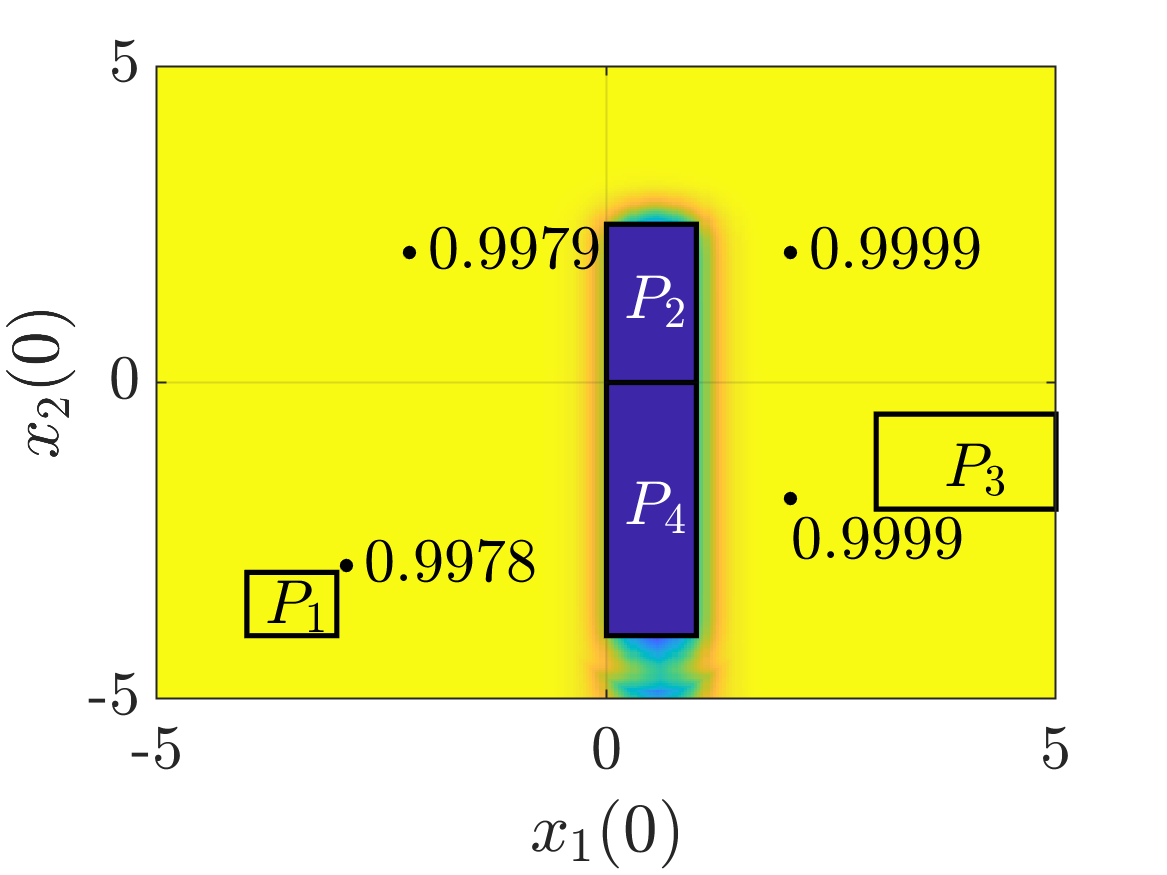}\label{fig:PD_satprob_R1_q1}}\hfill
\subfloat[
Single layer with $(\epsilon_2,\delta_2) = (0.3,0.0160)$. DFA state $q_1$.]  {\includegraphics[width=0.3\linewidth]{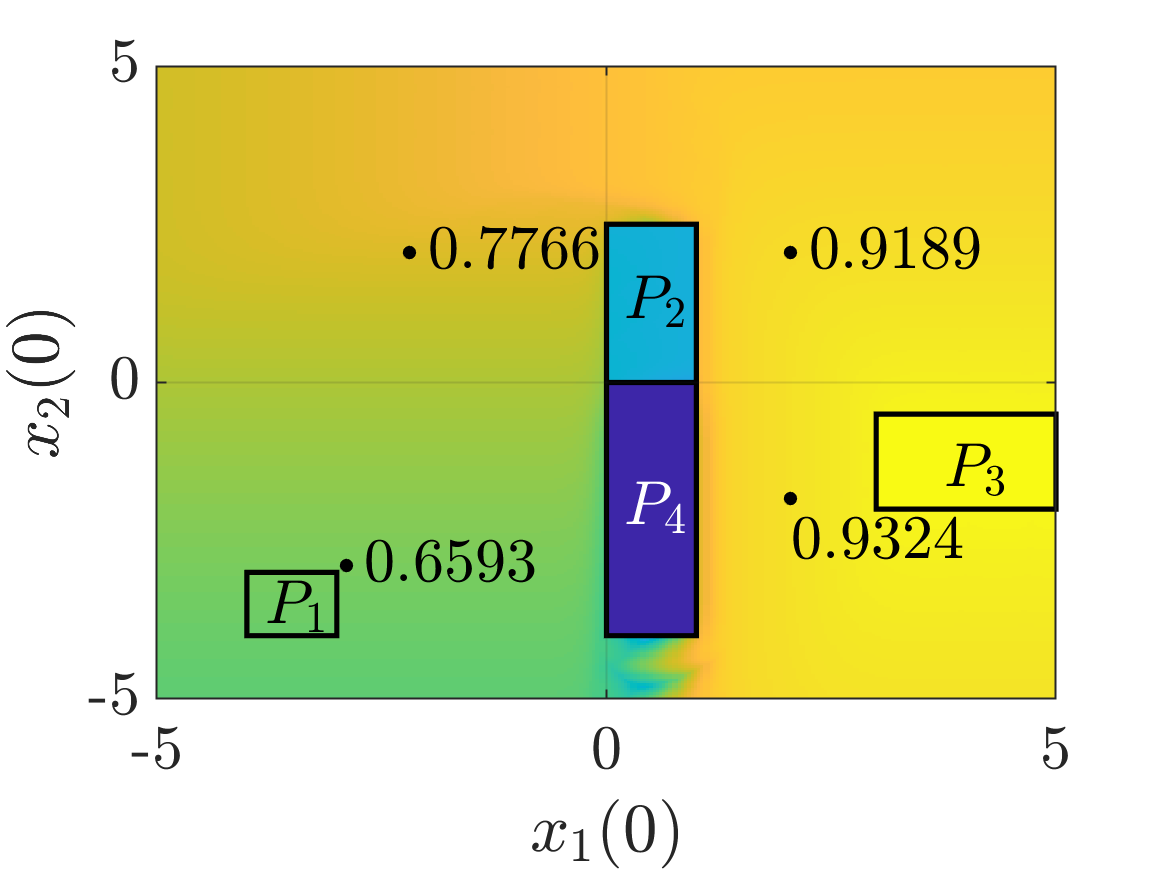}\label{fig:PD_satprob_R2_q1}} \hfill 
\subfloat[
Multiple layers with  simulation relation~$\boldsymbol{\mathscr{R}}$. 
DFA state~$q_1$.] {\includegraphics[width=0.3\linewidth]{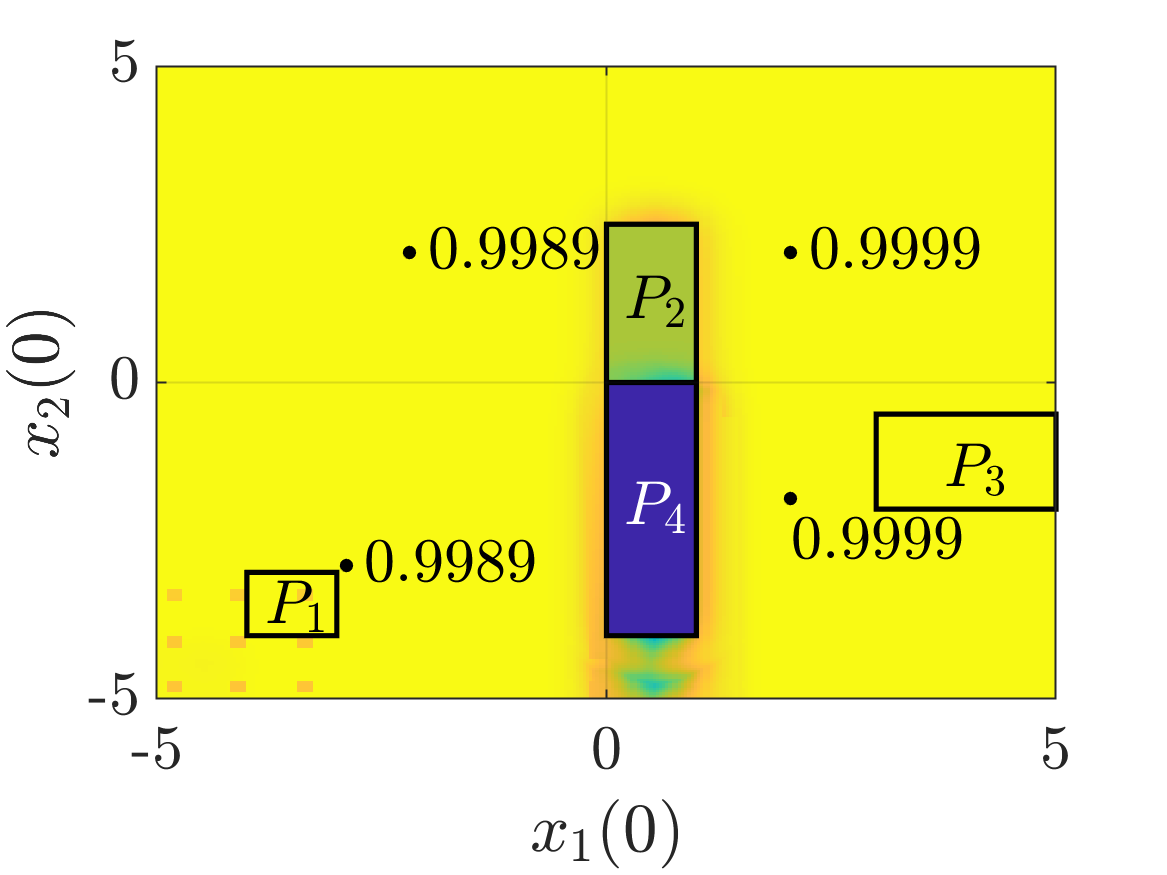}\label{fig:PD_satprob_ML_q1}} 
\hfill
\includegraphics[width=0.06\linewidth]{Figures/Results/colorbar}

\caption{Robust satisfaction probability of package delivery case for DFA state $q_0$ in (a)-(c) and for DFA state $q_1$ in (d)-(f). Here, a single layer is used in (a),(b), and (d),(e). A multi-layered simulation relation, with switching strategy as in Fig.~\ref{fig:SS_PD} is used in~(c) and (f).}  
\label{fig:PD_satprob}
\end{figure*}

Next, we 
revisit running example A by repeating the important steps. We also show the satisfaction probability and compare it to single-layered approaches.
%
As described throughout the paper, we consider a package delivery scenario with specification \eqref{eq:case3_specification},
with its corresponding DFA given in Fig.~\ref{fig:DFA_PD}. 
%
Based on Appendix~\ref{sec:Impl} and Lemma \ref{lem:epsdelBounds}, we found total deviation bounds 
\begin{align*}
\boldsymbol{\epsilon} = \begin{bmatrix}
0.5 & 0.3
\end{bmatrix}, \quad &  \boldsymbol{\delta} = \begin{bmatrix}
0 &0.1586  \\ 
0 & 0.0160
\end{bmatrix}.
\end{align*}
Next, we obtained a surrogate model by partitioning with $55 \times 55$ grid cells and found a switching strategy by following Alg.~\ref{alg:SS}. The optimal switching strategy corresponding to this grid is illustrated in Fig.~\ref{fig:SS_PD}. 
Then, we constructed abstract model $\Mh$ in the form of \eqref{eq:model_MLAbsLTI} by partitioning the state space 
with $283 \times 283$ regions and the input space with $3 \times 3$ regions. We 
obtained the robust satisfaction probability for DFA states $q_0$ and $q_1$ 
in respectively \mbox{Fig.~\ref{fig:PD_satprob_ML} and \ref{fig:PD_satprob_ML_q1}.}  The average 
computation time is $50$ seconds while using a memory of $340$ MB. The robust satisfaction probability of the multi-layered approach is higher everywhere compared to only using a single layer, as can be seen in Fig.~\ref{fig:PD_satprob}.

\subsection{Case study with heterogeneous layers}\label{sec:CaseHet}
To show the benefit of having heterogeneous layers, we apply it to a case study with the same complex reach-avoid specification as before, whose DFA is given in Fig.~\ref{fig:DFA_PD}. \newline 
The system dynamics are the same as of the running example. In terms of comparison to the system dynamics in Section~\ref{sec:caseB}, 
the state space is enlarged to $\mathbb{X}=[-20,5]^2$, the input space to $\mathbb{U}=[-5,5]^2$, and output space to $\mathbb{Y}=\mathbb{X}$. The full dynamics are given in \eqref{eq:running_example_original_model}.

For this case study, we implement heterogeneous layers with one DF and one DB layer, while we consider a different amount of waypoints in the DF layer. 
For the sampling-based layer, we chose $\Delta_w = 0.0001$. To obtain a good accuracy for the heterogeneous approach, $\Delta_w$ of the DF layer, should be substantially smaller than $\delta$ of the DB layer. It is in general, a good idea to choose $\Delta_w$ very small. We have matrix $D_w = I_2$ for the set $\mc{E}$ as in \eqref{eq:epsSet}. To decrease the size of the set $\mc{E}$, we used an interface function  \eqref{eq:interface} equal to $u_t = u_{w,t} + K(x_t - x_{w,t})$, with $K=-1.4I_2$. Following the derivation in the appendix, we computed $\epsilon_w = 2.6825$ for set $\mc{E}$ in  \eqref{eq:epsSet}.

For the discretization-based layer, we compute a finite-state abstraction as in \eqref{eq:model_MLAbsLTI} by gridding part of the state space $[-9,5] \times [-5,3] $ with $318 \times 180$ grid cells. We compute the similarity quantification between the original model $\M$ \eqref{eq:model_MLLTI} and its finite-state abstraction $\Mh$ \eqref{eq:model_MLAbsLTI} for interface function $u=\hat{u}$, and obtained $(\epsilon,\delta) = (0.18,0.1217)$.

\begin{figure*}[t]
	\centering
	\includegraphics[width=0.4\textwidth]{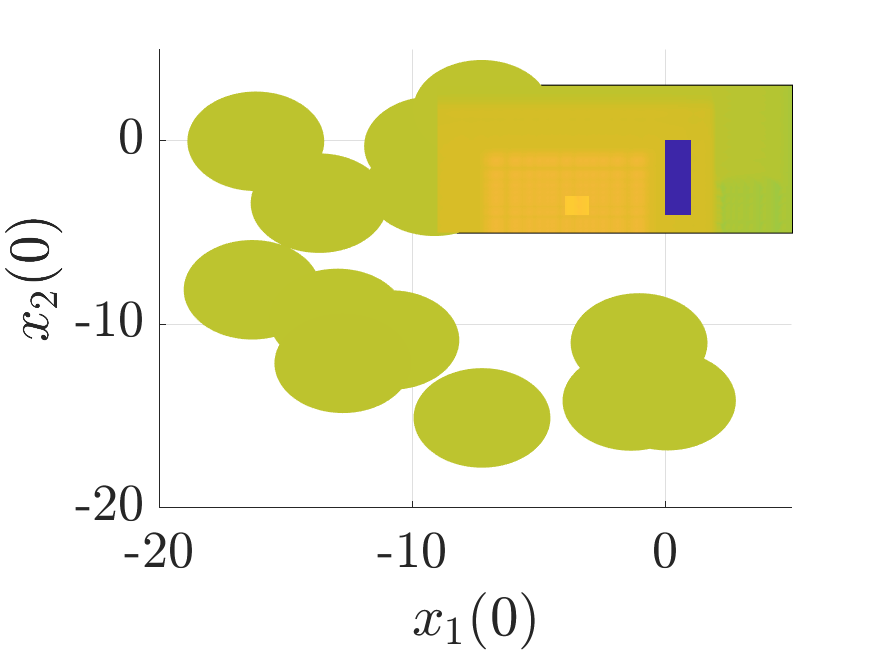}
	\hspace{-1em}
	\includegraphics[width=0.085\textwidth]{Figures/Results/colorbar}
	\caption{Robust satisfaction probability of package delivery case with enlarged spaces (Section~\ref{sec:CaseHet}) when using a multi-layered approach with heterogeneous layers. We used $(\epsilon,\delta)=(0.18,0.1217)$ for the DB layer with gridded area $[-9,5]\times [-5,3]$, and $\epsilon_w=2.6825,\Delta_w=0.0001$ for the DF layer.}
		\label{fig:1DB1SB_80}
\end{figure*}

Next, we perform a value iteration and synthesize a controller using the technique described in Section~\ref{sec:MM}. The robust satisfaction probability obtained with $48$ initial waypoints\footnote{The term \emph{initial waypoints} refers to the number of waypoints initially supplied to the algorithm. The number of waypoints in the end may be smaller due to checking well-posedness or larger due to re-sampling.} is shown in Fig.~\ref{fig:1DB1SB_80}. The satisfaction probability for the sample states is depicted as ellipsoids. The average 
computation time equals $5.05$ seconds, while using a memory of $653$ MB. When increasing the number of waypoints from $48$ to $180$, the average 
computation time increases from $5.05$ to $9.64$ seconds, while the memory usage is the same. The accuracy has not changed substantially, except for a larger coverage of the state space by the sampling-based layer. Hence, we omit the figure with the satisfaction probability. 

\begin{table*}[]
	\caption{Comparison between the multi-layered approach with heterogeneous layers and the corresponding single-layered approaches. The table includes satisfaction probabilities, computation time and memory usage. Here, \emph{DB, DF} denote respectively a single-layered \emph{discretization-based} or \emph{discretization-free} approach. \emph{DB+DF} denotes the multi-layered approach with heterogeneous layers. \emph{Sat. prob.} abbreviates \emph{satisfaction probability} and the minimal (nonzero) satisfaction probability is listed here. \emph{Comp. time} abbreviates \emph{Computation time}.}
	\label{tab:ruohan_comp}
	\scalebox{0.6}{
	\begin{tabular}{|l|l|l|l|l|l|l|l|}
		\hline
		
		 Approach                & Figure 	& Gridded area & Grid size         &  Grid cells & Sat. prob. & Comp. time (s) & Memory (MB)                      \\ \hline
		\textit{\textbf{DB}} & 		\ref{fig:rex_pureDB}		&  $[-20,5]^2$& $[0.0440 \times 0.0444]$ & $[568 \times 563]$ & $0.6846$ & $14.4$ & $2706$ \\ \hline
			\textit{\textbf{DF}} & \ref{fig:rex_pureSB} 		&   n.a     & n.a.              &     n.a.      & $0$& $0.3974$ & $0.0268$ \\ \hline
		 \textit{\textbf{DB+DF}} & \ref{fig:1DB1SB_80}	    & $[-9,5] \times [-5,3]$                       & $[0.0440 \times 0.0444]$ & $[318 \times 180]$           &$0.7343$  & $5.05$ & $653$  \\ \hline
		 
	\end{tabular}}
\end{table*}

\textbf{Comparison between a single-layered approach and a multi-layered approach with heterogeneous layers.} \newline 
Next, we make a fair comparison between the different approaches. First, we consider only a discretization-based approach, where we gridded the complete state space with $568 \times 563$ grid cells. Note that the grid size is now almost equal to the one we obtained before. We compute the similarity quantification between the original model $\M$ \eqref{eq:model_MLLTI} and its finite-state abstraction $\Mh$ \eqref{eq:model_MLAbsLTI} for interface function $u=\hat{u}$, and obtained $(\epsilon,\delta)=(0.18,0.1217)$. The robust satisfaction probability is shown in Fig.~\ref{fig:rex_pureDB}. The computation time equals $14.4$ seconds while using a memory of $2706$ MB.
Next, we consider only a sampling-based approach, where we used $\epsilon_w = 2.6825$ and obtained $48$ waypoints. The robust satisfaction probability is shown in Fig.~\ref{fig:rex_pureSB}. Here, we see that it gives $0$ for all robust satisfaction probabilities of waypoints $x_w \in \mathbb{X}_w$, due to the large set $\mc{E}$. The computation time equals $0.3974$ seconds while using a memory of $0.0268$ MB. It should be noted that due to the large set $\mc{E}$, we cannot obtain a strongly connected graph. 

Comparing the satisfaction probabilities in Fig.~\ref{fig:1DB1SB_80} and in Fig.~\ref{fig:rex_comparison} we see that the robust satisfaction probability of the heterogeneous-layered approach in Fig.~\ref{fig:1DB1SB_80} is higher than when using either only a discretization-based layer (see Fig.~\ref{fig:rex_pureDB}) or only a sampling-based layer (see Fig.~\ref{fig:rex_pureSB}). The results are summarized in Table~\ref{tab:ruohan_comp}. Here, we see that the heterogeneous approach outperforms the discretization-based approach on accuracy (satisfaction probability), computation time, and memory usage.
Therefore, we conclude that the multi-layered approach with heterogeneous layers outperforms the single-layered approach with either a sampling-based technique or a discretization-based technique.

\section{Conclusions}
\label{sec:conclusions}
In this paper, we proposed a multi-layered approach that allows us to switch between multiple simulation relations (homogeneous layers) and multiple abstraction-based techniques (heterogeneous layers). This approach makes it possible to use the advantages of each individual 
layer guided by the given specification, hence providing an efficient and less conservative computational approach for temporal logic control.
We illustrated the benefits of the proposed multi-layered approach with homogeneous and heterogeneous layers by applying it to multiple case studies. They all show the improved accuracy of the computations, and the case study with heterogeneous layers indicated a significant increase of $65\%$ in computational efficiency. 

In the future, we plan to integrate a comprehensive implementation of the approach within the tool \syscore~\cite{huijgevoort2023syscore}, which enhances its practical applicability and ease of use. An interesting topic for future research is to develop the theoretical details for allowing multiple discretization-based abstract models with different grid sizes and integrating abstraction-free methods that are based on the concepts of barrier certificates. 

\bibliographystyle{plain}
\bibliography{Sources} 

\appendix	
	\section{Implementation of similarity quantification for LTI systems}\label{sec:Impl} 
	In this section, we elaborate on how to compute the deviation bounds $(\boldsymbol{\epsilon},\boldsymbol{\delta})$ such that $\Mh \preceq_{\boldsymbol{\epsilon}}^{\boldsymbol{\delta}} \M$ holds (step 2 of Alg.~\ref{alg:controlSyn}). The derivation is given for linear time-invariant (LTI) systems but can be extended to nonlinear stochastic systems following the techniques described in \cite{huijgevoort2022piecewiseaffineabstraction}. The approach detailed next, follows the same reasoning as in \cite{huijgevoort2022similarity}, but we made adaptations such that it suits the multi-layered setting.
	
	Let the models $\M$ \eqref{eq:model_ML} and $\Mh$ \eqref{eq:model_MLAbs} be linear time-invariant (LTI) systems whose behavior is described by the following stochastic difference equations 
	\begin{equation}
		\M: \begin{cases}
			x(t+1) = Ax(t)+Bu(t)+B_ww(t) \\
			y(t) = Cx(t), \textrm{ and }
		\end{cases}
		\label{eq:model_MLLTI}
	\end{equation}
	\begin{equation}
		\Mh: \begin{cases}
			\hat{x}(t+1) = \Pi\left( A\hat{x}(t)+B\hat{u}(t)+B_w\hat{w}(t) \right) \\
			\hat{y}(t) = C\hat{x}(t),
		\end{cases}
		\label{eq:model_MLAbsLTI}
	\end{equation}
	with matrices $A, B, B_w$ and $C$ of appropriate sizes and with the disturbances $w(t),\hat w(t)$ having the standard Gaussian distribution, i.e., $w(t)\sim\mathcal{N}(0,I)=\mathbb P_{w}$ and $\hat w(t)\sim\mathcal{N}(0,I)=\mathbb P_{\hat w}$.  
	The abstract model is constructed by partitioning the state space $\X$ in a finite number of regions $\mathbb{A}_j \subset \X$ and operator $\Pi(\cdot): \X\rightarrow \Xh$ maps states $x\in \mathbb{A}_j$ to their representative points\footnote{In general, any point in the region $\mb{A}_j$ can be its representative point, but in practice the center has computational benefits.} $\hat{X}_j\in\Xh$. We assume that the  
	regions $\mathbb{A}_j$ are designed in such a way that the set $\mathscr{B}:= \bigcup_j\{\hat X_j-x_j | x_j\in \mathbb A_j\}$ is a bounded polytope and has vertices $\textrm{vert}(\mathscr{B})$. Details on constructing such an abstract LTI system can be found in \cite{huijgevoort2022similarity}.
	
	To compute the multi-layered simulation relations in 
	Definition \ref{def:HybSimRel},
	we choose the interface function $
	u(t) = \mathcal{U}_v(\hat{u}_t,\hat{x}_t,x_t) =\hat{u}(t)$ 
	and consider simulation relations $\mathscr{R}_i$
	\begin{equation}
		\mathscr{R}_i:=\left\{(\hat{x} , x)\in  \Xh\times  \X \mid ||x-\hat{x}||_D \leq \epsilon_i \right\},
		\label{eq:simrel}
	\end{equation}
	where 
	$||x||_D = \sqrt{x^TDx}$  with $D$ a symmetric positive definite matrix  $D=D^T \succ 0$. We use the same weighting matrix $D$ for all simulation relations $\mathscr{R}_i$, with $i\in \left\{1,2, \dots N_R \right\}$. The simulation relations in \eqref{eq:simrel} have an ellipsoidal shape 
	and for fixed precision such ellipsoids are illustrated in Fig.~\ref{fig:fixPrec}. A switch from one simulation relation to the other one is similarly illustrated in Fig.~\ref{fig:VarPrec}.
	
	For these relations \eqref{eq:simrel}, condition 1 in Definition \ref{def:HybSimRel} is satisfied by choosing weighting matrix $D \succ 0$, such that 
	\begin{equation}
		C^TC\preceq D.
		\label{eq:LMI_D}
	\end{equation}
	
	We can now construct kernels $\Wker_{ij}$ using a coupling compensator as introduced in \cite{huijgevoort2022similarity}. By doing so, condition 2 of Definition~\ref{def:HybSimRel} can be quantified via contractive sets for the error dynamics $x_{t+1}-\hat{x}_{t+1}$ based on the combined transitions \eqref{eq:combeqwiths}.
	We assume that there exist factors $\alpha_{ij} \in [0,1]$ with $\epsilon_j = \alpha_{ij}\epsilon_i$ that represent the set contraction between the different simulation relations.
	Now, we can describe the satisfaction of condition 2 as a function of $\delta_{ij}, \alpha_{ij}$ and $\epsilon_{i}$.
	\begin{lemma}\label{lem:epsdelBounds}
		Consider models $\M$ in \eqref{eq:model_MLLTI} and $\Mh$ in \eqref{eq:model_MLAbsLTI} for which simulation relations $\mathscr{R}_i$ 
		as in \eqref{eq:simrel}  are given with weighting matrix $D$ satisfying \eqref{eq:LMI_D}. Given $\delta_{ij}, \alpha_{ij},$ and $ \epsilon_i$, if there exist parameters $\lambda_{ij}$ and matrices $F_{ij}$ such that  the matrix inequalities  \begin{subequations}\label{eq:LMI_delta}\begin{align}
				& \hspace{-.5cm}\begin{bsmallmatrix}
					\frac{1}{\epsilon_i^2}D & F_{ij}^T\\
					F_{ij} & r_{ij}^2I
				\end{bsmallmatrix}\succeq 0,\hspace{2.5cm}\mbox{\itshape \small (input bound) } \\
				&  \hspace{-.5cm}\begin{bsmallmatrix}
					\lambda_{ij} D & \ast & \ast   \\
					0 & (\alpha_{ij}^2-\lambda_{ij})\epsilon_i^2 & \ast \\ 
					D(A+B_wF_{ij}) & D\beta_l & D
				\end{bsmallmatrix} \succeq 0  \mbox{  \itshape \small  (contraction) }
		\end{align}\end{subequations}
		are satisfied, then there exists a $\Wker_{ij}$ such that condition 2 in Definition \ref{def:HybSimRel} is satisfied. The matrix inequalities in  \eqref{eq:LMI_delta}  are parameterized with $\lambda_{ij}>0$ and should hold for all $\beta_l \in \textrm{vert}(\mathscr{B})$. Furthermore, $r_{ij}$  is computed as a function of $\delta_{ij}$, that is $r_{ij} = |2\idf \left( \frac{1-\delta_{ij}}{2} \right) |$, with $\idf$ denoting the inverse distribution function of the standard Gaussian distribution.
		\hfill\(\Box\)
	\end{lemma}
	
	This lemma allows us to conclude the following.
	\begin{theorem}\label{th:theorem}
		Consider models $\M$ in \eqref{eq:model_MLLTI} and $\Mh$ in \eqref{eq:model_MLAbsLTI}   
		for which simulation relations $\mathscr{R}_i$ as in \eqref{eq:simrel} are given with weighting matrix $D$ satisfying \eqref{eq:LMI_D}.
		If the inequalities \eqref{eq:LMI_delta} hold for all $i,j \in \left\{1,\dots, N_R\right\}^2$ and  
		there exists $i\in \left\{1, \dots, N_R\right\}$ such that $(\hat{x}_0,x_0)\in\mathscr{R}_i$
		then $\Mh$ is stochastically simulated by $\M$ in a multi-layered fashion as in Definition~\ref{def:HybSimRel}, denoted as $\Mh\preceq_{\boldsymbol{\epsilon}}^{\boldsymbol{\delta}} \M$.
	\end{theorem}
	
	\begin{proof}
		The proof of both Lemma \ref{lem:epsdelBounds} and Theorem \ref{th:theorem} build on top of the proofs of Theorem~8 and Theorem~9 in \cite{huijgevoort2022similarity} for controlled invariant sets. Instead of invariant sets, the proof uses contractive sets to handle the multi-layered simulation relation. 
		
		For the construction of the matrix inequalities in \eqref{eq:LMI_delta}, we follow \cite{huijgevoort2022similarity} and model the state dynamics of the abstract model \eqref{eq:model_MLAbsLTI} as
		$\hat{x}(t+1) = A\hat{x}(t)+B\hat{u}(t)+B_w(\hat{w}_{\gamma}(t)-\gamma(t))+\beta(t)$
		with disturbance $\hat{w}_\gamma \in\W \subseteq \mb{R}^{n_w}$, shift $\gamma \in \Gamma$ and deviation $\beta \in \mathscr{B}$. The disturbance is generated by a Gaussian distribution with a shifted mean, $\hat{w}_{\gamma}\sim\mathcal{N}(\gamma,I)$. The $\beta$-term pushes the next state towards the representative point of the grid cell. Based on \cite{huijgevoort2022similarity}, we choose stochastic kernels $\Wker_{ij}$ such that the probability of event $w-\hat{w}_\gamma =0$ is large. The error dynamics conditioned on this event equal
		$x_\Delta^+  = Ax_\Delta(t)+B_w\gamma_{ij}(t)-\beta(t)$,
		where state $x_\Delta$ and state update $x_\Delta^+ $ are the abbreviations of $x_\Delta(k):=x(t)-\hat{x}(t)$ and $x_\Delta(t+1)$, respectively. This can be seen as a system with state $x_\Delta$, constrained input $\gamma_{ij}$, and bounded \mbox{disturbance $\beta$.}
		
		For a given deviation $\delta_{ij}$, we compute a bound on the allowable shift as
		$\gamma_{ij} \in \Gamma_{ij} := \left\{ \gamma_{ij} \in \mb{R}^{n_w} \mid || \gamma_{ij} || \leq r_{ij}\right\}$ and we parameterize the shift $\gamma_{ij}=F_{ij}x_\Delta$ with the matrix $F_{ij}$. 
		In the exact same fashion as the proof of Theorem~9 in \cite{huijgevoort2022similarity}, 
		we can show that if there exists $\lambda_{ij}$ and $F_{ij}$ such that the matrix inequalities in \eqref{eq:LMI_delta} are satisfied, then the following implications also hold
		\begin{align*}
			& x_\Delta^\top D x_\Delta \leq \epsilon_i^2 \implies x_\Delta^\top F_{ij}^\top F_{ij} x_\Delta \leq r_{ij}^2 \hspace{1cm}\mbox{\itshape \small (input bound) } \\
			&  x_\Delta^\top D x_\Delta \leq \epsilon_i^2 \implies (x_\Delta^+)^\top D x_\Delta^+  \leq \alpha_{ij}^2\epsilon_i^2. \hspace{.8cm}\mbox{\itshape \small (contraction) }
		\end{align*}
		Therefore, we satisfy the bound $\gamma_{ij}\in\Gamma_{ij}$ and the simulation relation $\mathscr{R}_i$ describes an $\alpha_{ij}$-contractive set. Hence, using Lemma~6 in \cite{huijgevoort2022similarity}, we can conclude that there exists a kernel $\Wker_{ij}$, such that condition 2 in Definition~\ref{def:HybSimRel} is satisfied. 
		Since condition 1 in Definition~\ref{def:HybSimRel} is already satisfied by choosing $D$ appropriately, $\Mh \preceq_{\boldsymbol{\epsilon}}^{\boldsymbol{\delta}} \M$ holds as long as the conditions in Theorem~\ref{th:theorem} are satisfied.
		
		Concluding, since \eqref{eq:LMI_D} holds, condition 1 in Definition~\ref{def:HybSimRel} is satisfied for all $i,j$. If in addition $\lambda_{ij}$ and $F_{ij}$ satisfy \eqref{eq:LMI_delta}, then there exists a kernel $\Wker_{ij}$ such that condition 2 in Definition~\ref{def:HybSimRel} holds (Lemma~\ref{lem:epsdelBounds}). Once this does not only hold for a specific $i,j$, but for all $i,j \in [1, \dots, N_R]$ and there exists $i\in\left\{1,2,\dots,N_R\right\}$ with $(\hat{x}_0,x_0) \in \mathscr{R}_i$, then we have $\Mh \preceq_{\boldsymbol{\epsilon}}^{\boldsymbol{\delta}} \M$.
	\end{proof}
	
\section{Derivation for the ellipsoidal set of the waypoint model} \label{app:deriveEps}

We want to find $\epsilon_w$, or equivalently the ellipsoidal set $\mc{E}$ as in \eqref{eq:epsSet}, such that for all states corresponding to the ellipsoid around the waypoint, the next states remain inside the ellipsoid of the next waypoint, with a probability of $\Delta_w(x_w,x_w')$. Mathematically, this can be formulated as follows:
\begin{align*}
	& &	\forall x \in \mc{E}(x_w), \forall u_w \in \U_w: & x'\in \mc{E}(x_w'),
\end{align*} with probability $1-\Delta_w(x_w,x_w')$. This is equivalent to
\begin{align}\label{eq:simrelCond3_temp}
	& \forall x \in \mc{E}(x_w), \forall u_w \in \U_w: \\
	& || x' - x_w' ||_{D_w}  = ||\bar{A}(x-x_w) + B_ww||_{D_w
	} \leq \epsilon_w,
\end{align} with $\bar{A} = A+BK$. Here, the interface function $u = u_w + K(x-x_w)$ has been substituted. Using standard properties of the vector norm \cite[Chapter~1]{Belitskii1988matrix}, we find the following inequalities
\begin{align*}
	& 	||\bar{A}(x-x_w) + B_ww||_{D_w}  \leq 	||\bar{A}(x-x_w)||_{D_w} + ||B_ww||_{D_w} \notag \\
	&  \leq 	||\bar{A}||_{D_w}||(x-x_w)||_{D_w} + ||B_w||_{D_w} ||w||_{D_w}.
\end{align*} For all states $x_w \in \mc{E}(x_w)$ this implies that 
\begin{align}\label{eq:upperbound_temp1}
	& & || x' - x_w' ||_{D_w} & \leq ||\bar{A}||_{D_w}\epsilon_p + ||B_w||_{D_w} ||w||_{D_w}.
\end{align} For a given probability $\Delta_w(x_w,x_w')$ and weighting matrix $D_w = d_wI_{n_x}$, with scalar $d_w$, we can compute a bound on $w$ using the Chi-squared distribution with $n_w$ degrees of freedom \cite{Chew1966ConfidenceMVN}. Hence, we obtain
\begin{equation} \label{eq:wSet}
	||w||_{D_w} \leq \frac{1}{d_w} \sqrt{r_w}, \text{ with } r_w = \chi^{-1}(\Delta_w(x_w,x_w') | n_w),
\end{equation} with $\chi^{-1}(\cdot | n_w)$ denoting the inverse cumulative distribution function of the Chi-squared distribution with $n_w$ degrees of freedom. Intuitively, this means that for a random variable $w \sim \mc{N}(0,I)$, we compute the ellipsoidal set \eqref{eq:wSet} containing $w$ with a confidence equal to $\Delta_w(x_w,x_w')$. Substituting in \eqref{eq:upperbound_temp1}, we get 
\begin{align*}
	& & || x' - x_w' ||_{D_w} & \leq ||\bar{A}||_{D_w}\epsilon_p + ||B_w||_{D_w} \frac{1}{d_p} \sqrt{r_w}.
\end{align*}
Hence, to achieve \eqref{eq:simrelCond3_temp} 
$\epsilon_w$ should equal 
\begin{equation}
	\epsilon_w = |-(||\bar{A}||_{D_w}-1)^{-1}||B_w||_{D_w}\frac{1}{d_w} \sqrt{r_w}|.
\end{equation}

\end{document}